\documentclass{aastex} 

\usepackage{natbib}
\usepackage{lscape}
\usepackage[caption=false]{subfig} 
\begin{document}


\title{UV to FIR catalogue of a galaxy sample in nearby clusters: SEDs
  and environmental trends}


\author{Jonathan D. Hern\'andez-Fern\'andez\altaffilmark{1},
  J. Iglesias-P\'aramo\altaffilmark{1,2}, and
  J. M. V\'ilchez\altaffilmark{1}}

\altaffiltext{1}{Instituto de Astrof\'isica de Andaluc\'ia, Glorieta
  de la Astronom\'ia s/n, 18008 Granada; jonatan@iaa.es}

\altaffiltext{2}{Centro Astron\'omico Hispano Alem\'an C/ Jesús Durbán
  Rem\'on, 2-2 04004 Almer\'ia}

\begin{abstract}

In this paper, we present a sample of cluster galaxies devoted to
study the environmental influence on the star-formation activity. This
sample of galaxies inhabits in clusters showing a rich variety in
their characteristics and have been observed by the SDSS-DR6 down to
\mbox{$M_B$$\sim$-18} and by the GALEX AIS throughout sky regions
corresponding to several megaparsecs. We assign the broad-band and
emission-line fluxes from ultraviolet to far-infrared to each galaxy
performing an accurate spectral energy distribution for spectral
fitting analysis. The clusters follow the general X-ray luminosity
vs. velocity dispersion trend of
\mbox{$L_X$$\propto$$\sigma_{c}^{4.4}$}. The analysis of the
distributions of galaxy density counting up to the 5th nearest
neighbor $\Sigma_{5}$ shows: (1) the virial regions and the cluster
outskirts share a common range in the high density part of the
distribution. This can be attributed to the presence of massive galaxy
structures in the surroundings of virial regions (2) The virial
regions of massive clusters \mbox{($\sigma_{c}$$>$550 km s$^{-1}$)}
present a $\Sigma_{5}$ distribution statistically distinguishable
($\sim$96\%) from the corresponding distribution of low-mass clusters
\mbox{($\sigma_{c}$$<$550 km s$^{-1}$)}. Both massive and low-mass
clusters follow a similar density-radius trend, but the low-mass
clusters avoid the high density extreme. We illustrate, with Abell
1185, the environmental trends of galaxy populations. Maps of sky
projected galaxy density show how low-luminosity star-forming galaxies
appear distributed along more spread structures than their giant
counterparts, whereas low-luminosity passive galaxies avoid the
low-density environment. Giant passive and star-forming galaxies share
rather similar sky regions with passive galaxies exhibiting more
concentrated distributions.

\end{abstract}


\section{Introduction}

The clusters of galaxies are excellent laboratories to study the
influence of the environment on galaxies. This influence is formed by
environmental processes which are combinations of interaccions of
galaxies with other components of the Universe; galaxies, dark matter
and plasma. The highest peaks of density in the spatial distribution
of these components are in the cores of galaxy clusters. The galaxy
population in the centers of clusters reachs up to volume densities of
10$^3$ bright galaxies per Mpc$^{3}$ on spatial scales of $\sim$1 Mpc
and those galaxies have relative velocities of several hundreds of km
s$^{-1}$ \citep{Cox_2000}. The mass of dark matter haloes of clusters
is several orders of magnitude greater than the sum of masses of the
stellar component of galaxies with mass-to-light ratios that range
from 100 to \mbox{500 $M_{\odot}$/$L_{\odot}$} \citep{Cox_2000}
opposite the mass-to-light ratio for stellar component which cover the
range \mbox{1-10 $M_{\odot}$/$L_{\odot}$} \citep{Bell_et_al_2003}. The
pressure of the intracluster medium (ICM), which with
\mbox{n$_e$$\sim$10$^{-3}$ cm$^{-3}$} and temperatures goes from
10$^{7}$ to 10$^{8}$ K is enough high to acts on the gas component of
galaxies \citep{Gunn&Gott_1972}.

Each of the interactions of galaxies with these components (galaxies,
ICM, dark matter halo) has a contribution in the different
environmental processes. The interaction of galaxies with the ICM
dominates the gas stripping processes, where the interstellar medium
of galaxies is stripped via various mechanisms, including viscous and
turbulent stripping \citep{Toniazzo&Schindler_2001}, thermal
evaporation \citep{Cowie&Songaila_1977} and ram pressure stripping
\citep{Quilis_et_al_2000}. The tidal interactions among galaxies
dominates the galaxy mergers or strong galaxy-galaxy interactions
\citep{Mihos_2004} and the galaxy harassment
\citep{Moore_et_al_1996,Moore_et_al_1998,Moore_et_al_1999}. The
environmental process known as {\it{strangulation}},
{\it{starvation}}, or {\it{suffocation}} is dominated by the tidal
interaction with the dark matter halo of the cluster which removes the
hot gas halo of the galaxy \citep{Bekki_et_al_2002}. The environmental
processes act on the stellar and gas/dust components of a galaxy
modifying its gas content, the star formation level, the structural
and dynamical parameters, etc. In one side, the intensity of the
environmental processes depend on galaxy properties like the stellar
mass or the compactness of stellar component. Also, the environmental
influence depends on the environmental conditions and/or the cluster
properties as the density of cluster components (galaxies, ICM, dark
matter halo), the velocity field of the cluster, etc. Specifically,
there is a controversy about the dependence on global cluster
properties (e.g. $\sigma_{c}$) of the star formation activity of
cluster galaxy population. Numerous works point out there is no such
correlation
\citep{Smail_et_al_1998,Andreon&Ettori_1999,Ellingson_et_al_2001,Fairley_et_al_2002,De_Propris_et_al_2004,Goto_2005,Wilman_et_al_2005,Andreon_et_al_2006}
while other works claim the presence of a relation between the star
formation activity and the global cluster properties
\citep{Martinez_et_al_2002,Biviano_et_al_1997,Zabludoff&Mulchaey_1998,Margoniner_et_al_2001}.

The cores of galaxy cluster are located around the peaks of densities
of these components but the volume density of cluster components
converges to the field value towards regions outside the virial
regions in distances of some virial radii
\citep{Cox_2000,CAIRNS_I}. So, the transition between the cluster
centers and the surroundings samples a broad range in environmental
properties. The environmental processes act on galaxies with different
intensity depending on the galaxy (dynamical or stellar) mass or
luminosity \citep[see][for a review]{Boselli&Gavazzi_2006}, but in the
most of previous works the observed trends of galaxy properties are
restricted to giant L$\gtrsim$L${\ast}$ galaxies. The UV luminosity
has revealed as a good proxy of the recent star formation rate because
is a tracer of the more short-lived stars $\tau$$<$10$^8$ yr
\citep{Kennicutt_1998} and the UV-optical colors as an excellent
classifier between passive evolving galaxies and star-forming galaxies
\citep{Chilingarian&Zolotukhin_2011}. On the other hand, the optical
and near-infrared spectal ranges sample stellar populations with ages
which go from 10$^{9}$ to 10$^{10}$ years
\citep{Kennicutt_1998,Martin_et_al_2005}. This give us some important
insights into the global star formation history of a galaxy i.e. the
stellar mass, the time-scale of the star formation history, etc.

Following the former considerations, we will design a sample of
clusters nearby enough for their galaxies be observed around the
classical luminosity limit between giant and dwarf galaxies
$M_{B}$=-18 by the DR6 of SDSS. We stress the cluster galaxy
population must be observed by differents surveys from UV to FIR in
the central regions of each cluster and its surruondings up to several
times the size of virial region. This cluster sample allow us to study
the environmental behavior of different properties (current star
formation, stellar mass, attenuation, etc) of a galaxy population with
a broad luminosity range inhabits environments as different as the
center of galaxy clusters or their surroundings.

The remainder of the paper is organized as follows. In section
\ref{sec:clust_samp}, we describe the design of the cluster sample. In
section \ref{sec:spph_catalog}, we describe the compilation of
broad-band and emission line fluxes for the galaxy sample of the
cluster sample. In section \ref{sec:catalogue_sec}, we show the
compilation of fluxes for the galaxy sample, color-color distributions
and an example of the SED of a galaxy from the sample. In section
\ref{sec:discussion} we disccused three different items: the
bolometric X-ray luminosity vs. cluster velocity dispersion
$L_{X}$-$\sigma_{c}$ relation, the local density $\Sigma_{5}$
distribution of galaxy population split by their membership to virial
regions of low-mass/massive cluters and as a hint for future work and
the sky projected density of giant/low-luminosity and
passive/star-forming galaxy population in a massive cluster. We
summarized our findings in section \ref{sec:conclusions}

\section{Cluster sample selection}
\label{sec:clust_samp}

One of the purpose of the sample design is embrace a luminosity range
for the cluster galaxy sample as wide to contain the classical limit
between giant and dwarf galaxies, $M_B$=-18. This constrains the
redshift range of the cluster sample. The cluster sample is observed
in a sky area wihich is delimited by the intersection of observed sky
areas of SDSS and GALEX surveys. Both surveys are not completed (at
the moment of sample definition, March 2008) and have smaller observed
sky areas than the other ended surveys, 2MASS and IRAS. In order to
sample a broad range of environments, we select galaxy clusters
observed by these surveys up to regions several virial radius beyond
the virial region. So, we discard those clusters with a poor sky
coverage not only in the central regions but even in the outskirts of
clusters.

In the following, we describe the process to build the cluster
sample. In a first step, we take a compilation of Galaxy Clusters from
NED \footnote{NASA/IPAC Extragalactic Database,
  {\it{http://nedwww.ipac.caltech.edu/}}}. Thanks to this approach, we
take account all cluster selection criteria in the literature; visual
inspection, image-smoothing techniques, X-ray extended sources
detection, Red Sequence algorithm, surveys around cD galaxies,
etc. This avoid any kind of bias in the cluster selection. We have
selected all astrophysical objects with NED Object Type set to
{\bf{GClstr}}. 

We constrain the redhisft range to reach down the absolute magnitude
limit of dwarf galaxies. The Main Galaxy Sample of SDSS reach up to
$r'_{MGS}$$\sim$17.77 \citep{Strauss_et_al_2002}, while the absolute
magnitude limit for dwarf galaxy starts at $M_{B}^{Dwarf}$=-18
\citep{LF_review,Mateo_1998}, so:

\begin{displaymath}
\left\{ \begin{array}{l} 
R = r'_{MGS} - 0.1837(g-r) - 0.0971 ~~\textrm{\citep{Lupton_2005}} \nonumber \\
M_{R} \equiv M_{B}^{Dwarf} - (B-R) \nonumber \\
\mu \equiv R - M_{R} \nonumber \\ 
log{\it{z}} = 0.2 \mu - 8.477 + log{\it{h}} \textrm{ ( Local Universe,
  i.e. c{\it{z}}$\approx$H{\it{D}}=100h{\it{D}} )} \label{logz} 
\end{array} \right.
\end{displaymath}

with $B$, $R$ apparent Johnson magnitudes; $M_{B}$, $M_{R}$ absolute
Johnson magnitudes; $\mu$ distance modulus; $H$$\equiv$100$h$ with $H$
the Hubble's constant and {\it{z}} redshift. We assume $h$=0.7 in this
work. Assuming the ($B$-$R$) values observed by
\citet{Mobasher_et_al_2003} and the ($g$-$r$) values observed by
\citet{Blanton_et_al_2003} for red and blue galaxies, we obtain a
upper limit in redshift of:

\begin{displaymath}
\left\{ \begin{array}{l}
\textrm{Blue galaxies: }(B-R)\approx0.8, \; ($g-r$)\approx0.2 \Rightarrow {\it{z}}\approx0.044 \\
\textrm{Red galaxies: } (B-R)\approx2.0, \; ($g-r$)\approx1.0 \Rightarrow {\it{z}}\approx0.071 \\  
\end{array} \label{zcut}  \right.
\end{displaymath}

Then, we choose $z$=0.05 as the upper limit in redshift as a
compromise between red and blue galaxies and initially start with a
cluster sample from $z$=0 to $z$=0.05. This initial sample contains
1575 clusters.

We check by eye the distribution of the SDSS plates and the GALEX
fields for this cluster sample over a sky region up to a projected
radius of some Abell radius \citep{Abell_1958} from the center of each
cluster. After that, we set the lower limit in redshift for the
cluster sample to $z$=0.02 because this redshift limit is enough to
cover the sky area of a typical galaxy cluster with only a few SDSS
plates (1.5 deg radius) or GALEX fields (0.5 deg radius). In a second
step, we crosscorrelated the coordinates of cluster centers reported
by NED with the position of the SDSS plates and the GALEX fields, in
order to know what the cluster centers are, at least, in one SDSS
plate and one GALEX field. This gives a cluster sample of 373 galaxy
cluster with redshift from 0.02 to 0.05. In order to get a good SDSS
sky coverage of clusters, we check by eye the sky coverage of SDSS
Main Galaxy Sample up to a projected radius of 2.2 R$_{Abell}$ from
the cluster center. For a subsequent procedure, we need an
spectroscopic galaxy sample covering an sky region with this specific
radius or more extensive. We select only those clusters with a good
SDSS sky coverage over an sky area with this size. This selection
gives a sample of 230 clusters for 0.02$<$$z$$<$0.05.

The clusters from different catalogs have different selection and
detection criteria and we do not control whether there are spurious
clusters in some of these catalogs. On the one hand, we have to clean
our cluster sample from non confident clusters and possible
artifacts. On the other hand, we need a reliable measure of the
cluster velocity dispersion, $\sigma_{c}$ in order to characterize a
cluster sample with a broad range in $\sigma_{c}$, from poor to rich
clusters. We solve this two issues using the procedure proposed by
\citet{Poggianti_et_al_2006} in their Appendix C but assuming cluster
center reported by NED instead of Bright Cluster Galaxy (BCG) as the
center of galaxy cluster.

In the first step, we select the galaxies inside 2.2 Abell radii from
the NED center and within a redshift range defined by
$\Delta$$z$=$\pm$0.015 from the cluster redshift given by NED. From
these galaxies, we estimate the cluster redshift $z_{c}$ and the
cluster redshift dispersion $\sigma_{z}$ as the median and the median
absolute deviation, respectively.  If $\sigma_{z}$ is higher than
0.0017 ($\approx$$\sigma_{c}$=500 km s$^{-1}$ at $z$=0), we set
$\sigma_{z}$ to this value. This step is useful to avoid too much
contamination from surrounding galaxy structures. Then, we computed
the radius r$_{200}$ from $z_{c}$ and $\sigma_{z}$ using the following
equation:

\begin{equation}
r_{200} = 1.73 \frac{\sigma_{c}}{1000~\textrm{km s$^{-1}$}} \frac{1}{\sqrt{\Omega_{\lambda} + \Omega_{0}(1+z_{c})^{3} }} h^{-1} ~Mpc
\end{equation} 

which is taken from \citet{Finn_et_al_2005}. First, we recomputed
z$_{c}$ and then $\sigma_{z}$ from those galaxies within
$\pm$3$\sigma_{z}$ from z$_{c}$ and more nearby to cluster center than
1.2 $r_{200}$. This process iterates until it reaches the
convergence. After each iteration, every galaxy in the initial sample
can reenter to the cluster sample whether it meets the constraints on
redshift and position. If the process does not converge, we discard
that cluster. The error of the final $\sigma_{c}$ is computed using a
bootstrap algorithm applies to the galaxy sample in the cluster.

In this procedure, there are clusters which reach the convergence and
show a final z$_{c}$ far away from NED cluster redshift or with
$\sigma_{z}$ $\gg$ 1000 km s$^{-1}$. After a visual check to radial
velocity histograms of these structures, we conclude those galaxy
structures are far from be real clusters. In order to discard those
structures, we add two constrains to the cluster sample:

\begin{eqnarray}
|z_{c} - z_{NED}|  \le 0.0033  \nonumber   \\ 
\sigma_{z}  \le 1300 \textrm{ km s$^{-1}$}  \nonumber       
\end{eqnarray}

After applying the procedure from \cite{Poggianti_et_al_2006} and
including this constrain to the former sample, the resulting sample is
composed by 86 clusters. At the end of this procedure we still impose
a further condition related to the presence of clusters with more than
one NED identifier: NED only classifies two clusters from different
catalogues as being the same cluster if their angular separation is
less than 2 arcmin (Marion Schmitz - NED team, private
communication). Using this clue, we take the cluster name from the
most ancient catalogue to identify those clusters with more than one
NED identifier.

As a final step, we visually check the GALEX AIS coverage of each
cluster up to some Abell radius. We end up with 16 clusters in the
redshift range \mbox{0.02$<$$z$$<$0.05}. Their basic properties are
listed in Table \ref{CS}. Their appearance in the sky and their radial
velocity distributions are shown in figures \ref{radec_CS1} and
\ref{czh_CS2}, respectively. Figure \ref{fig:SDSS_images1} are the
color-composite images of the central regions of clusters retrieved
from the SDSS Navigate Tool {\it
  http://skyserver.sdss.org/public/en/tools/chart/navi.asp}.

\begin{table}[p]
\caption{Main properties of the cluster sample.}
\begin{center}
\resizebox{!}{7cm}{ 
\begin{tabular}{c        c              c            c           c            c          c          c             c            c         c         }
\hline
ID$_{NED}$     & $\alpha$(J2000) & $\delta$(J2000) & $z_{med}$  & $\sigma_{c}$& $r_{200}$ & $n_{200}$ & $n_{tot}$ & $\theta_{tot}$& log(L$_X$) \\
                       & deg          & deg        &           & km s$^{-1}$ & Mpc       &         &          &  deg         & L$_{\odot}$ \\
 (1)                   & (2)          & (3)        & (4)       & (5)        & (6)       & (7)     &  (8)     &  (9)         &  (10)     \\
\hline
 UGCl 141              & 138.499      & 30.2094    & 0.0228    & 501.8      & 1.21      &   48    &  413     &  4.159       &  42.12  \\
 WBL 245               & 149.120      & 20.5119    & 0.0255    &  86.7      & 0.20      &    2    &  88      &  3.720       &  ...    \\
 UGCl 148 NED01        & 142.366      & 30.2139    & 0.0263    & 316.7      & 0.76      &   21    &  354     &  3.606       &  ...    \\
 ABELL 2199            & 247.154      & 39.5244    & 0.0303    & 756.2      & 1.83      &  313    &  1104    &  3.125       &  44.85  \\
 WBL 213               & 139.283      & 20.0403    & 0.0290    & 537.1      & 1.29      &   62    &  548     &  3.266       &  $\le$41.9  \\
 WBL 514(*)            & 218.504      & 3.78111    & 0.0291    & 633.7      & 1.52      &   88    &  580     &  3.257       &  43.18  \\
 WBL 210               & 139.025      & 17.7242    & 0.0287    & 433.3      & 1.06      &   56    &  402     &  3.298       &  43.22  \\
 WBL 234               & 145.602      & 4.27111    & 0.0291    & 243.6      & 0.58      &    6    &  87      &  3.262       &  ...    \\
 WBL 205               & 137.387      & 20.4464    & 0.0288    & 679.8      & 1.60      &   37    &  527     &  3.289       &  ...    \\
 UGCl 393              & 244.500      & 35.1000    & 0.0314    & 637.9      & 1.52      &  121    &  529     &  3.016       &  43.60  \\
 UGCl 391              & 243.352      & 37.1575    & 0.0330    & 407.0      & 0.97      &    8    &  637     &  2.874       &  ...    \\
 B2 1621+38:[MLO2002]  & 245.583      & 37.9611    & 0.0311    & 607.3      & 1.46      &   95    &  1053    &  3.046       &  43.19  \\
 UGCl 271              & 188.546      & 47.8911    & 0.0305    & 323.2      & 0.72      &   23    &  181     &  3.104       &  ...    \\
 ABELL 1185            & 167.699      & 28.6783    & 0.0328    & 789.3      & 1.90      &  228    &  754     &  2.894       &  43.58  \\
 ABELL 1213            & 169.121      & 29.2603    & 0.0469    & 565.7      & 1.35      &   98    &  305     &  2.021       &  43.77  \\
 UGCl 123 NED01        & 127.322      & 30.4828    & 0.0499    & 849.0      & 2.00      &  113    &  260     &  1.900       &  44.32  \\
\hline
\label{CS}
\end{tabular}} 
\end{center}

The description of this table is in the following page.

\end{table}

\newpage
\addtocounter{table}{-1}
\begin{table}[ht]
\caption{Main properties of the cluster sample.}

(1) NED identifier, (2) and (3) Celestial coordinates of cluster
center from NED webpage, (4) Cluster average redshift, (5) Cluster
velocity dispersion, (6) Radius 200, (7) No. of galaxies inside virial
region with SDSS redshift, (8) No. of galaxies associated to each
cluster selected by criteria exposed in section \ref{sec:sdss_data},
(9) Half size of sky square region retrieved for each cluster,
computed assuming the Local Universe approximation c$z$=H$D$, the
small-angle approximation $D_{P}$=$D$$\times$$\theta$[rad] and a
projected radius $R_{P}$=7.1 Mpc (10) Bolometric X-ray luminosity from
\citet{Mahdavi&Geller_2001} except for WBL 213
\citep{Mahdavi_et_al_2000}. (*) The historical criterion is not
applied. In the case of WBL 514, we have selected WBL 514 instead of
MKW07 because this object is split in two clusters by a late reference
\citep{Struble&Rood_1991}. The source of the data is
specified. Otherwise, the data are results from this work. The cluster
compilation was carry out from NED updated at March 28, 2008.

\end{table}

\newpage


\begin{figure}[p]
\centering
\resizebox{0.75\hsize}{!}{\includegraphics{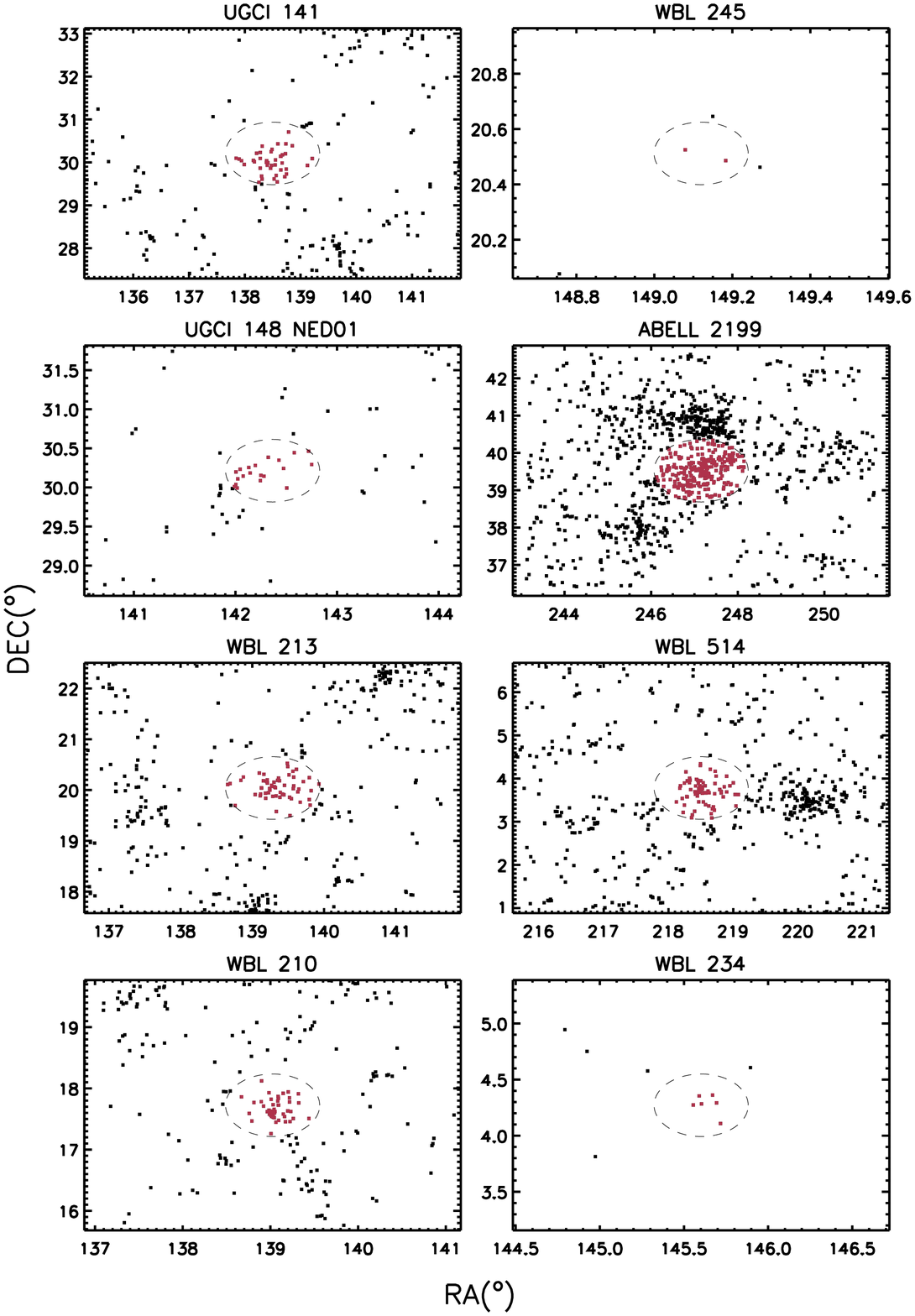}}
\caption{RA-DEC projection of the cluster sample. Ordinate axis is for
  declination and the abscissa axis is for right ascension. The red
  points correspond to galaxies in the virial region and the black
  points to the rest. All galaxies in the panels come from the DR6 of
  SDSS and are included in the cluster galaxy sample. In each panel,
  the dashed circle has a radius set to the $r_{200}$ of each
  cluster. The size of each panel is set to
  8$r_{200}$$\times$8$r_{200}$.}
\label{radec_CS1}
\end{figure}

\addtocounter{figure}{-1}
\begin{figure}[p]
\centering
\resizebox{0.75\hsize}{!}{\includegraphics{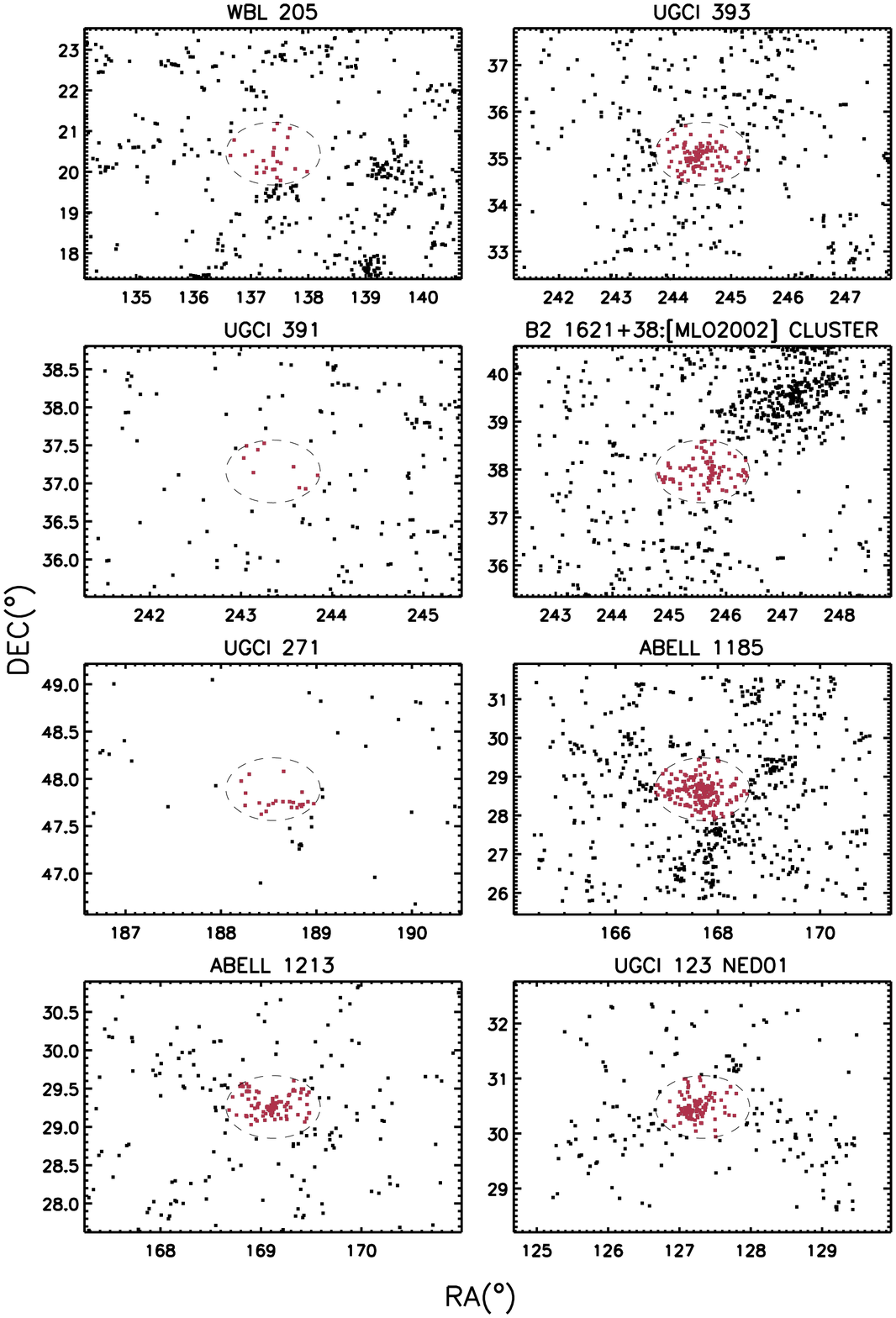}}
\caption{Continued.}
\label{radec_CS2}
\end{figure}


Taking a look to the sky distribution of clusters from the sample in
the figure \ref{radec_CS1}, it can be seen the wide variety of the
cluster sample in cluster richness and spatial structure and in some
cases, the presence of galaxy structures around the virial regions of
clusters. The richness goes from the poor cluster WBL 245 or WBL 234
with only a few galaxies in their central regions to the massive
cluster ABELL 2199, which is assembled in the supercluster ABELL 2197
- ABELL 2199 - B2 1621+38:[MLO2002] or the cluster ABELL 1185, with
clear evidence of galaxy structures as filaments. There are apparently
"isolated'' clusters as UGCl 271 or UGCl 148 NED01 opposite the
example of WBL 514 with a close "twin'' cluster, WBL 518
\citep{Beers_et_al_1995}.

\begin{figure}[p]
\centering
\resizebox{0.75\hsize}{!}{\includegraphics{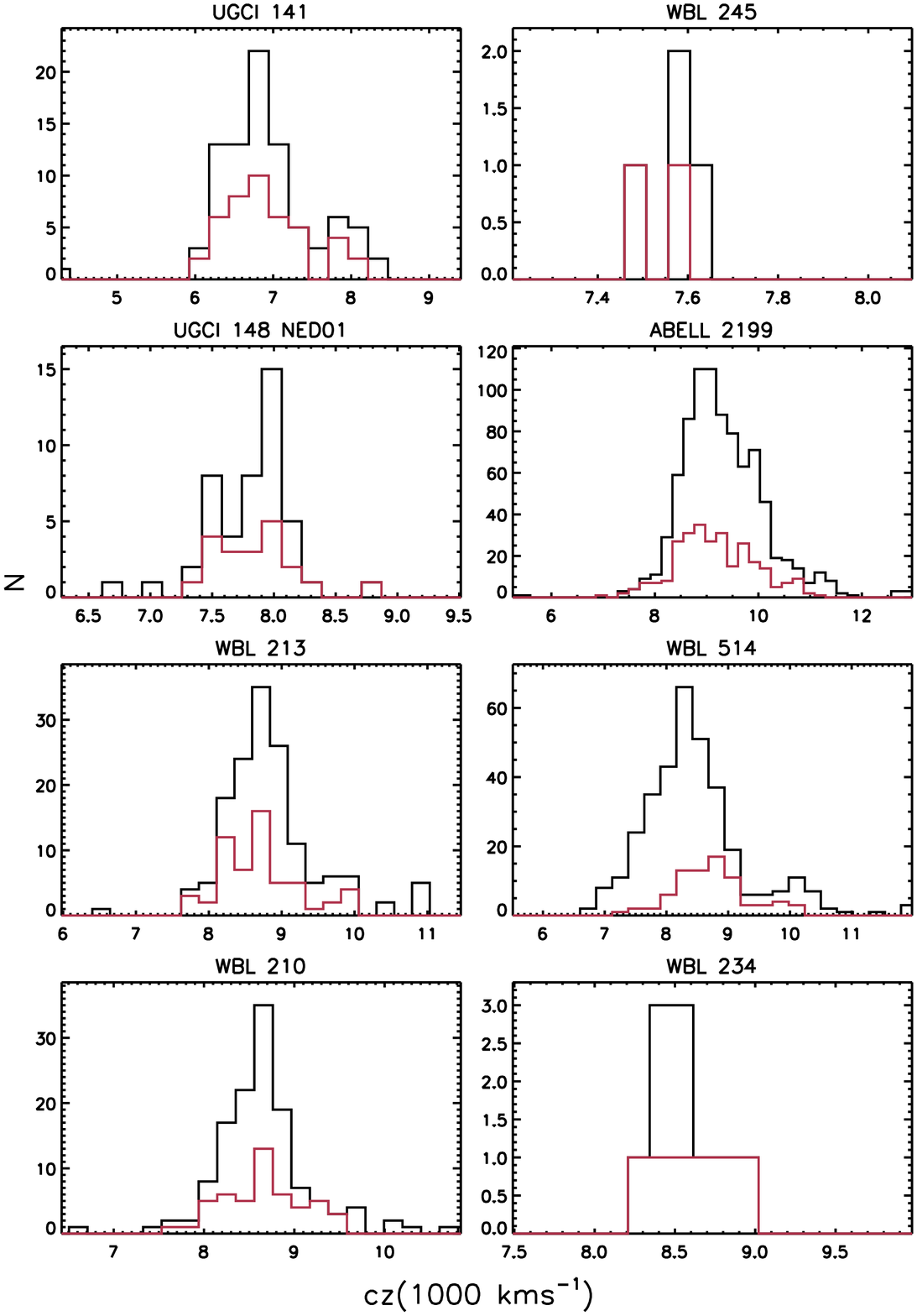}}
\caption{Radial velocity histograms for the cluster sample. The black
  histograms represent the galaxy sample inside a projected radius
  $R_{P}$ three times the virial radius $R_{P}$$<$3$r_{200}$ and the
  red histograms correspond to those galaxies inside a projected
  radius set to one virial radius $R_{P}$$<$$r_{200}$. The range of
  abscissa in each panel is set to
  c$z_{c}$-5$\sigma_{c}$$<$c$z$$<$c$z_{c}$+5$\sigma_{c}$ of each cluster.}
\label{czh_CS1}
\end{figure}

\addtocounter{figure}{-1}
\begin{figure}[p]
\centering
\resizebox{0.75\hsize}{!}{\includegraphics{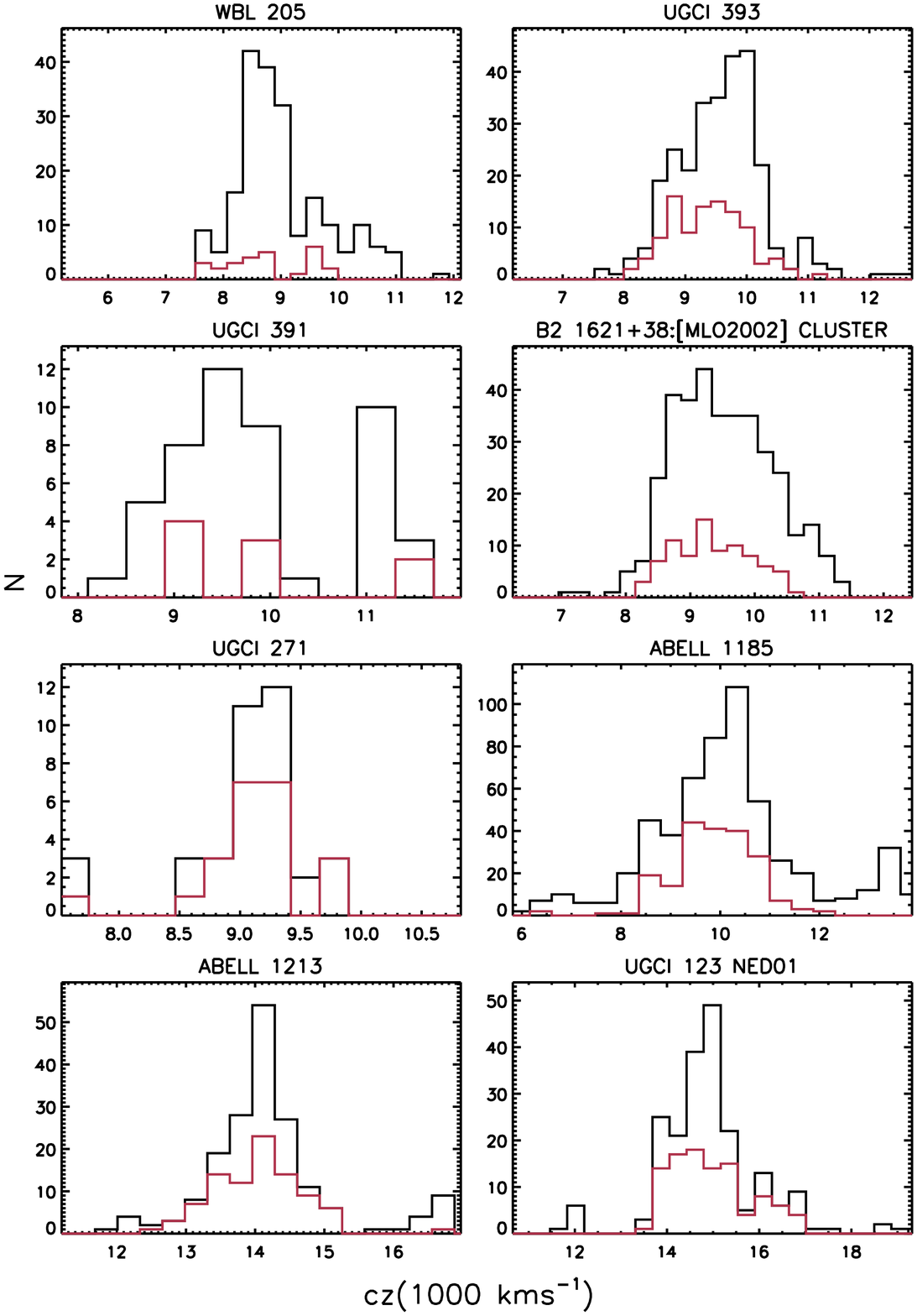}}
\caption{Continued.}
\label{czh_CS2}
\end{figure}


\newpage
\begin{figure}[p]
\centering

\subfloat[UGCl 141]{\label{fig:UGCl_141_SDSS}\includegraphics[height=0.20\textheight]{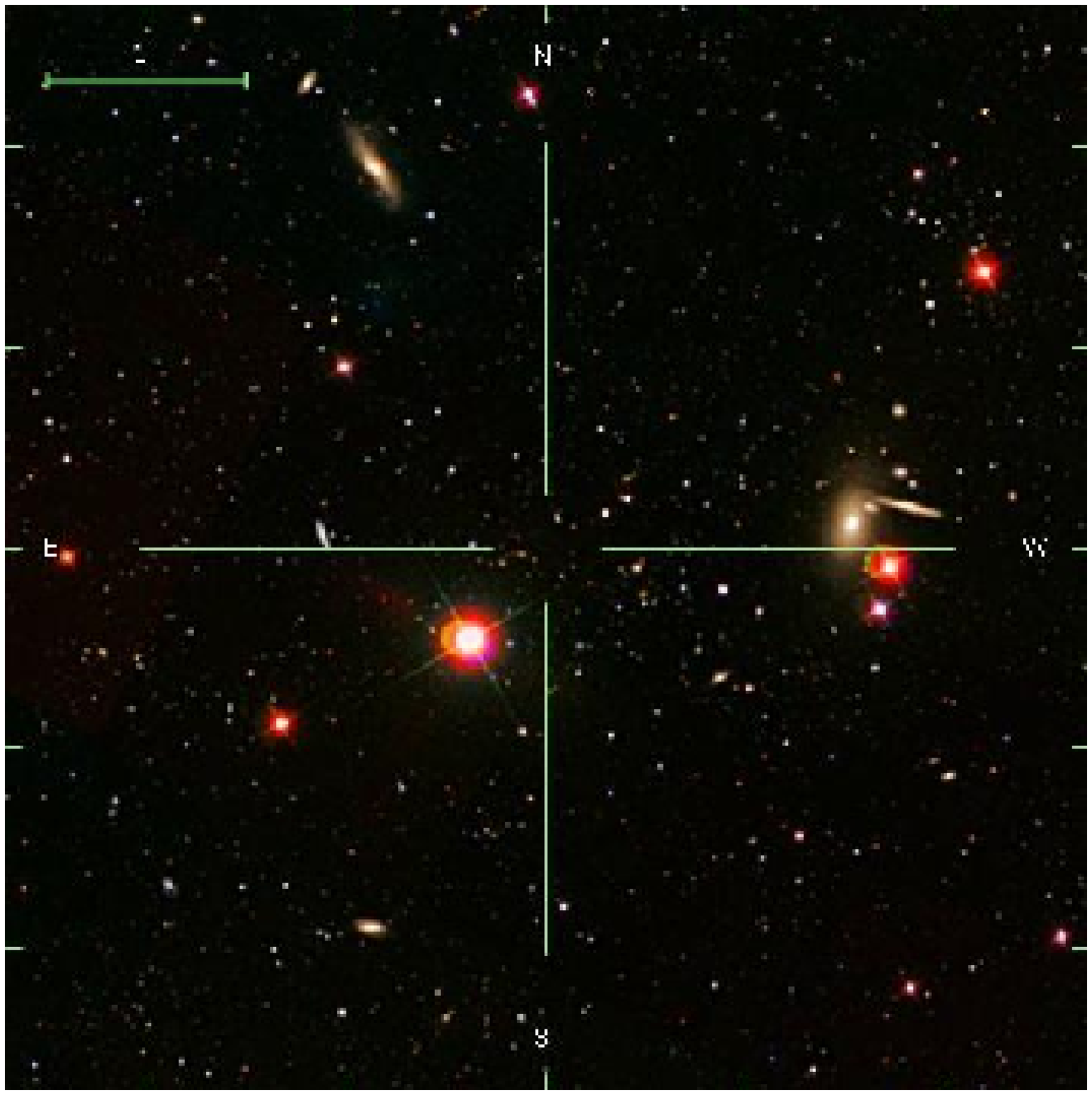}}
\hspace{0.05\textwidth} 
\subfloat[WBL 245]{\label{fig:WBL_245_SDSS}\includegraphics[height=0.20\textheight]{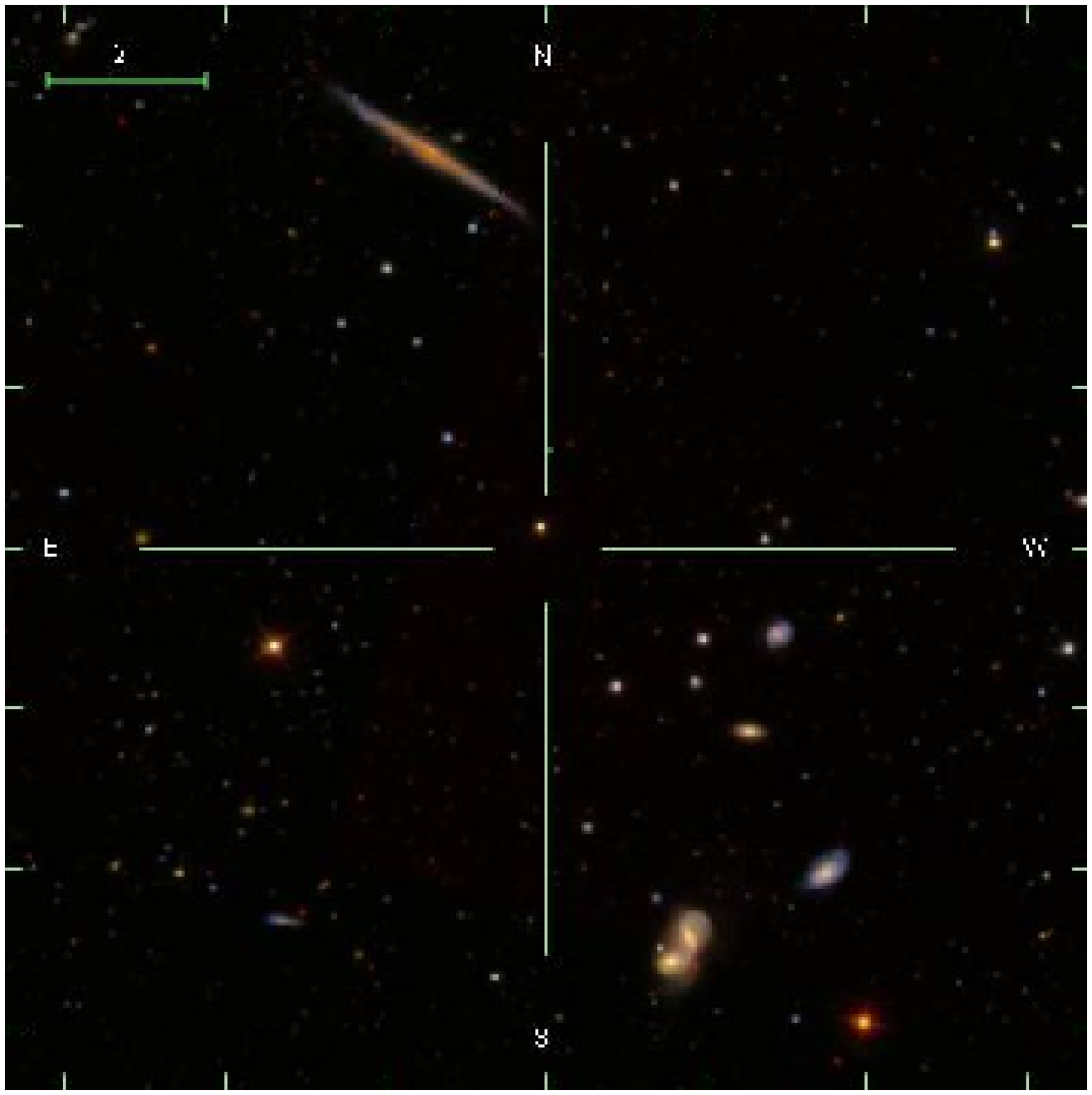}}

\subfloat[UGCl 148]{\label{fig:UGCl_148_NED1_SDSS}\includegraphics[height=0.20\textheight]{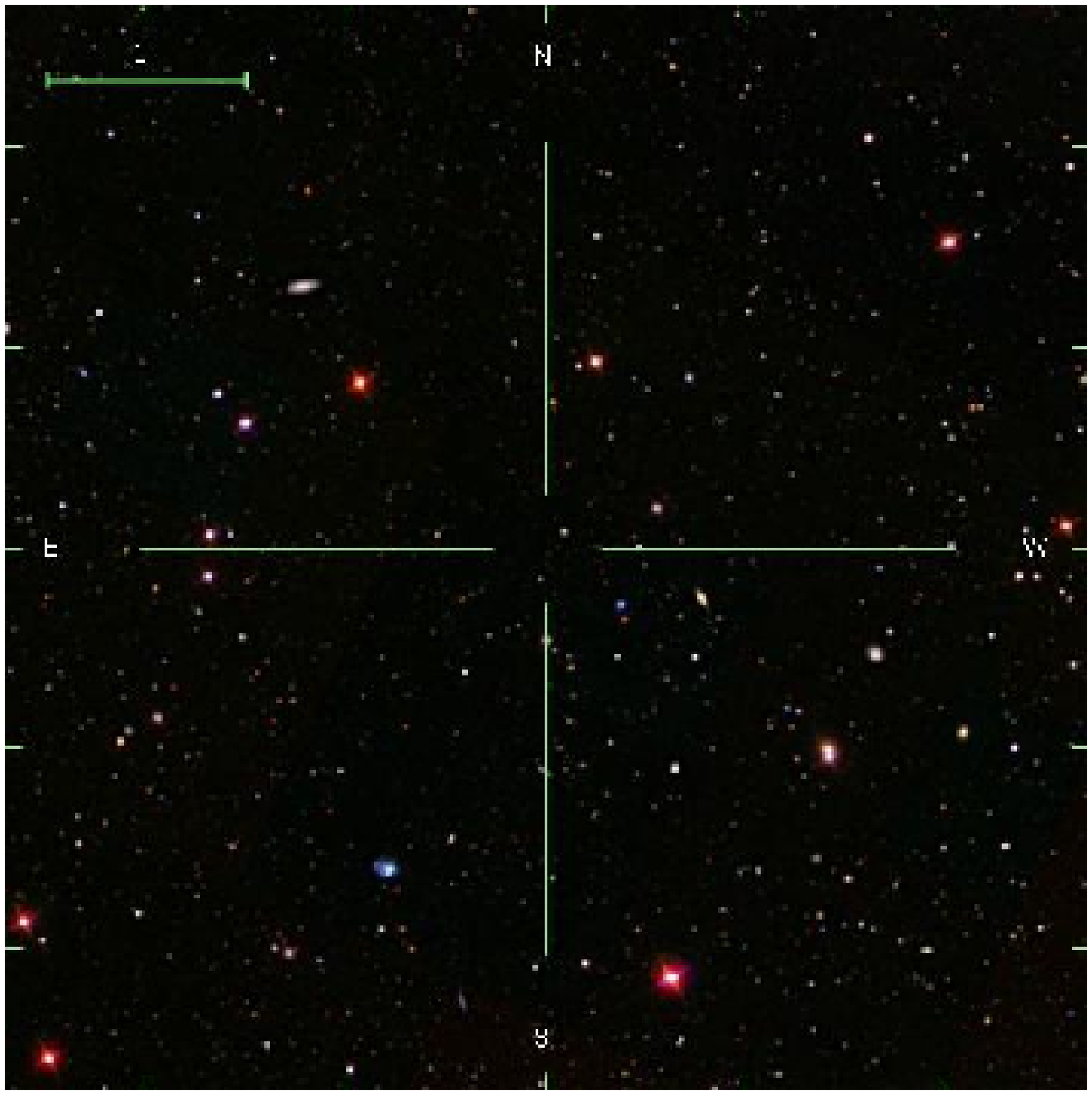}}
\hspace{0.05\textwidth} 
\subfloat[ABELL 2199]{\label{fig:ABELL_2199_SDSS}\includegraphics[height=0.20\textheight]{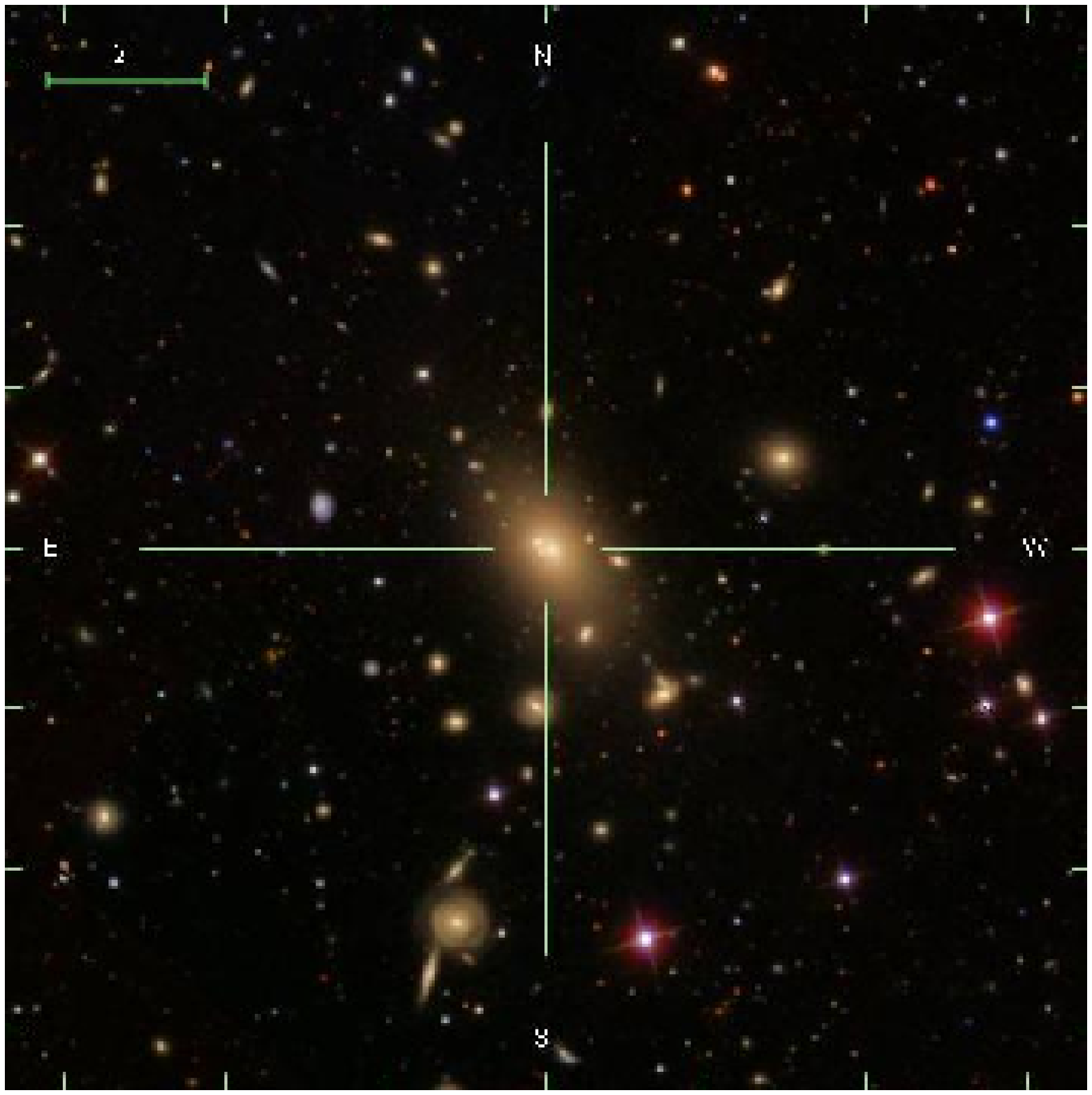}}

\subfloat[WBL 213]{\label{fig:WBL_213_SDSS}\includegraphics[height=0.20\textheight]{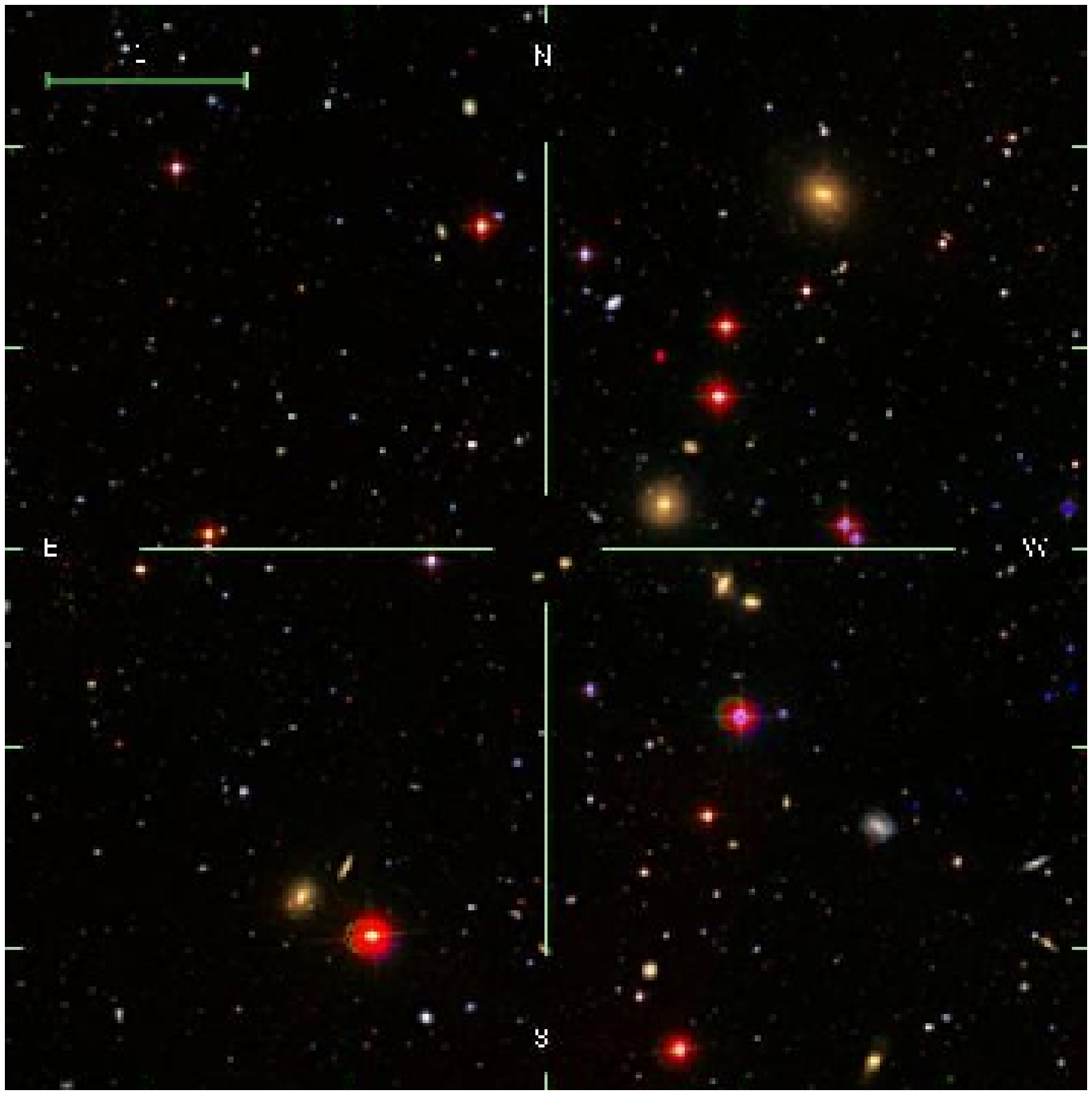}}
\hspace{0.05\textwidth} 
\subfloat[WBL 514]{\label{fig:WBL_514_SDSS}\includegraphics[height=0.20\textheight]{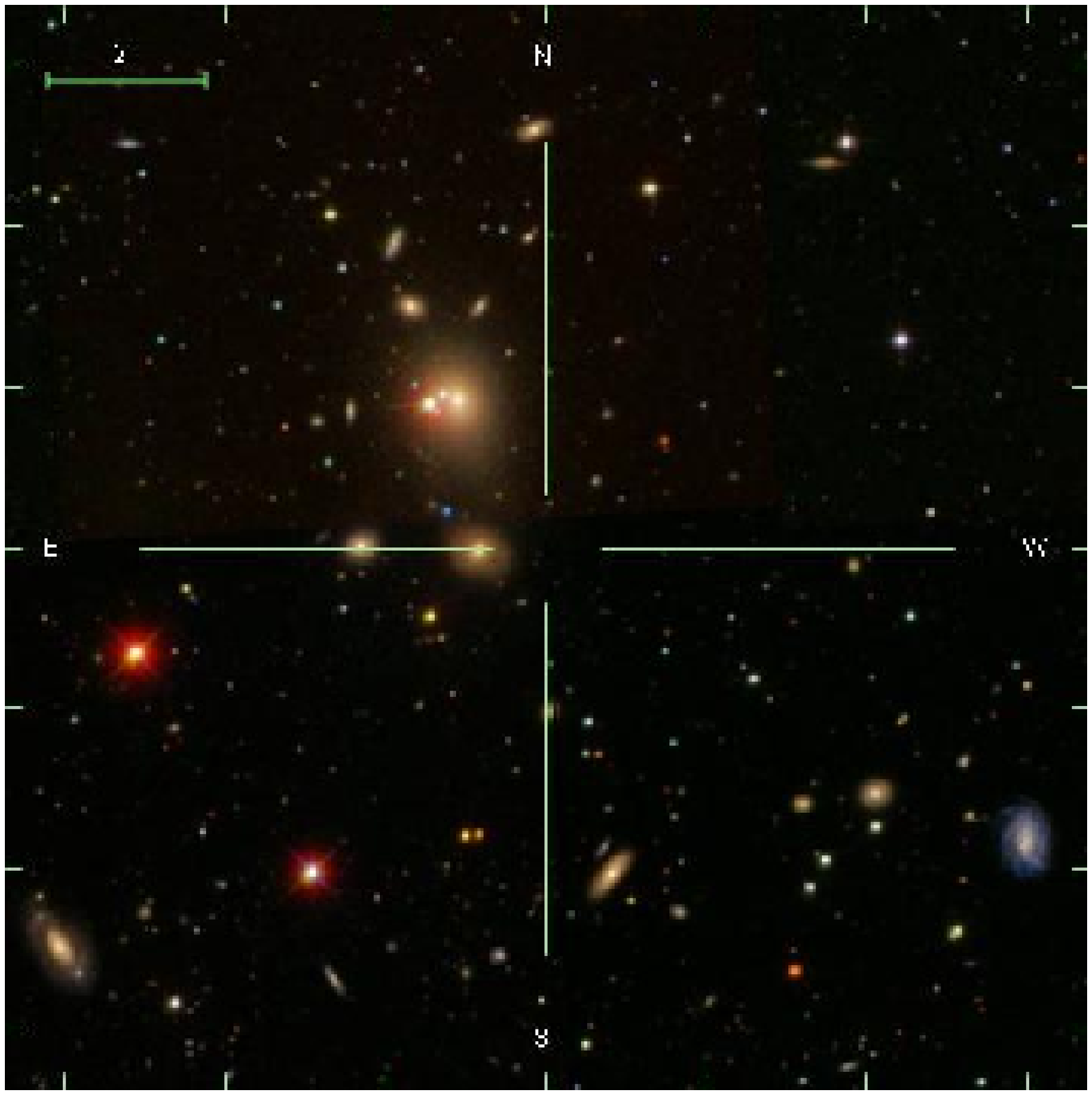}}

\subfloat[WBL 210]{\label{fig:WBL_210_SDSS}\includegraphics[height=0.20\textheight]{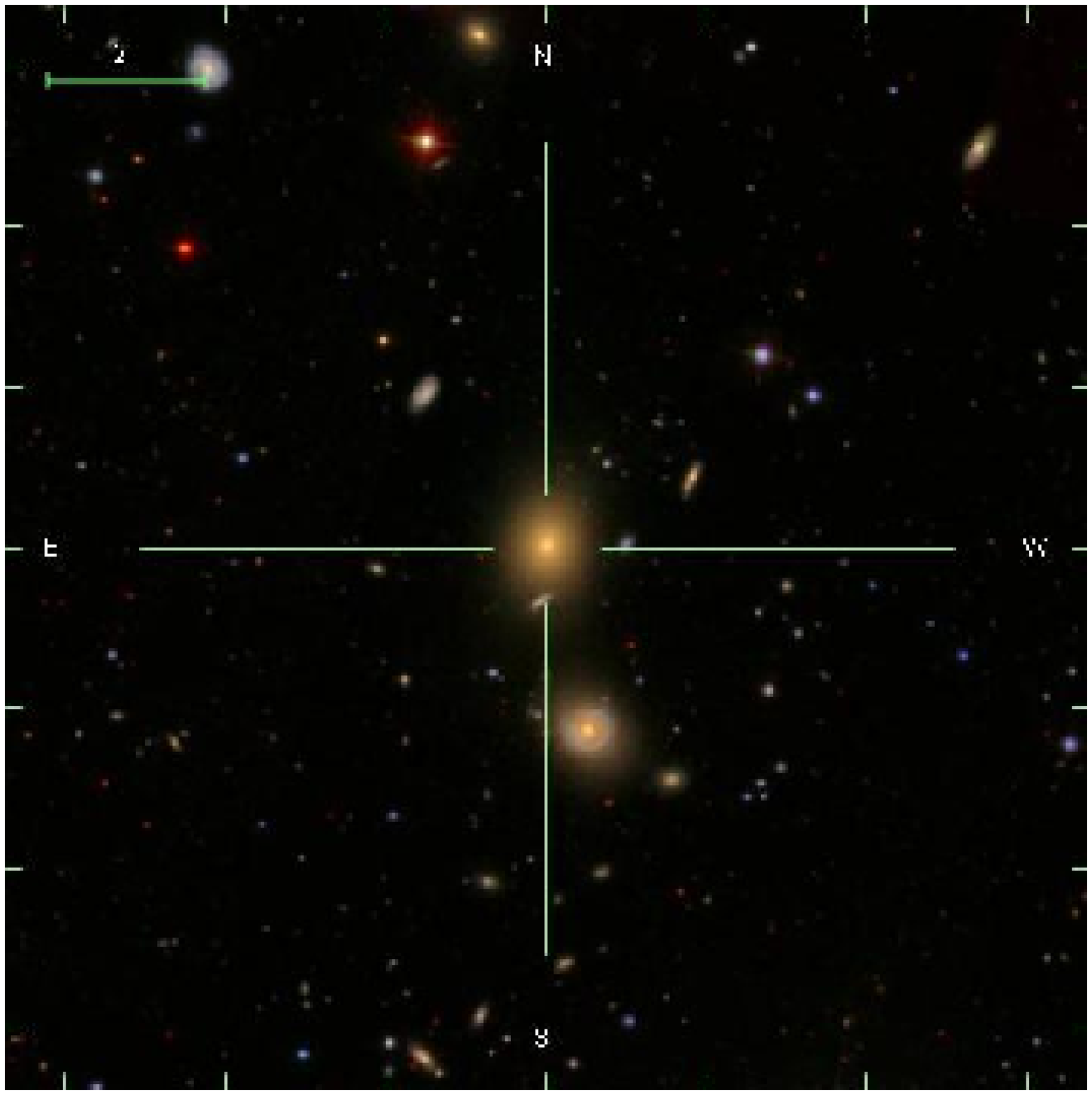}}
\hspace{0.05\textwidth} 
\subfloat[WBL 234]{\label{fig:WBL_234_SDSS}\includegraphics[height=0.20\textheight]{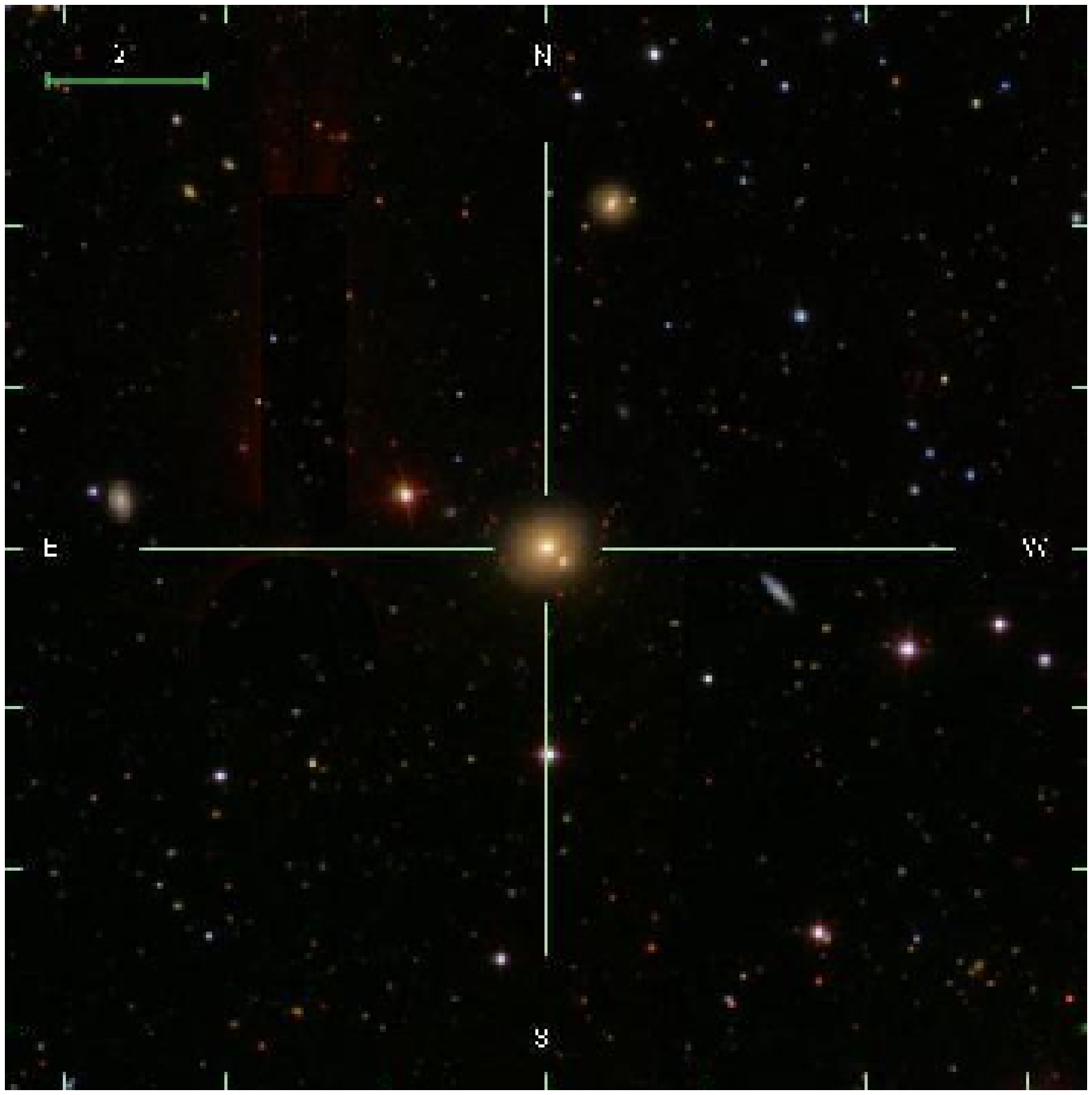}}

\caption{SDSS color-composite images of the central regions of
  clusters. The horizontal line in the upper left corner indicates the
  pixel-scale of the image.}
\label{fig:SDSS_images1}
\end{figure}

\newpage
\addtocounter{figure}{-1}
\begin{figure}[p]
\centering

\subfloat[WBL 205]{\label{fig:WBL_205_SDSS}\includegraphics[height=0.20\textheight]{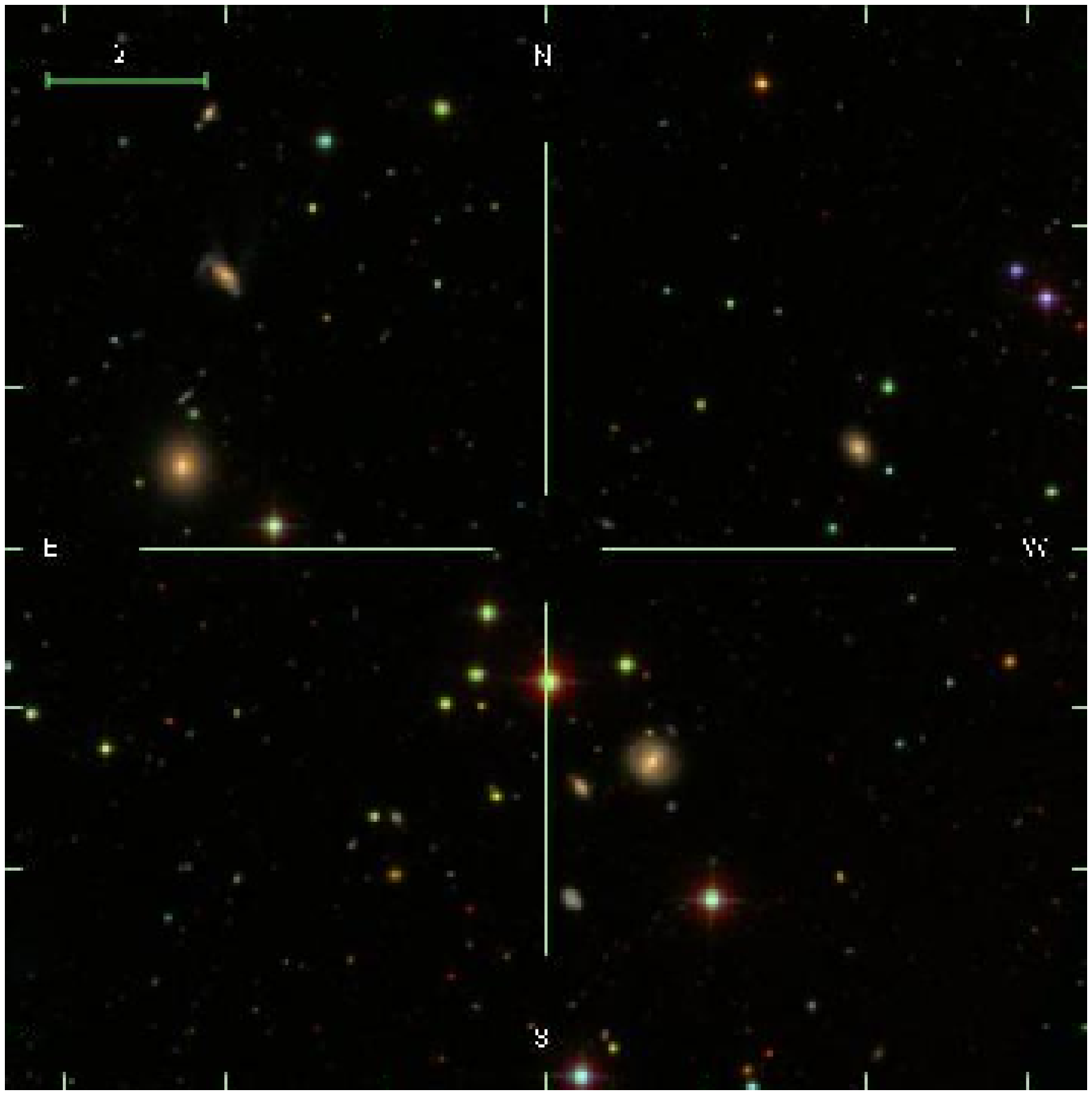}}
\hspace{0.05\textwidth} 
\subfloat[UGCl 393]{\label{fig:UGCl_393_SDSS}\includegraphics[height=0.20\textheight]{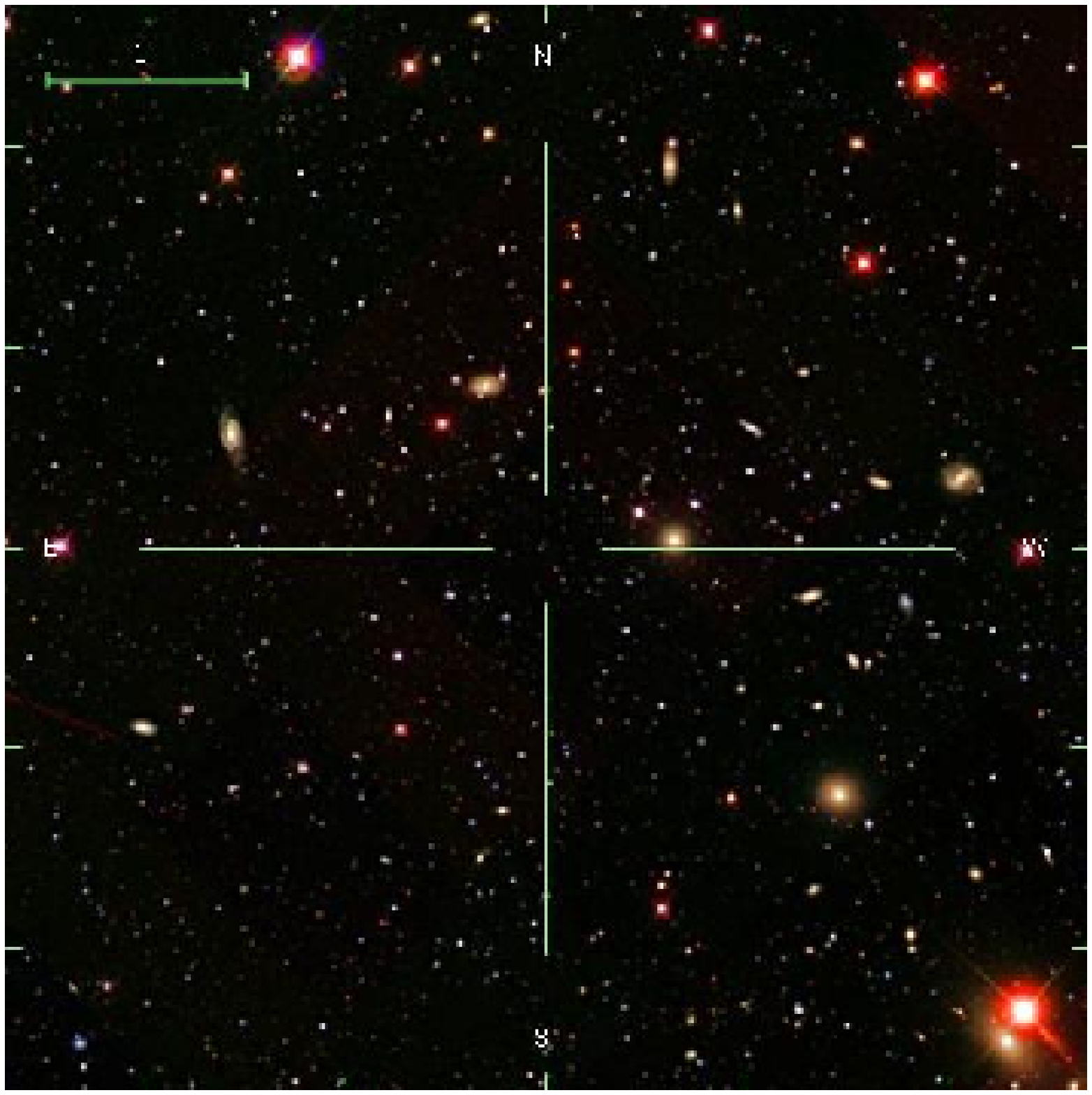}}

\subfloat[UGCl 391]{\label{fig:UGCl_391_SDSS}\includegraphics[height=0.20\textheight]{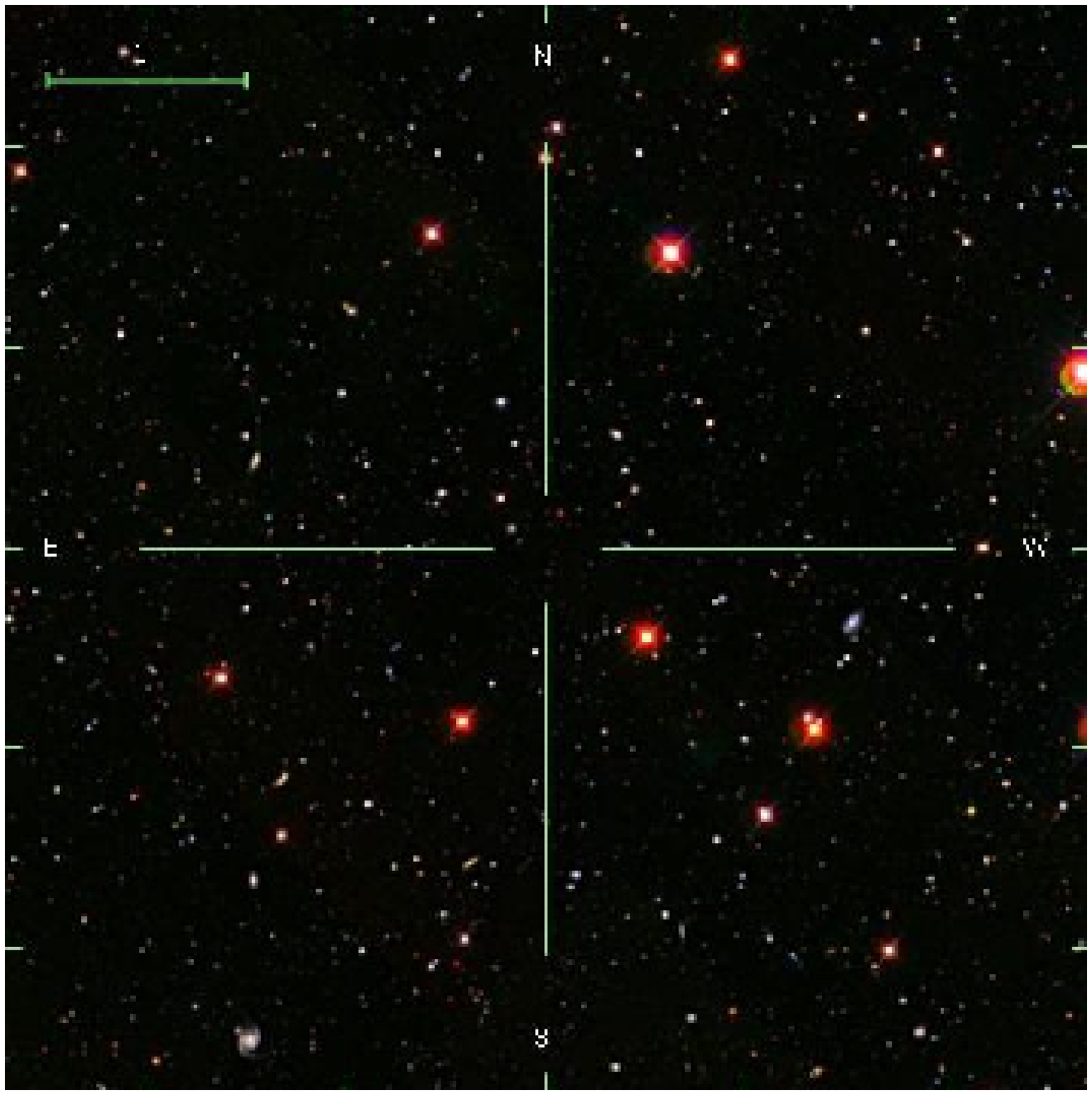}}
\hspace{0.05\textwidth} 
\subfloat[B2 1621+38:$\lbrack$MLO2002$\rbrack$]{\label{fig:B2_SDSS}\includegraphics[height=0.20\textheight]{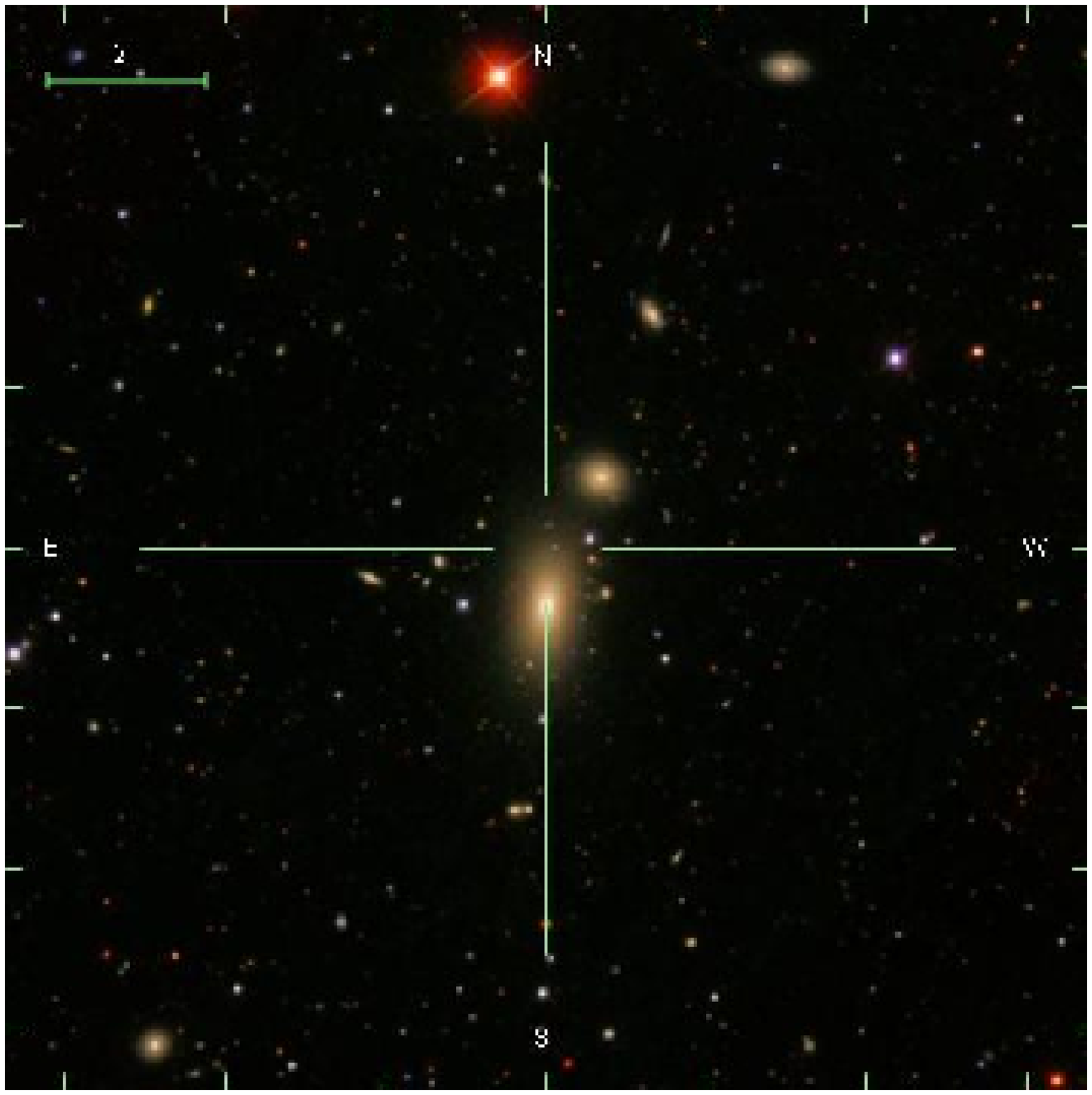}}

\subfloat[UGCl 271]{\label{fig:UGCl_271_SDSS}\includegraphics[height=0.20\textheight]{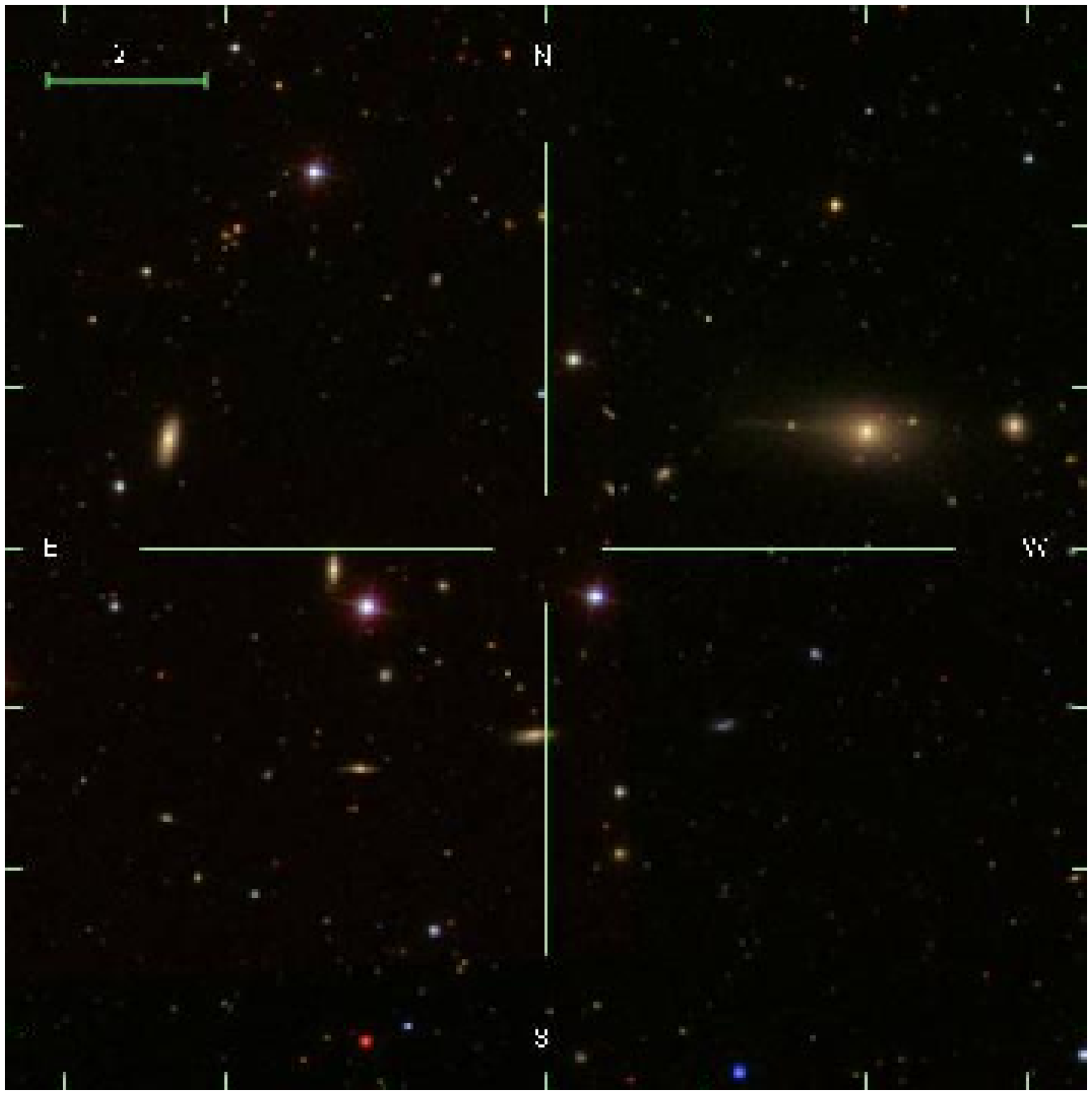}}
\hspace{0.05\textwidth} 
\subfloat[ABELL 1185]{\label{fig:ABELL_1185_SDSS}\includegraphics[height=0.20\textheight]{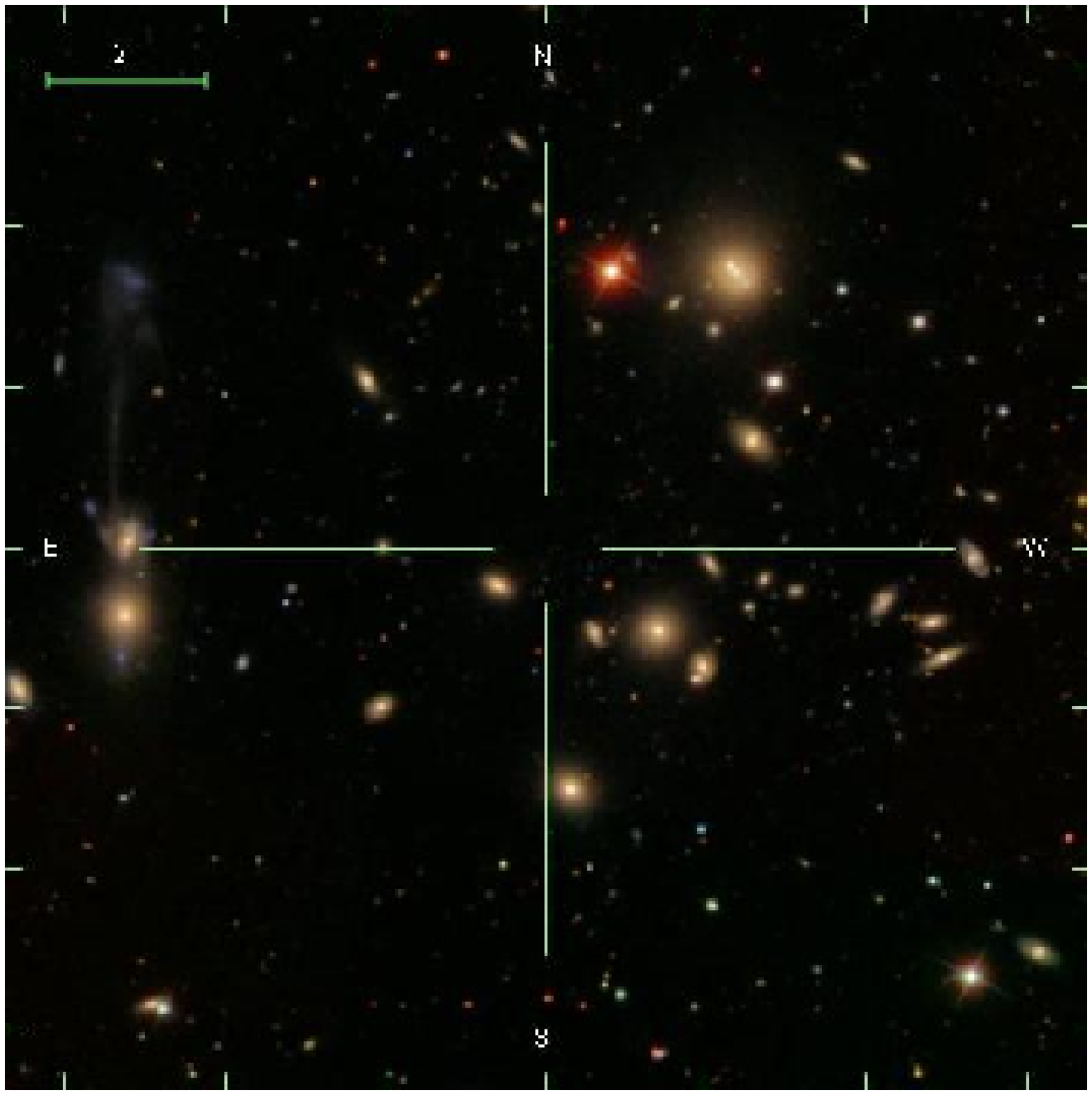}}

\subfloat[ABELL 1213]{\label{fig:ABELL_1213_SDSS}\includegraphics[height=0.20\textheight]{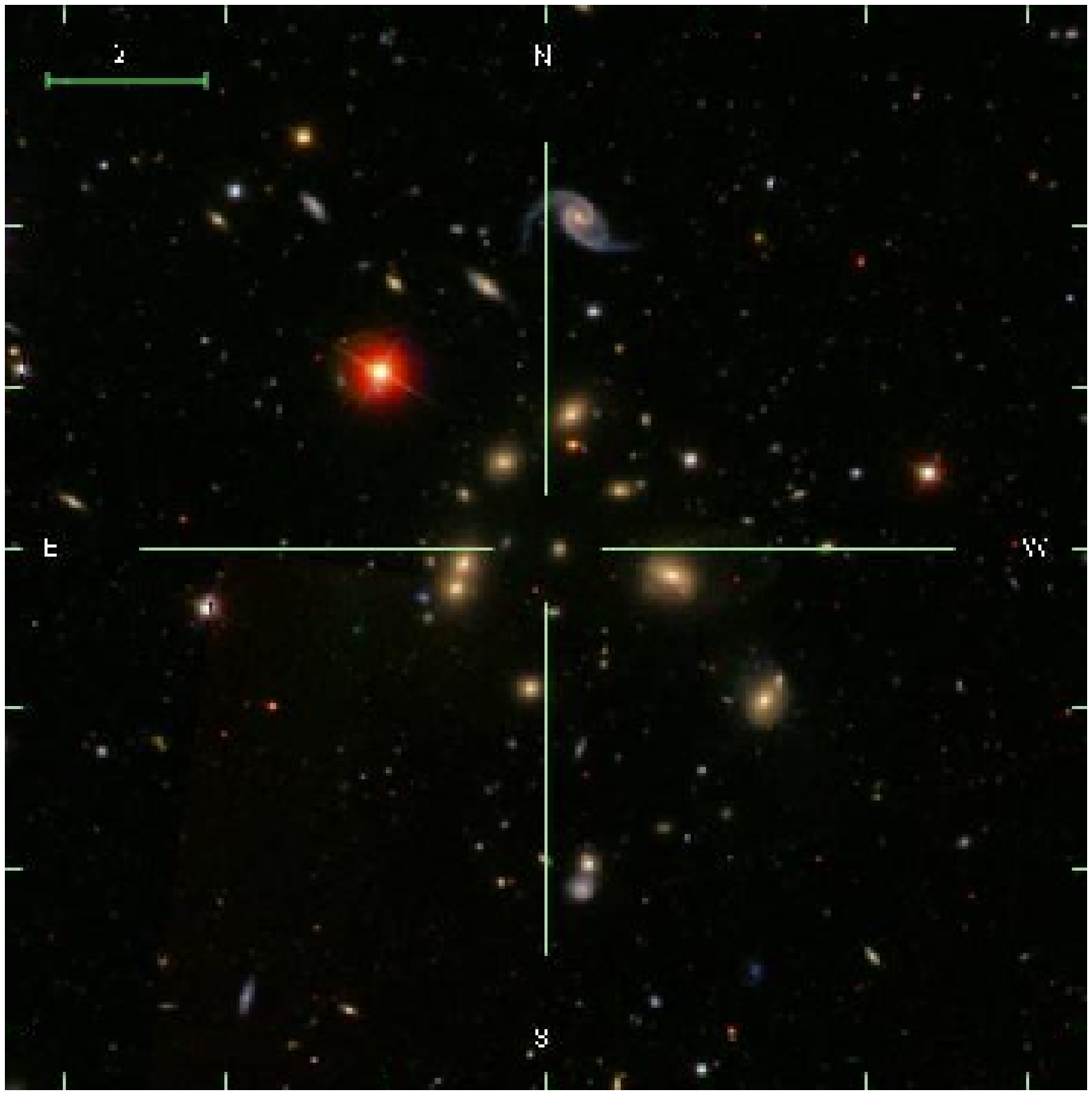}}
\hspace{0.05\textwidth} 
\subfloat[UGCl 123 NED1]{\label{fig:UGCl_123_NED1_SDSS}\includegraphics[height=0.20\textheight]{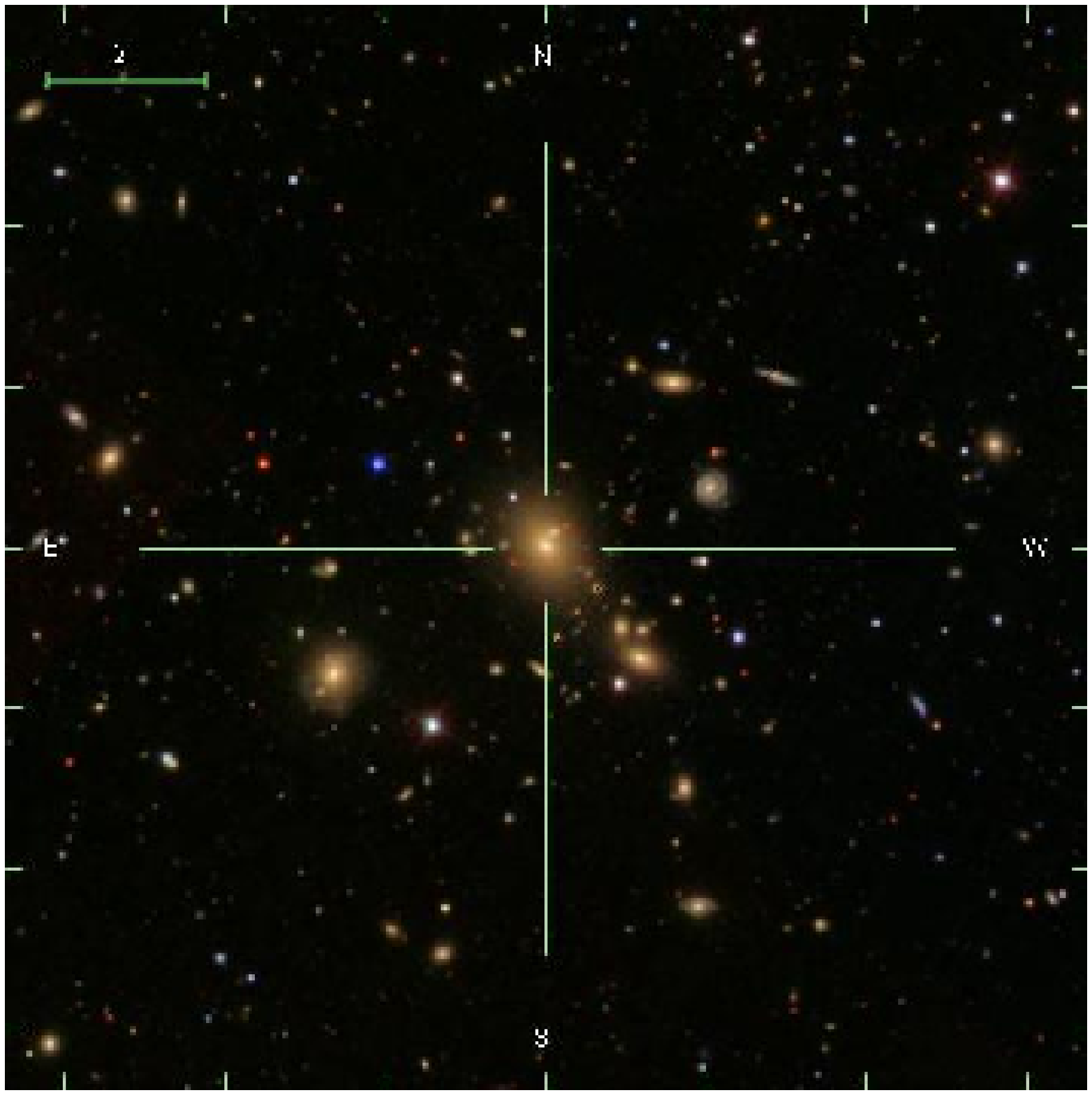}}

\caption{Continued.}
\label{fig:SDSS_images2}
\end{figure}

\section{Crossmatching of galaxy catalogues and compilation of spectrophotometric data}
\label{sec:spph_catalog}

One purpose of this work is the compilation of broad-band and emission
line fluxes from the ultraviolet around 1350 \AA\ to the far-infrared
around 100 $\mu$m for the cluster galaxy sample. In the last decades,
this task becomes possible thanks to several sky surveys covering
large areas of sky from UV to FIR. We present a brief summary about
the main galaxy surveys from we retrieve spectrophotometric fluxes for
the cluster galaxy sample and summarize the main figures of each
survey in Table \ref{surveys} :

\begin{itemize}

\item Galaxy Evolution Explorer \citep[GALEX,][]{Martin_et_al_2005}
  was launched to, among others surveys, cover all sky at different
  depth and areas in two UV filters, the far-ultraviolet ($FUV$) band
  (1350-1750 \AA) and the near-ultraviolet ($NUV$) band (1750-2750
  \AA). The AIS plans to survey the entire sky down to a sensitivity
  of m$_{AB}$$\approx$20.5, comparable with the sensitivity of the
  SDSS Main Galaxy Sample, $r'_{MGS}$=17.77
  \citep{Strauss_et_al_2002}.

\item The SDSS Project \citep[6th Data Release
  in][]{Adelman-McCarthy_et_al_2008} retrieved spectra from, among
  other astronomical objects, all galaxies with $r'$$<$17.77 from the
  SDSS Imaging Catalog. The SDSS photometric system
  \citep{Fukugita_et_al_1996} cover from 3000 to 11000 \AA\ in five
  broad band filters ($u'$, $g'$, $r'$, $i'$ and $z'$).

\item The Two Micron All Sky Survey \citep[2MASS,][]{Cutri_et_al_2001}
  has uniformly scanned the majority of the sky in three near-infrared
  (NIR) bands, J (1.25 $\mu$m), H (1.65 $\mu$m), and K$_{s}$ (2.17
  $\mu$m).

\item The Infrared Astronomical Satellite
  \citep[IRAS,][]{Neugebauer_et_al_1984} was a project to perform an
  unbiased, sensitive all sky survey at 12, 25, 60 and 100 $\mu$m,
  down to a limiting flux of 0.2 Jy at 60 $\mu$m. This mission
  produced two main catalogues; the Point Source Catalog catalogue
  \citep[PSC,][]{IRAS_PSC} and the Faint Source Catalogue
  \citep[FSC,][]{Moshir_1993}.

\end{itemize}

\begin{table}[p]
\caption{Main figures of galaxy surveys.}
\begin{center}
\begin{tabular}{l                              l           l              l                     l }
\hline
 SURVEY                                      & band      & $\lambda_{c}$ & $\Delta\lambda_{c}$ & m$_{lim}$   \\         
                                             &           &  $\mu$m       & $\mu$m             & AB mag     \\         
 (1)                                         & (2)       & (3)           & (4)                & (5)        \\         
\hline
 GALEX $^{(a)}$                               & $FUV$     &    0.1550     &   0.400            &   20.5     \\        
                                             & $NUV$     &    0.2250     &   1.000            &   20.5     \\        
 SDSS $^{(b)}$                                & $u'$      &    0.3551     &   0.599            &   22.0     \\        
                                             & $g'$      &    0.4686     &   1.379            &   22.2     \\        
                                             & $r'$      &    0.6165     &   1.382            &   22.2     \\        
                                             & $i'$      &    0.7481     &   1.535            &   21.3     \\        
                                             & $z'$      &    0.8931     &   1.370            &   20.5     \\        
 2MASS $^{(c)}$                               & $J$       &    1.25       &   1.620            &   16.39    \\        
                                             & $H$       &    1.65       &   2.510            &   16.37    \\        
                                             & $K_{s}$    &    2.17       &   2.620            &   16.34    \\        
 IRAS(PSC+FSC) $^{(d)}$                       &  12$\mu$m &   12          &   7.00             &   10.64    \\        
                                             &  25$\mu$m &   25          &   11.15            &   10.64    \\        
                                             &  60$\mu$m &   60          &   32.5             &   10.64    \\        
                                             & 100$\mu$m &  100          &   32.5             &    8.9     \\        
\hline
\end{tabular}
\end{center}
(1) Survey, (2) Spectral band, (3) Central wavelength, (4) Spectral
bandwidth and (5) Completeness limit. $^{(a)}$
\citet{Martin_et_al_2005}, $^{(b)}$
\cite{Adelman-McCarthy_et_al_2008}, $^{(b)}$
\citet{Finlator_et_al_2000} and $^{(d)}$ \citet{IRAS_PSC} +
\citet{Moshir_1993}. Some galaxies in the FSC have upper limits with
fluxes greater than these nominal values.
\label{surveys}
\end{table}

\subsection{SDSS data}
\label{sec:sdss_data}

The cross-correlation of celestial coordinates from different
catalogues have been accomplished using the SDSS celestial coordinates
as the fiducial coordinates. For each cluster, we retrieve all
galaxies from the DR6 of SDSS with the following criteria:

\begin{itemize}
\item $R_{P}$ $\le$ 7.1 Mpc  
\item $z_{c}$ - 5$\sigma_{c}$ $\le$ $z$ $\le$ $z_{c}$ + 5$\sigma_{c}$ 
\item $z$ $\ge$ 10$^{-3}$ (In order to avoid stars in the lowest
  redshift clusters)
\end{itemize}

We retrieve photometric and spectroscopic data from SDSS database for
this galaxy sample. The photometric fluxes come from the five
broad-band filters of SDSS. We select the {\it{``composite flux''}}
magnitude \citep{SDSS_II} as the suitable way to retrieve the total
flux from each galaxy with the minimum uncertainty in color. We summed
to the error reported by SDSS photometric pipeline
\citep[{\texttt{photo}},][]{Lupton_et_al_2001} and the calibration
errors reported in the DR6 of SDSS
\citep{Adelman-McCarthy_et_al_2008}, the standard deviation (based on
the interquartile range) of distribution of the difference
$r_{composite}$-$r_{petrosian}$ to account uncertainties in color
which are not present in the standard accurate-color photometry
\citep[i.e. petrosian magnitude,][]{SDSS_II}.

We include spectroscopic data regarding to spectroscopic redshift and
the fluxes for the four emission lines of the BPT diagram
\citep{BPT_1981}; [OIII] ($\lambda$=5007 \AA), H$\beta$
($\lambda$=4861 \AA), [NII] ($\lambda$=6584 \AA) and H$\alpha$
($\lambda$=6563 \AA). \citet{Moustakas_et_al_2006} claim the
extinction-corrected H$\alpha$ luminosity is a reliable Star Formation
Rate (SFR) tracer, even in highly obscured star-forming galaxies. We
derive galaxy SFRs from the extinction-corrected H$\alpha$
luminosity. The extinction correction is applied using the Balmer
decrement method and the \citet{Cardelli_et_al_1989} extinction law
with R$_{V}$=3.1. We take a HI recombination line ratio in the
theoretical case B nebulae at T=10$^4$ K as
H$\alpha$/H$\beta$=2.87. We apply the scaling law between SFR and
H$\alpha$ luminosity proposed by \citet{Kennicutt_1998}.

SDSS project has a pair of fiber-fed double spectrographs with 3
arcsec of fiber diameter on sky. This produces a loss of light from
external parts of the largest galaxies. In order to reduce systematic
and random errors from this ``aperture effect'' in SFR estimation
\citet{Kewley_et_al_2005} recommend selecting galaxy samples with the
fiber capturing more than the 20\% of the galaxy B$_{445nm}$-light. We
assume a SDSS spectrum as representative of a galaxy when the fiber
contains, at least, one fifth of the total $g$-band flux of the
galaxy. So, we select these galaxies with:

($g_{fiber}$-$g_{model}$)$\le$-2.5 alog$_{10}$(0.2)

$g_{fiber}$ is the $g$-band magnitude measured inside an aperture
similar to those produce by the SDSS fiber and $g_{model}$ is the
$g$-band {\it ``model''} magnitude. In this case, we scale H$\alpha$
fiber flux to H$\alpha$ total flux using
10$^{-0.4(g_{model}-g_{fiber})}$ as scaling factor. Otherwise, we set
H$\alpha$ fiber flux (without any scaling) as the lower limit for the
H$\alpha$ total flux of these galaxies

\subsection{SDSS-GALEX crosscorrelation}

Following the criterium proposed by \citet{Obric_et_al_2006}, we
choose a matching radius of 6 arcsec between the SDSS and GALEX AIS
celestial coordinates. We accomplish the source matching using the
GALEX application
{\it{GalexView}} \footnote{{\it{http://galex.stsci.edu/GalexView/\#}}}. In
the case there is not a GALEX source in the matching circle, we do not
assign an UV flux to SDSS source. The fraction of SDSS sources without
GALEX detection is less than 20\%. There are two options for the case
of a non matched source; or this sky region is not observed by GALEX
AIS, or the UV flux for the SDSS source is under GALEX AIS detection
limit. The first case does not introduce a biased selection of
galaxies i.e. there is no correlation between the celestial
coordinates and the galaxy properties. In the second case, we have a
completeness limit for the SDSS Main Galaxy Sample of
$r'_{MGS}$$<$17.77 and the GALEX AIS reach down to $NUV_{lim}\sim$22
for Galactic extinction-corrected magnitudes, while the UV-optical
color separation between blue and red galaxies is
$NUV$-$r$$\sim$4. So, this case only affects to red galaxies in the
lowest flux bin $r'$$\gtrsim$16.

We choose the elliptical aperture photometry \citep[\texttt{MAG\_AUTO}
  option in SExtractor code, ][]{Bertin&Arnouts_1996} for GALEX
sources in order to have the complete UV flux for each source. These
magnitudes are corrected from Galactic extinction using the excess
color E(B-V) reported in GALEX tables for each UV source and assuming
the Cardelli extinction law \citep{Cardelli_et_al_1989}.

\subsection{SDSS-2MASS crosscorrelation}

The 2MASS project has enough image quality \citep[FWHM$\sim$2.5-2.7
  arcsec,][]{Cutri_et_al_2001} to discriminate point-like sources
(i.e. stars) from the extended ones (i.e. galaxies); the angular
distance at $z$=0.05 is 0.977 kpc arcsec$^{-1}$. So, we only
crosscorrelate the galaxy sample with the 2MASS All-Sky Extended
Source Catalog (XSC) and not the 2MASS All-Sky Point Source Catalog
(PSC). We follow \citet{Blanton_et_al_2005} and set the matching
radius to 3 arcsec.

The NIR magnitudes for each SDSS source without 2MASS counterpart are
fixed, as a lower limiting flux, to the completeness limit in each
2MASS band \citep{Finlator_et_al_2000}. In this case, we set the error
for the lower limit to a nominal value of $\Delta$m=1 mag, which is
the magnitude interval along the NIR galaxy counts decrease from the
100\% completeness down to
zero\footnote{{\it{http://www.ipac.caltech.edu/2mass/releases/allsky/doc/sec2\_3d3.html}}}. The
matching rates vary from cluster to cluster and are around 40-60\%.

We choose the photometry named {\it{total magnitude}} for the three
NIR bands which is obtained from the integral between the lowest
elliptical radius with a surface brightness of $\mu$=20 mag
arcsec$^{-2}$ \citep[this corresponds to $\sim$1$\sigma$ of the sky
  background,][]{Cutri_et_al_2001} and a elliptical Sérsic profile
\citep{Sersic_1963} fitted to the surface brightness profile of the
galaxy \citep{Jarrett_et_al_2000}. We apply the magnitude conversion
from Vega system to AB system from \citet{Finlator_et_al_2000}.

\subsection{SDSS-IRAS crosscorrelation}

Owing to the low angular resolution of IRAS
telescope\footnote{{\it{http://irsa.ipac.caltech.edu/IRASdocs/exp.sup/ch2/C3.html}}},
the galaxies resemble IRAS point-like sources. So, we crossmatch the
galaxy sample with a joint catalogue of PSCz$\oplus$FSC; {\it{Point
    Source Catalogue}} \citep{IRAS_PSC} $\oplus$ {\it{Faint Source
    Catalog}} \citep{Moshir_1993}. The FSC is $\sim$2.5 times deeper
in limiting flux than the PSCz catalogue and the approximated flux
frontier between this two catalogues is around 0.4 Jy. We set the
matching radius to r=30 arcsec, the value proposed by
\citet{Blanton_et_al_2005}. Anyway, the matching rate is quite low
$\sim$1-5\%. The upper limit in IRAS flux for galaxies without IRAS
counterpart is set to the values proposed for the FSC at each IRAS
band \citep{Moshir_1993}. We set the upper limit error to the nominal
(absolute+relative) error reported in PSCz catalogue: 11\% + 0.06 Jy.

\section{The spectrophotometric catalogue}
\label{sec:catalogue_sec}

The format of the spectrophotometric catalogue is presented in Table
\ref{catalogue_table}. It contains 53 columns that are described
below, including the relevant observational parameters,
spectrophotometric fluxes from UV to FIR and SFR
estimates\footnote{The catalogue will be presented in its entirety in
  the online version of the paper.}:

Columns (1). ID: number associated to the position of the galaxy
inside the cluster galaxy sample as a identifier.

Columns (2) and (3). ObjID and specObjID: SDSS Imaging Catalog and Main Galaxy Sample identifier of the galaxy. 

Columns (4) and (5). RA and DEC: SDSS right ascension and declination (J2000) in degrees.               

Columns (6) and (7).  z and $\epsilon_{z}$: SDSS spectroscopic redshift and its uncertainty.                       

In sets of three elements, the following columns show the AB magnitude
of galaxy, its uncertainty and the detection identifier$^{(a)}$ for
the following spectral bands: 

Columns ( 8), ( 9) and (10). AB$_{FUV}$, $\sigma_{FUV}$ and i$_{FUV}$: the GALEX $FUV$ band.       

Columns (11), (12) and (13). AB$_{NUV}$, $\sigma_{NUV}$ and i$_{NUV}$: the GALEX $NUV$ band.       

Columns (14), (15) and (16). AB$_{u'}$, $\sigma_{u'}$ and i$_{u'}$: the SDSS $u'$ band.          

Columns (17), (18) and (19). AB$_{g'}$, $\sigma_{g'}$ and i$_{g'}$: the SDSS $g'$ band.          

Columns (20), (21) and (22). AB$_{r'}$, $\sigma_{r'}$ and i$_{r'}$: the SDSS $r'$ band.          

Columns (23), (24) and (25). AB$_{i'}$, $\sigma_{i'}$ and i$_{i'}$: the SDSS $i'$ band.          

Columns (26), (27) and (28). AB$_{z'}$, $\sigma_{z'}$ and i$_{z'}$: the SDSS $z'$ band.          

Columns (29), (30) and (31). AB$_{J}$, $\sigma_{J}$ and i$_{J}$: the SDSS $J$ band.           

Columns (32), (33) and (34). AB$_{H}$, $\sigma_{H}$ and i$_{H}$: the SDSS $H$ band.           

Columns (35), (36) and (37). AB$_{Ks}$, $\sigma_{Ks}$ and i$_{Ks}$: the SDSS $K_{s}$ band.        

Columns (38), (39) and (40). AB$_{12\mu m}$, $\sigma_{12\mu m}$ and i$_{12\mu m}$: the IRAS 12 $\mu$m band.     

Columns (41), (42) and (43). AB$_{25\mu m}$, $\sigma_{25\mu m}$ and i$_{25\mu m }$: the IRAS 25 $\mu$m band.     

Columns (44), (45) and (46). AB$_{60\mu m}$, $\sigma_{60\mu m}$ and i$_{60\mu m }$: the IRAS 60 $\mu$m band.     

Columns (47), (48) and (49). AB$_{100\mu m}$, $\sigma_{100\mu m}$ and i$_{100\mu m }$: the IRAS 100 $\mu$m band.    

Columns (50), (51) and (52). SFR, $\sigma_{SFR}$ and i$_{SFR}$:
H$\alpha$-derived star formation rate (SFR), its uncertainty and
detection identifier in SFR.

Column (53). Cluster: identifier for the parent cluster of the
galaxy. The cluster identifiers are codified in the following way:
A=ABELL, B2=B2 1621+38:[MLO2002] CLUSTER, N=NED, U=UGCl, W=WBL.

$^{(a)}$ Code for detection identifiers:  \\ 
{\bf{~1}} $\equiv$ Source detected on this band, \\ 
{\bf{~0}} $\equiv$ Source undetected on this band (upper limit in flux),  \\ 
{\bf{-1}} $\equiv$ Source not observed on this band and \\ 
{\bf{-2}} $\equiv$ Lower limit in flux.

\begin{landscape}
\begin{table}[p]
\caption{Spectrophotometric catalogue of cluster galaxy sample.}
\begin{center}
\resizebox{22cm}{!}{ 
\begin{tabular}{c                c                        c                    c             c           c         c          c             c             c        c             c             c        }
\hline
       ID  &     ObjID              &     specObjID          & RA         &  DEC       & z       &$\epsilon_{z}$&AB$_{FUV}$&$\sigma_{FUV}$&i$_{FUV}$&AB$_{NUV}$ &$\sigma_{NUV}$& i$_{NUV}$  \\  
      (1)  &       (2)              &     (3)                &     (4)    & (5)        & (6)     & (7)        & (8)       &  (9)        &   (10) &    (11)   &  (12)      &   (13)   \\ 
           &                        &                        & deg        &  deg       &         &         &  AB mag    &   AB mag   &         &   AB mag   &  AB mag    &         \\ 
\hline
         1 &     587735239565377792 &     357982217034530816 & 134.655685 &  31.482407 & 0.02662 & 0.00009 &  19.049200 &   0.133216 &     1    &  18.683599 &   0.082258 &     1     \\ 
         2 &     587735240639381760 &     357982219328815104 & 134.670593 &  32.448460 & 0.02233 & 0.00009 &  -1.000000 &   1.000000 &    -1    &  -1.000000 &   1.000000 &    -1     \\ 
         3 &     587735043615096960 &     358263757475938304 & 135.079346 &  32.780834 & 0.02231 & 0.00009 &  -1.000000 &   1.000000 &    -1    &  -1.000000 &   1.000000 &    -1     \\ 
         4 &     587735043078946944 &     358263758667120640 & 136.960205 &  33.468132 & 0.02638 & 0.00017 &  21.178200 &   0.279031 &     1    &  19.302299 &   0.095046 &     1     \\ 
         5 &     587735239567474944 &     358545248592330752 & 139.522614 &  33.917965 & 0.02438 & 0.00016 &  -1.000000 &   1.000000 &    -1    &  21.215200 &   0.279032 &     1     \\ 
         6 &     587735239567540352 &     358545248617496576 & 139.730331 &  34.034649 & 0.02227 & 0.00018 &  -1.000000 &   1.000000 &    -1    &  21.621300 &   0.384734 &     1     \\ 
         7 &     587735239567540224 &     358545248625885184 & 139.611649 &  34.036049 & 0.02317 & 0.00020 &  -1.000000 &   1.000000 &    -1    &  21.976801 &   0.469288 &     1     \\ 
         8 &     587735239567737088 &     358545248667828224 & 140.089035 &  34.238449 & 0.02461 & 0.00008 &  -1.000000 &   1.000000 &    -1    &  18.375999 &   0.050605 &     1     \\ 
         9 &     587735239567605888 &     358545248823017472 & 139.744232 &  34.133747 & 0.02226 & 0.00015 &  -1.000000 &   1.000000 &    -1    &  -1.000000 &   1.000000 &    -1     \\ 
        10 &     587735042543124608 &     358545248831406080 & 139.641464 &  34.293934 & 0.02166 & 0.00014 &  -1.000000 &   1.000000 &    -1    &  20.270000 &   0.189778 &     1     \\ 
\hline
\label{catalogue_table}
\end{tabular} }  
\end{center}
\textsc{NOTE:} This table will be presented in its entirety in the
online version of the paper.
\end{table}
\end{landscape}

\addtocounter{table}{-1}
\begin{landscape}
\begin{table}[p]
\caption{Continued.} 
\begin{center}
\resizebox{22cm}{!}{ 
\begin{tabular}{cc   c             c        c             c             c        c             c             c        c             c             c        c             c             c      }
\hline
       ID  &   AB$_{u'}$ &$\sigma_{u'}$&   i$_{u'}$ & AB$_{g'}$ &$\sigma_{g'}$&    i$_{g'}$ & AB$_{r'}$ &$\sigma_{r'}$& i$_{r'}$ &  AB$_{i'}$   &$\sigma_{i'}$&  i$_{i'}$ & AB$_{z'}$  &$\sigma_{z'}$&  i$_{z'}$   \\  
       (1) &      (14)  & (15)       &  (16)      &   (17)     &    (18)  &  (19)      &   (20)     &    (21)  &  (22)      &   (23)     &    (24)  &  (25)      &   (26)     &    (27)  &  (28)      \\ 
           &  AB mag    &   AB mag   &         &   AB mag   &  AB mag    &          &  AB mag   &   AB mag   &         &   AB mag   &  AB mag    &         &  AB mag    &   AB mag   &           \\ 
\hline
         1 &       18.517775 &   0.108053 &     1    &  17.700541 &   0.066168 &     1    &  17.592373 &   0.066663 &     1    &  17.379465 &   0.070090 &     1    &  17.194235 &   0.108876 &     1   \\ 
         2 &       19.067827 &   0.105126 &     1    &  18.039906 &   0.066950 &     1    &  17.741817 &   0.063210 &     1    &  17.584496 &   0.065851 &     1    &  17.514194 &   0.092795 &     1   \\ 
         3 &       19.101942 &   0.120221 &     1    &  17.946131 &   0.064513 &     1    &  17.827415 &   0.065841 &     1    &  17.913485 &   0.068607 &     1    &  17.935137 &   0.115220 &     1   \\ 
         4 &       16.447611 &   0.084428 &     1    &  14.826661 &   0.066261 &     1    &  14.076668 &   0.066026 &     1    &  13.676976 &   0.066025 &     1    &  13.416355 &   0.077062 &     1   \\ 
         5 &       17.861683 &   0.097150 &     1    &  16.316339 &   0.068362 &     1    &  15.610915 &   0.067196 &     1    &  15.253843 &   0.067416 &     1    &  15.014357 &   0.081505 &     1   \\ 
         6 &       19.127745 &   0.143379 &     1    &  17.663599 &   0.073838 &     1    &  16.970215 &   0.070757 &     1    &  16.692688 &   0.071775 &     1    &  16.589867 &   0.093554 &     1   \\ 
         7 &       19.035378 &   0.141855 &     1    &  17.719177 &   0.066303 &     1    &  17.065872 &   0.063679 &     1    &  16.616327 &   0.064424 &     1    &  16.577681 &   0.089601 &     1   \\ 
         8 &       17.831083 &   0.089150 &     1    &  16.744196 &   0.060600 &     1    &  16.188751 &   0.060178 &     1    &  16.101572 &   0.061675 &     1    &  16.413498 &   0.083945 &     1   \\ 
         9 &       18.705692 &   0.112631 &     1    &  17.124117 &   0.071256 &     1    &  16.368345 &   0.071518 &     1    &  16.046734 &   0.069902 &     1    &  15.803750 &   0.090045 &     1   \\ 
        10 &       16.759426 &   0.089591 &     1    &  15.192950 &   0.066689 &     1    &  14.493539 &   0.066356 &     1    &  14.086758 &   0.066258 &     1    &  13.676766 &   0.077599 &     1   \\ 
\hline
\end{tabular} } 
\end{center}
\end{table}
\end{landscape}

\addtocounter{table}{-1}
\begin{landscape}
\begin{table}[p]
\caption{Continued.} 
\begin{center}
\resizebox{22cm}{!}{ 
\begin{tabular}{cc            c             c        c             c             c        c             c             c        c             c             c }
\hline
       ID  &  AB$_{J}$    &$\sigma_{J}$ & i$_{J}$   & AB$_{H}$  &$\sigma_{H}$&  i$_{H}$    & AB$_{Ks}$ &$\sigma_{Ks}$& i$_{Ks}$   & AB$_{12\mu m}$ &$\sigma_{12\mu m}$& i$_{12\mu m}$    \\ 
       (1) &     (29)     & (30)     &      (31)  &   (32)   &    (33)    &  (34)      &   (35)     &    (36)  &  (37)      &   (38)     &    (39)    &    (40)      \\ 
           &     AB mag   &  AB mag    &         &   AB mag   &  AB mag    &         &  AB mag    &   AB mag   &         &  AB mag    &   AB mag   &         \\                                                                                                                                                                                                                                                                                                                                                                                                                                                                                                                                                                                                                                                                                                                                                                                                                                                                                                                                                                      
\hline
         1 &    16.389999 &   0.500000 &     0    &  16.370001 &   0.500000 &     0    &  16.340000 &   0.500000 &     0    &  10.647425 &   0.410000 &     0         \\
         2 &    16.389999 &   0.500000 &     0    &  16.370001 &   0.500000 &     0    &  16.340000 &   0.500000 &     0    &  10.647425 &   0.410000 &     0         \\
         3 &    16.389999 &   0.500000 &     0    &  16.370001 &   0.500000 &     0    &  16.340000 &   0.500000 &     0    &  10.647425 &   0.410000 &     0         \\
         4 &    13.371000 &   0.024000 &     1    &  13.155000 &   0.036000 &     1    &  13.294000 &   0.041000 &     1    &  10.647425 &   0.410000 &     0         \\
         5 &    15.266000 &   0.056000 &     1    &  14.981000 &   0.064000 &     1    &  15.249000 &   0.089000 &     1    &  10.647425 &   0.410000 &     0         \\
         6 &    16.389999 &   0.500000 &     0    &  16.370001 &   0.500000 &     0    &  16.340000 &   0.500000 &     0    &  10.647425 &   0.410000 &     0         \\
         7 &    16.389999 &   0.500000 &     0    &  16.370001 &   0.500000 &     0    &  16.340000 &   0.500000 &     0    &  10.647425 &   0.410000 &     0         \\
         8 &    16.389999 &   0.500000 &     0    &  16.370001 &   0.500000 &     0    &  16.340000 &   0.500000 &     0    &  10.647425 &   0.410000 &     0         \\
         9 &    15.926000 &   0.086000 &     1    &  15.654000 &   0.107000 &     1    &  15.919000 &   0.156000 &     1    &  10.647425 &   0.410000 &     0         \\
        10 &    13.681000 &   0.026000 &     1    &  13.454000 &   0.035000 &     1    &  13.626000 &   0.051000 &     1    &  10.647425 &   0.410000 &     0         \\

\hline
\end{tabular} } 
\end{center}
\end{table}
\end{landscape}

\addtocounter{table}{-1}
\begin{landscape}
\begin{table}[p]
\caption{Continued.} 
\begin{center}
\resizebox{22cm}{!}{ 
\begin{tabular}{cccc          c        c             c             c         c            c            c             c                   c                c       }
\hline
       ID &AB$_{25\mu m}$&$\sigma_{25\mu m}$&i$_{25\mu m }$&AB$_{60\mu m}$&$\sigma_{60\mu m}$&i$_{60\mu m}$&AB$_{100\mu m}$&$\sigma_{100\mu m}$&i$_{100\mu m}$& SFR &$\sigma_{SFR}$ &  i$_{SFR}$    & Cluster     \\ 
  (1)      &  (41)     & (42)           &   (43)     &    (44)  & (45)         &   (46)    &   (47)      &   (48)         &      (49)  &    (50)&  (51)           &      (52)  &   (53)   \\                                    
           &  AB mag     &  AB mag    &         &   AB mag   &  AB mag    &         &  AB mag    &   AB mag   &         & $M_{\odot}$yr$^{-1}$ & $M_{\odot}$yr$^{-1}$  &            &            \\                         
\hline                                                                                                                                                                                                               
         1 &   10.647425 &   0.410000 &     0    &  10.647425 &   0.410000 &     0    &   8.900000 &   0.170000 &     0    &      0.022868      &      0.012458      & -2      &     U141    \\   
         2 &   10.647425 &   0.410000 &     0    &  10.647425 &   0.410000 &     0    &   8.900000 &   0.170000 &     0    &      0.128186      &      0.052203      &  1      &     U141    \\   
         3 &   10.647425 &   0.410000 &     0    &  10.647425 &   0.410000 &     0    &   8.900000 &   0.170000 &     0    &      0.208392      &      0.081589      &  1      &     U141    \\   
         4 &   10.647425 &   0.410000 &     0    &  10.647425 &   0.410000 &     0    &   8.900000 &   0.170000 &     0    &      0.000000      &      0.590124      & -2      &     U141    \\   
         5 &   10.647425 &   0.410000 &     0    &  10.647425 &   0.410000 &     0    &   8.900000 &   0.170000 &     0    &      0.000000      &      0.074153      & -2      &     U141    \\   
         6 &   10.647425 &   0.410000 &     0    &  10.647425 &   0.410000 &     0    &   8.900000 &   0.170000 &     0    &      0.000000      &      0.007125      & -2      &     U141    \\   
         7 &   10.647425 &   0.410000 &     0    &  10.647425 &   0.410000 &     0    &   8.900000 &   0.170000 &     0    &      0.000000      &      0.003340      & -2      &     U141    \\   
         8 &   10.647425 &   0.410000 &     0    &  10.647425 &   0.410000 &     0    &   8.900000 &   0.170000 &     0    &      0.081989      &      0.027873      & -2      &     U141    \\   
         9 &   10.647425 &   0.410000 &     0    &  10.647425 &   0.410000 &     0    &   8.900000 &   0.170000 &     0    &      0.000000      &      0.021264      & -2      &     U141    \\   
        10 &   10.647425 &   0.410000 &     0    &  10.647425 &   0.410000 &     0    &   8.900000 &   0.170000 &     0    &      0.000000      &      0.183563      & -2      &     U141    \\   
\hline
\end{tabular} } 
\end{center}
\end{table}
\end{landscape}

\begin{figure}[p]
\centering
\resizebox{0.85\hsize}{!}{\includegraphics{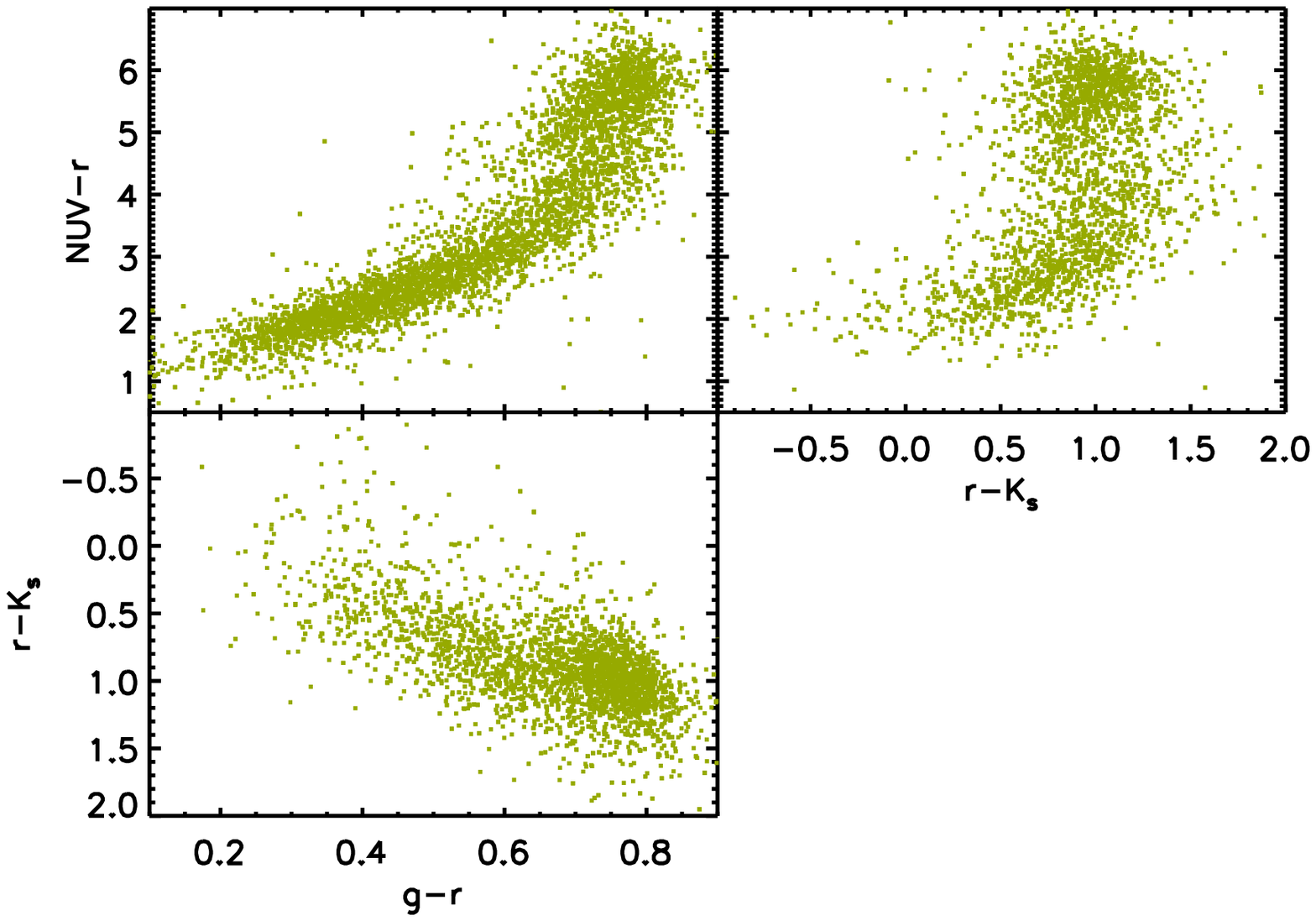}}
\caption{Color-color diagrams. From top to bottom, and from left to
  right: ($NUV$-$r$) vs. ($g$-$r$), ($NUV$-$r$) vs. ($r$-$K_{s}$) and
  ($r$-$K_{s}$) vs. ($g$-$r$) color-color diagrams.}
\label{cc_diagrams}
\end{figure}

In Figure \ref{cc_diagrams}, we show the cluster galaxy sample in
three UV-optical-NIR color-color diagrams; in each panel we show only
galaxies with detection in the three corresponding spectral bands.
Figure \ref{cc_diagrams} shows how the galaxy sample traces the color
distribution of the two main spectral types of galaxies; the passive
galaxies and star-forming galaxies. The ``red sequence'' which is
constituted by the family of passive galaxies becomes a ``red clump''
around ($NUV$-$r$)$\sim$5.75, ($g$-$r$)$\sim$0.75 and
($r$-$K_{s}$)$\sim$1.0 while the ``blue cloud'' of the star-forming
galaxies turns into a sort of ``blue sequence'' which is more clearly
visible in the UV-optical color diagram. We stress that the spectral
information from UV bands allow us a more accurated selection of
star-forming galaxies based on UV-optical color diagrams,
cf. subsection \ref{sec:gal_proj_dist} and figure \ref{NUVr_ur}. This
is especially important for the study of a genuine sample of
star-forming galaxies carried out in this and subsequent works.

\begin{figure}[p]
\centering
\resizebox{0.85\hsize}{!}{\includegraphics{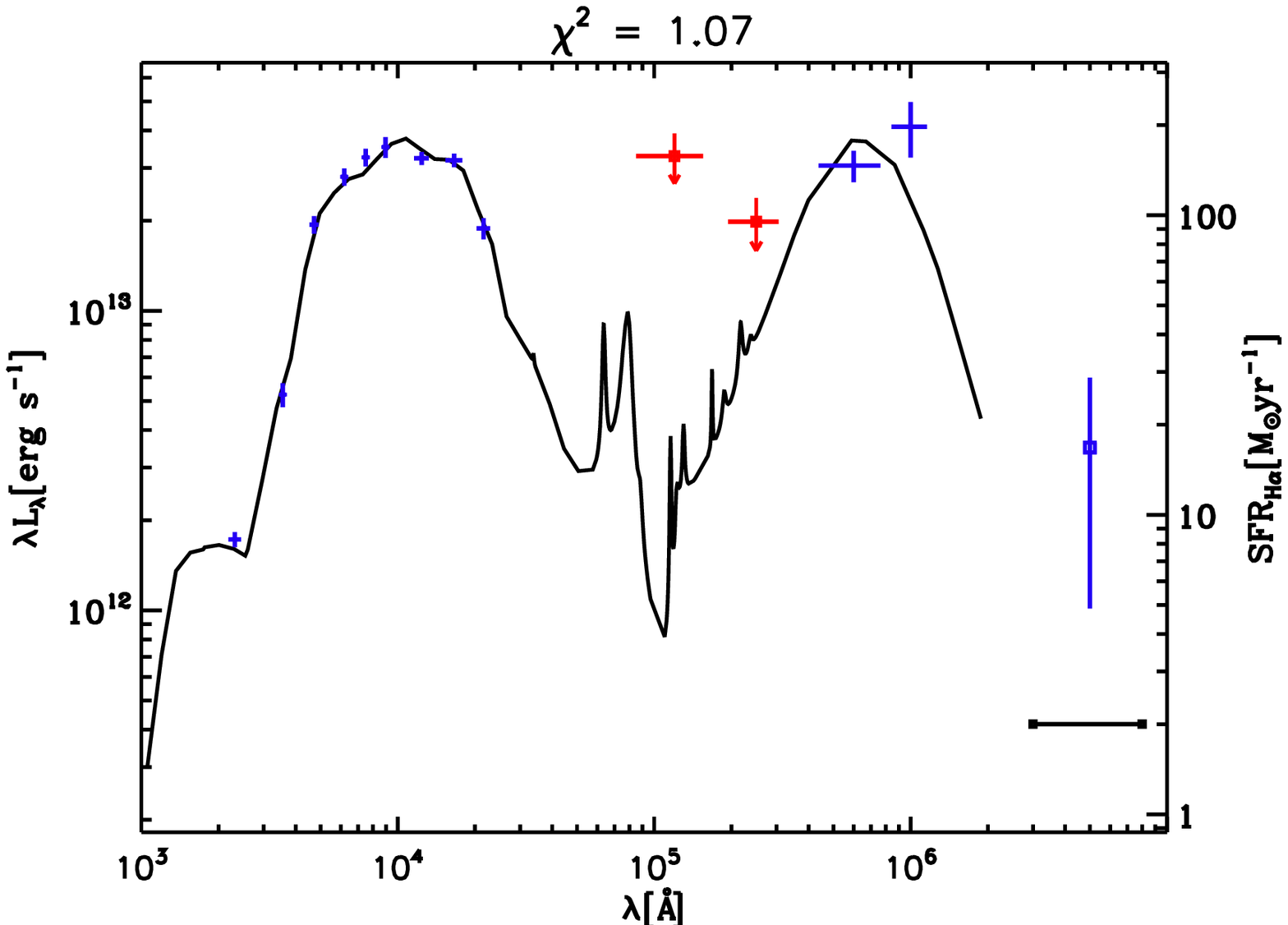}}
\caption{Example of a galaxy SED. The left ordinate axis presents the
  broad-band luminosity and the right ordinate axis the SFR. The solid
  line is the best fitted spectral template from a synthetic spectral
  library in \citet{Hernandez-Fernandez_Tesis}. In the top of the
  graph, we show the value of chi-square for this fit. From left to
  right, blue data are the $NUV$ band from GALEX, the five optical
  bands from SDSS, the three NIR bands from 2MASS, the 60 and 100
  $\mu$m IRAS bands and the H$\alpha$-SFR. Red data correspond to the
  upper limits in 12 and 25 $\mu$m IRAS bands.}
\label{SED_fit}
\end{figure}

The figure \ref{SED_fit} shows an example of SED from the cluster
galaxy sample composed by the broad-band fluxes and the SFR derived
from H$\alpha$ luminosity which covers three dex in wavelength and one
dex in luminosity spectral density. The figure \ref{SED_fit}
highlights the importance of a consistent photometry capturing the
total flux in each band along the SED in order to apply an accurate
spectral fitting analysis. The figure \ref{SED_fit} also illustrates
the comparison of this SED with its best fitted spectral template from
a synthetic spectral library in \citet{Hernandez-Fernandez_Tesis}.

\section{Discussion}
\label{sec:discussion}

In this work, we build up an extended catalogue of galaxies belonging
to a sample of nearby clusters carefully selected to minimize cluster
selection bias and to include a large diversity of cluster
properties. Especial care has been exercised to follow a appropriate
methodology producing a self consistent spectrophotometry along the
SED. In this section we discuss the general properties of the selected
clusters, together with the spectral characterization of their
galaxies and paying especial attention to the environmental trends of
the sample.

\subsection{X-ray luminosity vs. velocity dispersion} 

\begin{figure}[p]
\centering
\resizebox{0.85\hsize}{!}{\includegraphics{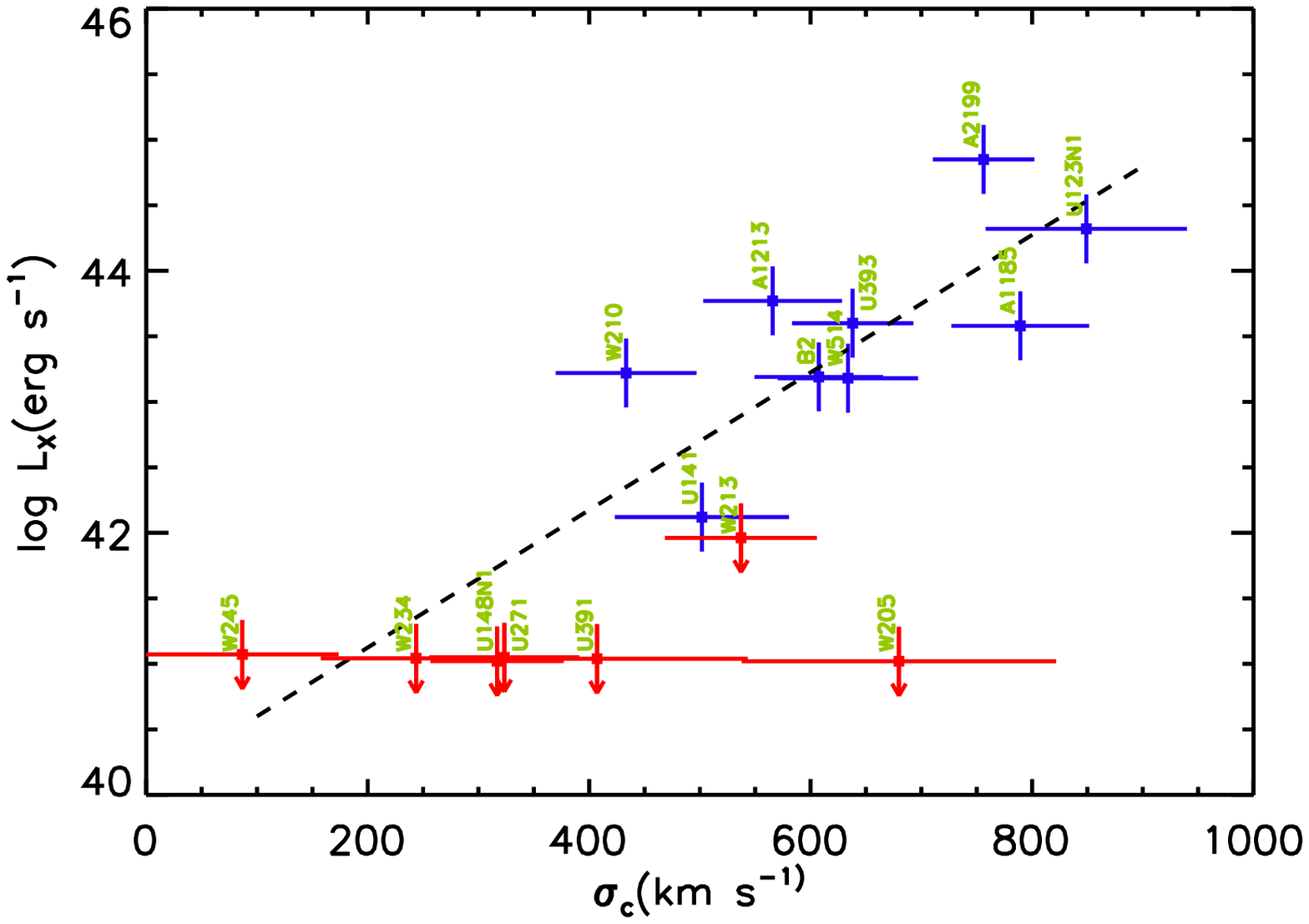}}
\caption{$L_{X}$-$\sigma_{c}$. Bolometric X-ray luminosity vs. cluster
  velocity dispersion. Blue data points indicate X-ray detections and
  the red data point with $L_{X}$$\sim$10$^{42}$ erg s$^{-1}$
  indicates a confident upper limit in X-ray luminosity. Red data
  points set to X-ray luminosities $\sim$10$^{41}$ erg s$^{-1}$ are
  associated to undetected X-ray sources. These data points are
  slightly displaced from $L_{X}$=10$^{41}$ erg s$^{-1}$ for the sake
  of clarity. The dashed line represent the $L_{X}$-$\sigma_{c}$
  relation from \citet{Mahdavi&Geller_2001}. The cluster identifier in
  the plot are codified in the following way: A=ABELL, B2=B2
  1621+38:[MLO2002] CLUSTER, N=NED, U=UGCl, W=WBL.}
\label{Lx_sigma}
\end{figure}

In figure \ref{Lx_sigma}, we plot bolometric X-ray luminosity
vs. cluster velocity dispersion, the $L_{X}$-$\sigma_{c}$ relation,
for the cluster sample. The velocity dispersion and the associated
errors are computed assuming the procedure proposed by
\citet{Poggianti_et_al_2006}. The bolometric X-ray luminosity values
are taken from \citet{Mahdavi_et_al_2000} and
\citet{Mahdavi&Geller_2001}, assigning an uncertainty of 30\% to the
X-ray luminosity in the same way as \citet{Mahdavi&Geller_2001}.

The $L_{X}$-$\sigma_{c}$ relation for the galaxy clusters with
associated X-ray detection (nine clusters) or an associated upper
X-ray flux limit (WBL 213) follow in a consistent way the
$L_{X}$$\propto$$\sigma_{c}^{4.4}$ relation found by
\citet{Mahdavi&Geller_2001} for a sample of 280 galaxy clusters. For
some clusters of the sample, we did not find an associated X-ray
source in \citet{Mahdavi&Geller_2001} catalogue neither one NED object
with X-ray associated flux ({\bf{GGroups}}, {\bf{GClusters}} or
{\bf{Xray}} source) clearly associated to these clusters. Also, we
know there is no sources with X-ray bolometric luminosities under
10$^{41}$ erg s$^{-1}$ in \citet{Mahdavi&Geller_2001}
catalogue. Assuming these clusters are around or under this X-ray
luminosity (with the typical uncertainties for these X-ray
luminosities) this group would show a locus consistent with
$L_{X}$$\propto$$\sigma_{c}^{4.4}$ trend, except for the cluster WBL
205. In this cluster, $\sigma_{c}$ is overestimated due to WBL 205 is
clearly formed by two dynamical substructures (see figure
\ref{czh_CS2}).

\subsection{Distribution and radial trend of the local galaxy density $\Sigma_{5}$}

\begin{figure}[p]
\centering
\resizebox{0.85\hsize}{!}{\includegraphics{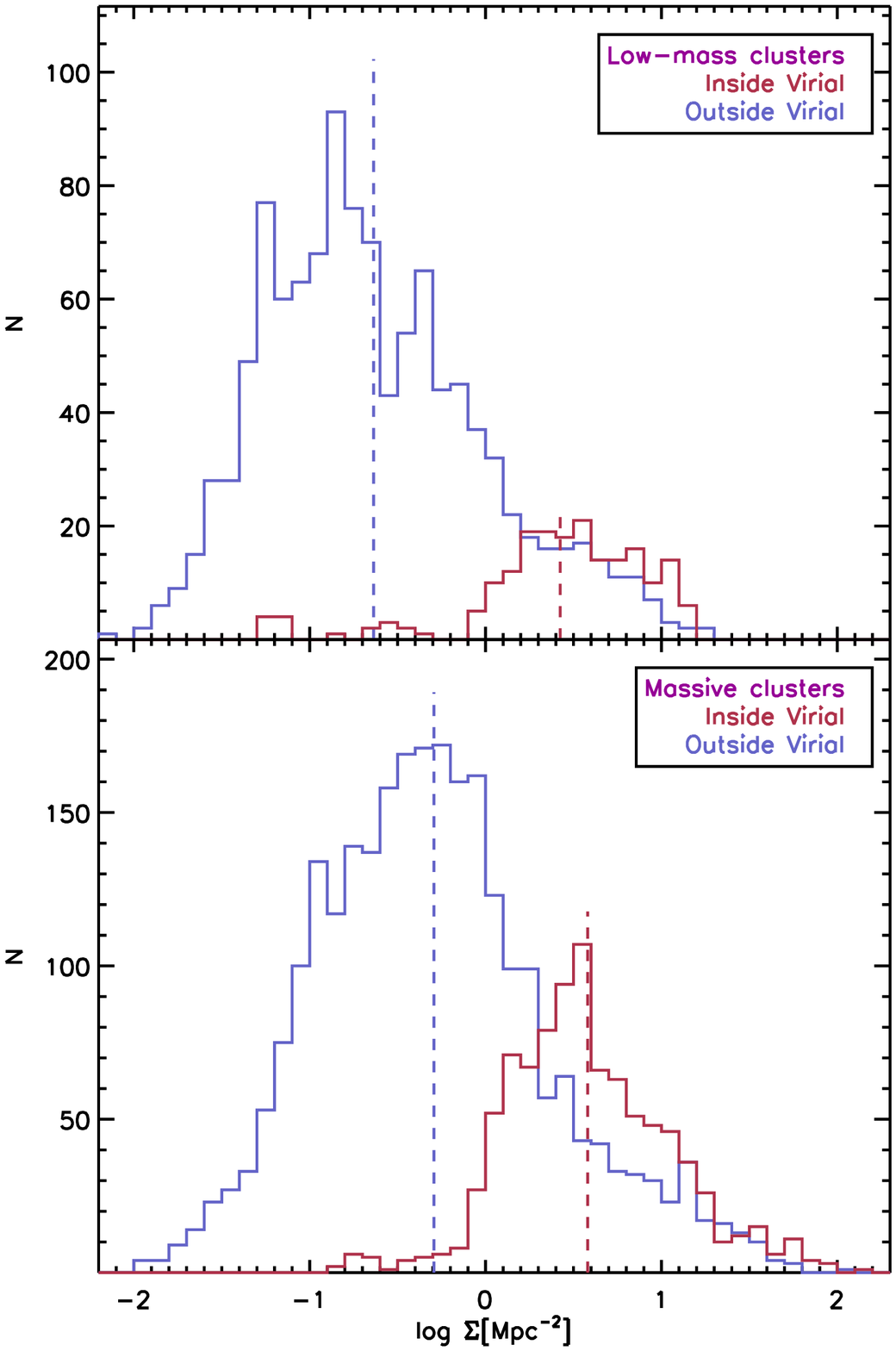}}
\caption{$\Sigma_{5}$ distribution. Reddish/bluish histograms
  correspond to galaxies inside/outside virial regions. Top panel
  show low-mass clusters $\Sigma_{5}$ distribution and the bottom
  panel, massive clusters $\Sigma_{5}$ distribution. Vertical dashed
  lines show the mean value of $\Sigma_{5}$ distribution in each
  case.}
\label{s5_dist}
\end{figure}

In Figure \ref{s5_dist}, we plot the distribution of local galaxy
density of the cluster galaxy sample. We choose $\Sigma_{5}$ as local
density estimator following \cite{Balogh_et_al_2004}; this density is
computed for each galaxy inside a circle containing up to the fifth
neighboring galaxies more luminous than $M_{r}$=-20.6 with radial
velocities not farther than 1000 km s$^{-1}$ from the radial velocity
of each galaxy:

\begin{equation}
\Sigma_{5} = \frac{5}{4 \pi r^{2}_{5}}
\end{equation}

with $r_{5}$ the distance to the fifth neighboring galaxy more
luminous than $M_{r}$=-20.6 within $\pm$ 1000 km s$^{-1}$ in radial
velocity. We reject from $\Sigma_{5}$ distributions galaxies with
``edge effects''; those galaxies which some of their fifth first
neighbors is placed far from the radial limits of galaxy sample (7
Mpc) or with a radial velocity out of the limits given by
$\pm$5$\sigma_{c}$ around the cluster redshift. We consider four
galaxy subsamples in two intervals of velocity dispersion of the
parent cluster ($\sigma_{c}$$<$550 km s$^{-1}$ - low-mass clusters and
$\sigma_{c}$$>$550 km s$^{-1}$ - massive clusters) and segregated by
their membership to virial regions. The threshold for the cluster
velocity dispersion $\sigma_{c}$=550 km s$^{-1}$ between the low-mass
and the massive clusters approximately matches a gravitational mass of
2$\cdot$10$^{14}$$M_{\odot}$ \citep{Cox_2000}, a similar value to the
characteristic mass of the distribution of cluster mass
\citep{Henry&Arnaud_1991}. Also, \citet{Poggianti_et_al_2006} choose a
similar value for $\sigma_{c}$ as a boundary between two distinct
cluster environments with regard to their star formation activity; the
massive clusters (those with a high $\sigma_{c}$) are extremely
hostile environments for star formation activity. They found a
different trend of the [OII] emission-line fraction with the
$\sigma_{c}$ in these two cluster environments. The membership to the
virial regions is assigned to galaxies inside a projected radius of
$r_{200}$ of each cluster and under the general caustic profile in a
phase diagram obtained by \citet{CAIRNS_I} for a sample of clusters in
the Local Universe.

In a first look to figure \ref{s5_dist}, the $\Sigma_{5}$ ranges from
$\sim$10$^{-2}$ to $\sim$10$^{2}$, a more broad range than the range
of $\Sigma_{5}$ distribution shown by \citet{Balogh_et_al_2004} for
two galaxy sample from SDSS DR1 \citep[Sloan Digital Sky Survey Data
  Release I,][]{SDSS_I} and the $\Sigma_{5}$ distribution of 2dFGRS
\citep[Two degrees Field Galaxy Redshift Survey,][]{2dFGRS} that go
from $\sim$3$\cdot$10$^{-2}$ to $\sim$30. In the higher density side,
this difference comes from the lower statistics of this two samples
($\sim$186240 galaxies for SDSS DR1 and $\sim$220000 for 2dFGRS)
versus the SDSS DR6 with $\sim$790220 galaxies i.e. this release
contains a higher number of galaxies from the highest density regions,
the clusters.

The $\Sigma_{5}$ distribution of virial regions occupy the range
-1$\lesssim$log$\Sigma_{5}$$\lesssim$1.2 in both cases, massive
clusters and low-mass clusters. Although, the high density tail of
massive clusters (log $\Sigma_{5}$$>$1.2) is absent in the low-mass
clusters. In addition, the mean of $\Sigma_{5}$ for massive clusters
(log $\Sigma_{5}$$\approx$0.6) is $\approx$0.2 dex higher than the
mean of $\Sigma_{5}$ for low-mass clusters. We apply a
kolmogorov-Smirnov test to the $\Sigma_{5}$ distributions of virial
regions from the low-mass clusters and the massive clusters. They have
a probability of $\sim$4\% to come from the same parent population, so
they are statistically distinguishable.

The $\Sigma_{5}$ distribution of galaxies from the outskirts present a
common range \linebreak
(-2$\lesssim$log$\Sigma_{5}$$\lesssim$1.3). Further, two differences
are noticed: (1) the presence of a high density tail
(log$\Sigma_{5}$$\gtrsim$1.3) in massive clusters and (2) the mean of
$\Sigma_{5}$ in the outskirts of massive clusters (log
$\Sigma_{5}$$\approx$-0.3) is $\approx$0.35 dex higher than the
corresponding mean for the low-mass clusters.

The difference between the mean of $\Sigma_{5}$ for galaxies in virial
regions and galaxies from the outskirts is more than one dex for the
low-mass clusters versus the difference for massive clusters which is
$\approx$0.9 dex. The overlapping in the high density side of
$\Sigma_{5}$ distributions between virial regions and the outskirts
can be explained in the following way. The sample is designed
following a set of observational constrains described in section
\ref{sec:clust_samp} but the galaxy substructures around the virial
region of selected clusters in the sample may not fulfill those
constrains. So, there may be galaxy structures in the outskirts of
virial regions as massive as their parent cluster, the way one would
expect from the similarity of the high density tails between virial
regions and outskirts. Anyway, there is a $\Sigma_{5}$ interval below
log$\Sigma_{5}$$\sim$1 where the galaxy subsample from the outskirts
prevails over the galaxies from virial regions. Also, the absence of
the highest density tail in the low-mass clusters is a clear evidence
of the local density reach up their higest values in the more massive
galaxy structures, the richest clusters.

\begin{figure}[p]
\centering
\resizebox{0.85\hsize}{!}{\includegraphics{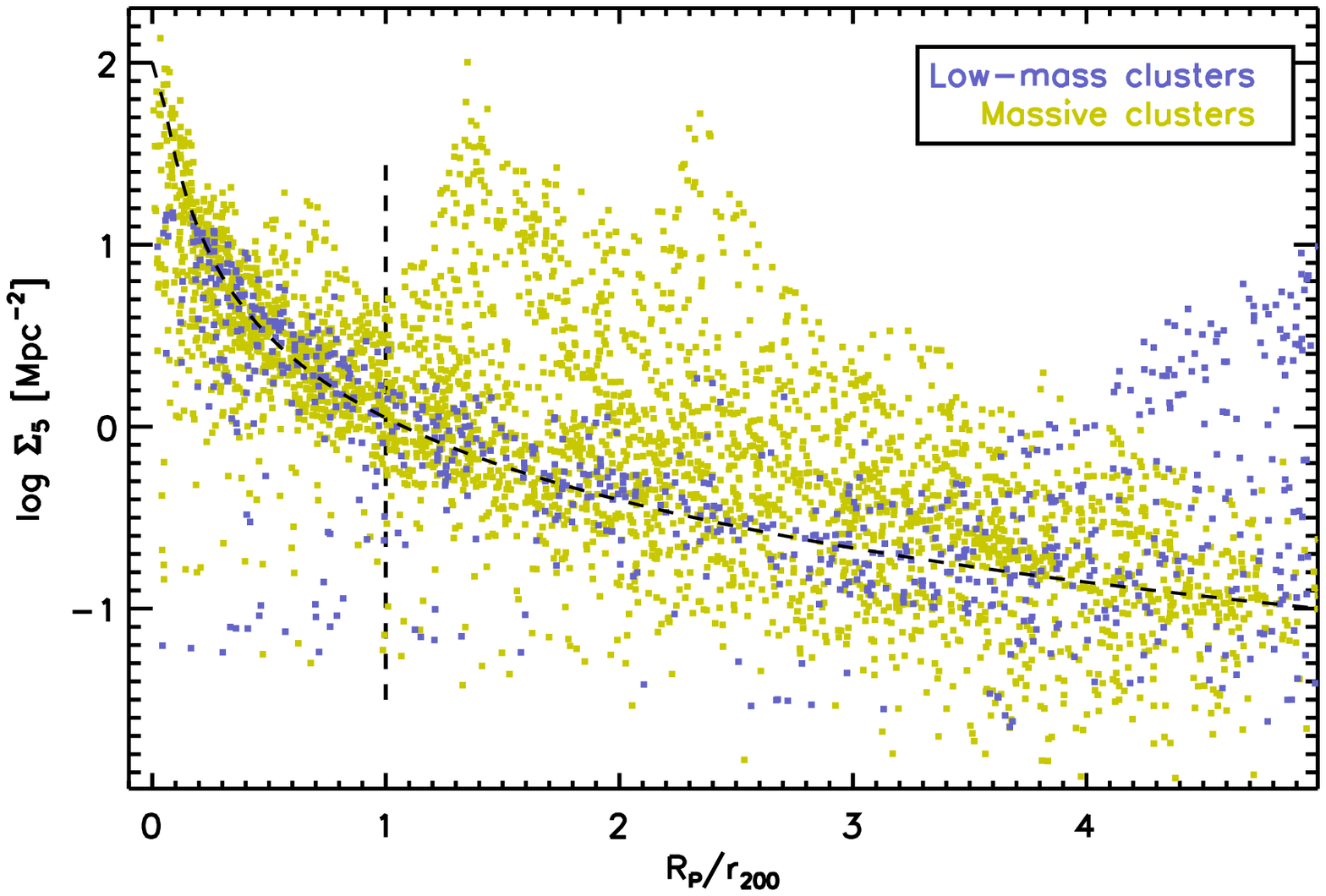}}
\caption{$\Sigma_{5}$ vs. $R_{P}$/$r_{200}$. The projected density to
  fifth neighbor versus projected radius normalized to radius 200,
  $r_{P}$$\equiv$$R_{P}$/$r_{200}$. The legend identifies the
  subsample of clusters. The vertical dashed line delimit the a
  projected radius equal to $r_{200}$. The dashed curve is a King
  profile fit by eye to the main trend.}
\label{s5_rr200}
\end{figure}

In Figure \ref{s5_rr200}, it can be seen a broad trend for the
$\Sigma_{5}$-$r_{P}$ relation ($r_{P}$$\equiv$$R_{P}$/$r_{200}$), with
the highest densities near to cluster centers at the top of a
correlation in the virial region and the lowest densities far from the
virial regions in the same way as found by \citet{CAIRNS_III}. We find
the $\Sigma_{5}$-$r_{P}$ relation is biased in $\sim$0.5 dex toward
lower densities regarding the $\Sigma_{5}$-$r_{P}$ relation obtained
by \citet{CAIRNS_III}. This bias would come from a deeper luminosity
cut for neighboring galaxies which is set to $M_{K}$=-22.7, enlarging
the sample of neighboring galaxies devoted to compute the local
density. The density-radius trend shows a more broad relation outside
the virial region than the trend for the virial region. This came from
the presence of galaxy structures which have peaks of density similar
to those in the center of virial regions (e.g. ABELL 2197 or B2
1621+38:[MLO2002]). The massive clusters show galaxy structures with
higher densities in the outskirts of virial regions than the low-mass
clusters. Both the massive and low-mass clusters follow a similar
trend inside the virial region, but the low-mass clusters reach only
up to log$\Sigma_{5}$$\sim$1.2 avoiding the highest density tail while
the massive clusters reach up to log$\Sigma_{5}$$\sim$2. In the
outskirts, the major concentration of galaxies in the lower side of
the relation traces a common trend for both massive and low-mass
clusters.

In Figure \ref{s5_rr200}, we plot a King profile \citep{King_1966} fit
by eye to the major concentration of galaxy points along the
$\Sigma_{5}$-$r_{p}$ relation:

\begin{equation}
log \Sigma = log \Sigma^{0} - \beta log \left[ 1 + \left(
  \frac{r_{p}}{r_{c}} \right)^{2} \right] , ~~ \textrm{with} ~\Sigma^{0} = 2,
~\beta=0.75, ~r_{c}=0.05
\label{eq:king}
\end{equation} 

The king profile was initially applied to the projected galaxy density
of Coma cluster by \cite{King_1972}. The fit from equation
\ref{eq:king} in the $\Sigma_{5}$-$r_{p}$ relation seems to reconcile
the narrow relation inside the virial region with the concentration in
the lower side of the relation for the surroundings. Both, the massive
and low-mass clusters seem to follow the same relation along the
clustercentric radius, with the massive clusters occupying the top of
the density-radius fit.


\subsection{Galaxy projected distribution}
\label{sec:gal_proj_dist}

In this section we stress the relevance of a detailed mapping of the
sky distribution of different galaxy populations as a tool for the
study of environmental trends of galaxy properties. Such study is
illustrated here for Abell 1185, a massive cluster of our sample. A
similar analysis extended to the complete cluster sample is out of the
scope of this paper and will be presented elsewhere
\citep{Hernandez-Fernandez_et_al_2011}. We segregate galaxy
populations according to their luminosity between giant galaxies
$M_{r}$$<$-19.5 and low-luminosity galaxies -19.5$<$$M_{r}$$<$-18, and
also to their spectral type between passive galaxies and star-forming
galaxies. In order to differentiate passive galaxies from star-forming
galaxies, we take advantage of the ($NUV$-$r$) vs. ($u$-$r$)
color-color diagram. We assume a galaxy is a passive galaxy whether
its colors fulfill the following prescription:

\begin{displaymath}
\left\{ \begin{array}{ll}
$NUV$-$r$ > 4.9                & \textrm{for } $u$-$r$ < 2.175 \\
$NUV$-$r$ > -2($u$-$r$) + 9.25 & \textrm{for } $u$-$r$ > 2.175 \\
  $u$-$r$ > 2.22               & \textrm{whether there is no GALEX counterpart} \label{eq:NUVr_ur_cut}
\end{array} \right.
\end{displaymath}

\begin{figure}[p]
\centering
\resizebox{1.00\hsize}{!}{\includegraphics{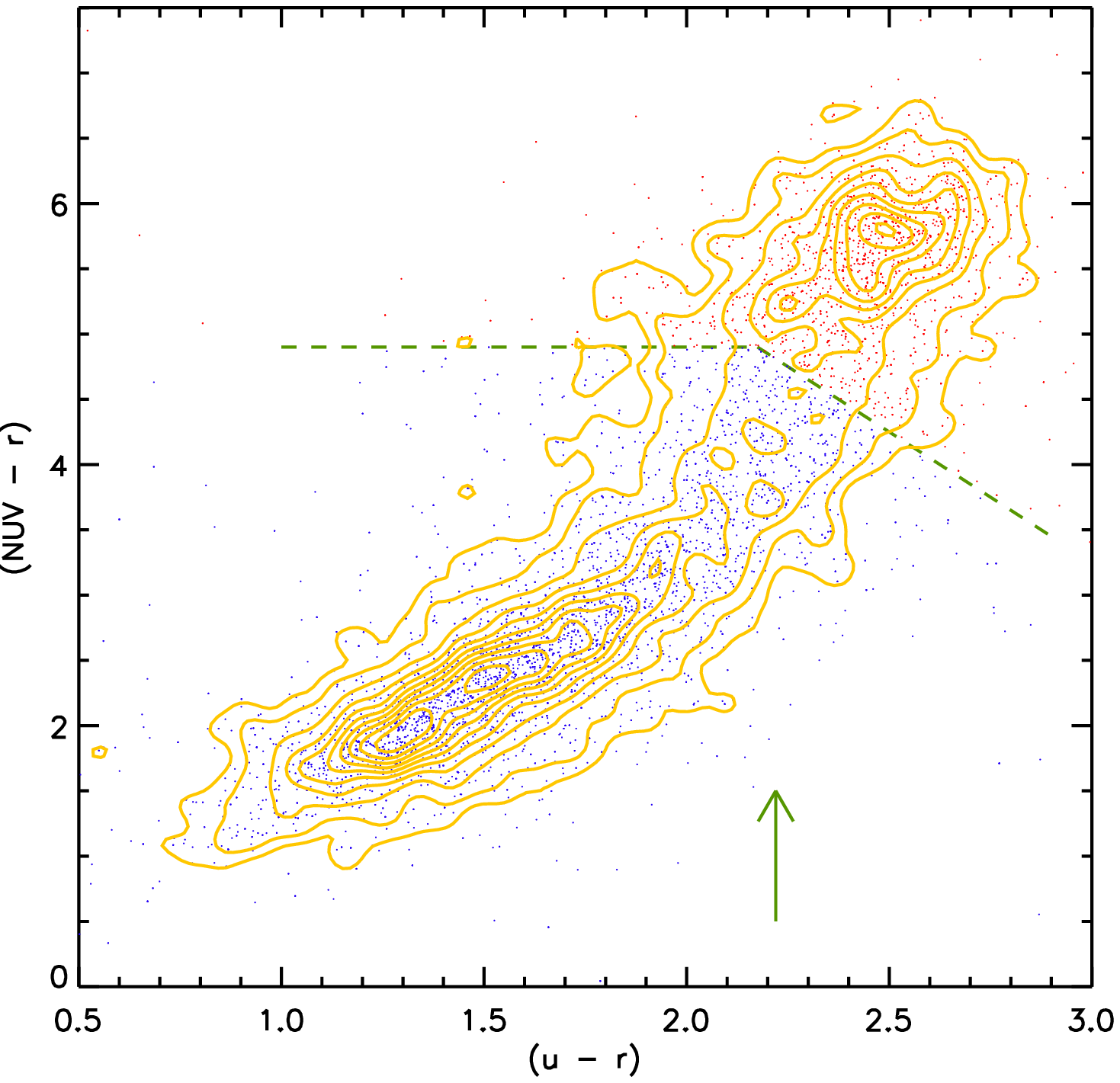}}
\caption{($NUV$-$r$) vs. ($u$-$r$). Yellow isocontours represents the
  isodensity contours of galaxies. Green dashed broken line is the
  color-color cut for galaxies with UV detection. Green vertical arrow
  points out to the $u$-$r$ cut for galaxies without UV data. Blue and
  red points represent, respectively, star-forming and passive
  galaxies under the prescription shown in the section
  \ref{eq:NUVr_ur_cut}.}
\label{NUVr_ur}
\end{figure}

As can be seen in Figure \ref{NUVr_ur}, this selection seems more
accurated to differentiate star-forming galaxies from passive galaxies
than the $u$-$r$ color cut proposed by
\citet{Strateva_et_al_2001}. The broken line trace the minimum in the
density of data points of ($NUV$-$r$) vs. ($u$-$r$) diagram between
the maximum of density regarding the "red sequence'' and the more
extended maximum tracing the "blue cloud''. The left side of the
frontier tries to include in the passive galaxy side the locus of
evolved "E+A'' galaxies in a UV-optical diagram
\citep{Kaviraj_E+A}. In the case there is no UV data for a galaxy, we
apply the Strateva's $u$-$r$ cut.

In a forthcoming paper \citep{Hernandez-Fernandez_et_al_2011_NUVr}, we
take advantage of this UV-optical color frontier in order to make up a
sample of star-forming galaxies in clusters. We analyze the spatial
variation of distributions of spectral properties for this sample of
star-forming galaxies. We find statistically significant differences,
applying a Kolmogorov-Smirnov test, in those distributions throughout
different environments i.e. virial regions, infall regions and field
environment.

\begin{figure}[p]
\centering
\resizebox{0.75\vsize}{!}{\includegraphics{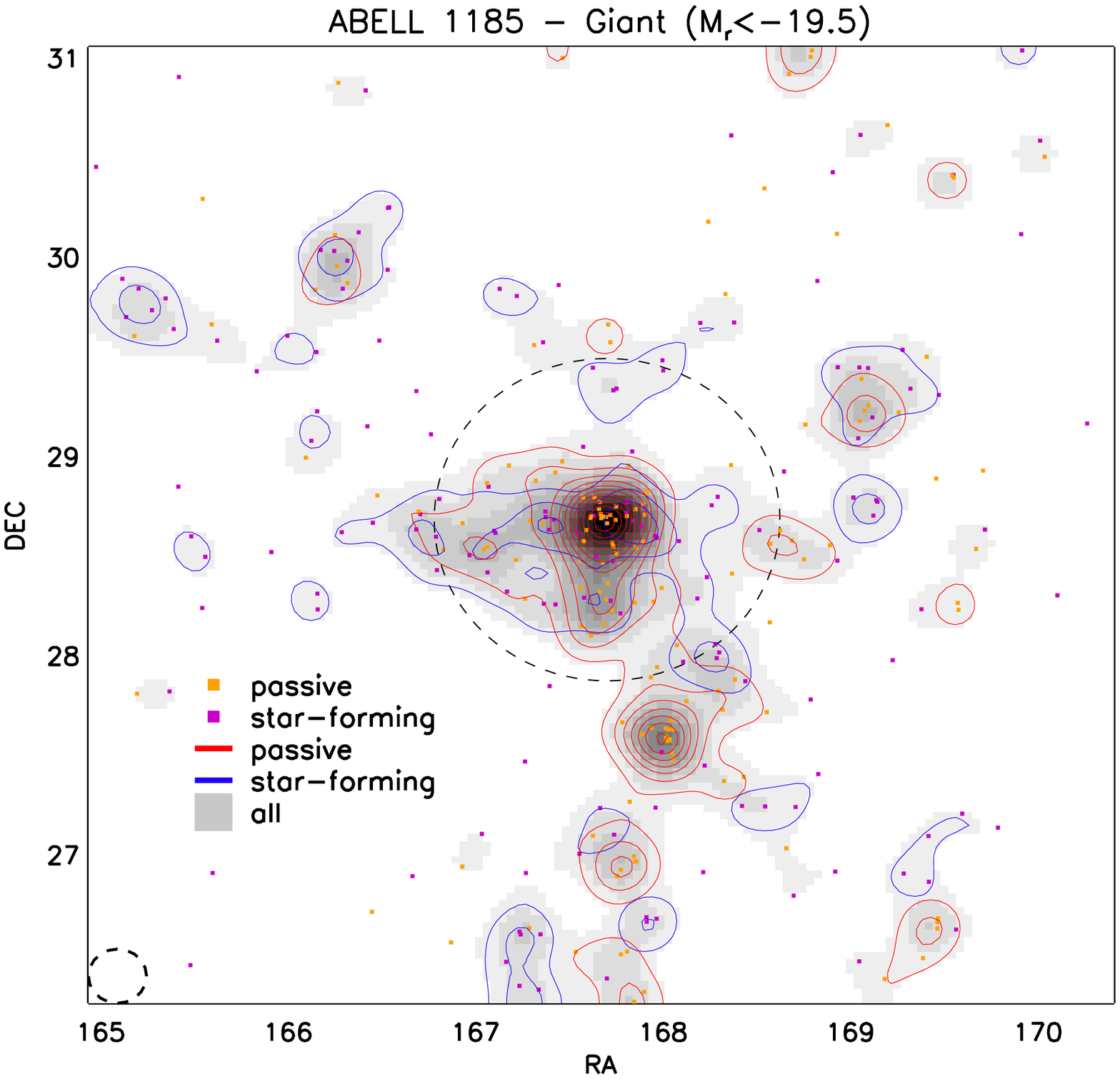}}
\caption{Sky projected density of giant $M_{r}$$<$-19.5 galaxies
  around ABELL~1185. The grey intensity map corresponds to sky
  projected density of giant galaxies (both passive and star-forming
  galaxies). Orange/Magenta points represent sky position and red/blue
  contours represent isodensity lines of the sky projected density of
  giant galaxies classified as passive/star-forming galaxies. The
  lowest density contour correspond to a $\Sigma$=3 gal/Mpc$^{2}$ and
  the contours are equispaced in $\Delta$$\Sigma$=3 gal/Mpc$^{2}$ up
  to the maximum in density. The circle in the lower-left corner shows
  the FWHM size of gaussian kernel to compute the density map.}
\label{A1185_G}
\end{figure}

\begin{figure}[p]
\centering
\resizebox{0.75\vsize}{!}{\includegraphics{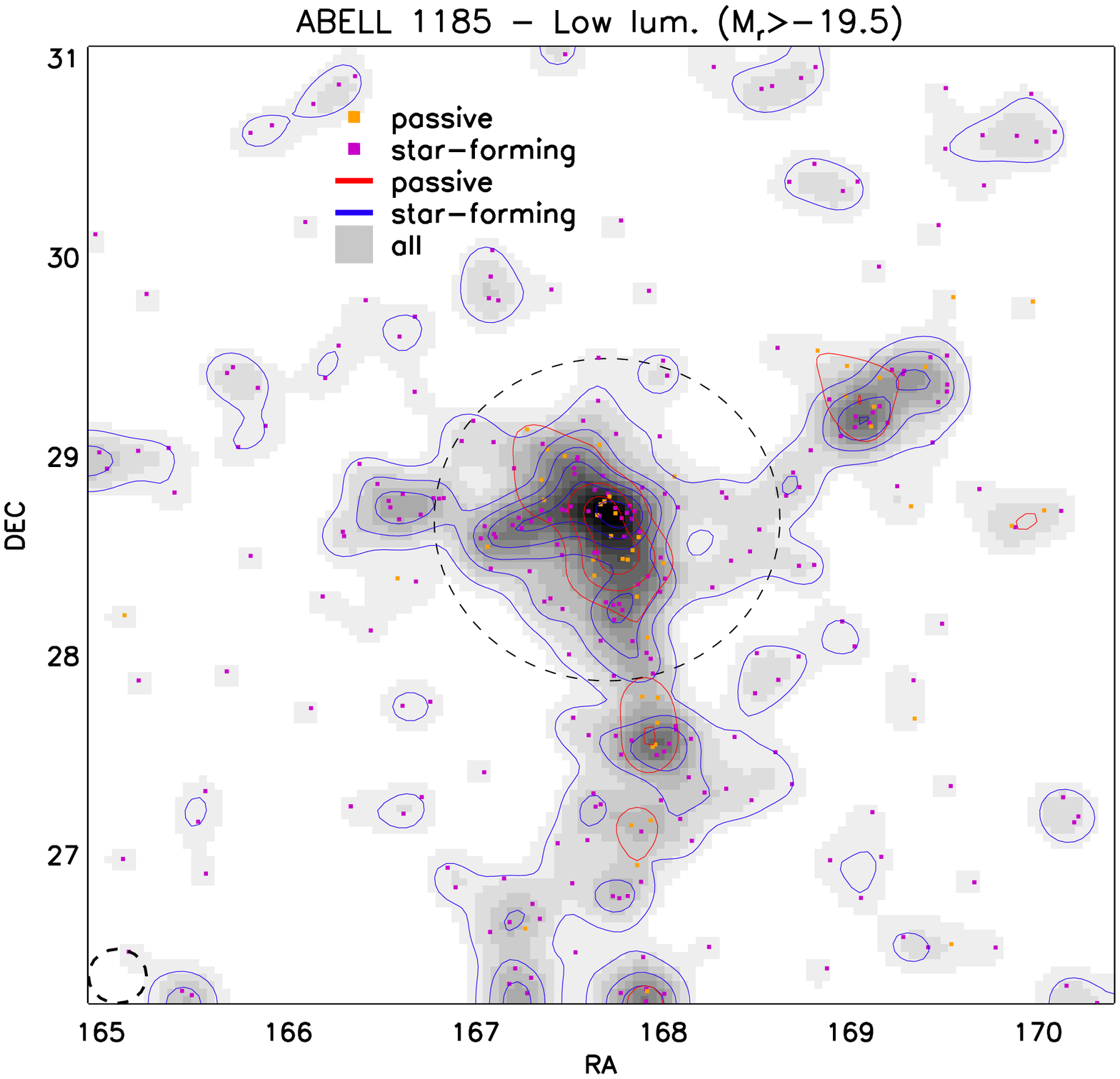}}
\caption{Sky projected density of low-luminosity
  \hbox{-19.5$<$$M_{r}$$<$-18} galaxies around ABELL~1185. Color code,
  isodensity lines and the rest of elements of the figure are defined
  in the same way as figure \ref{A1185_G}.}
\label{A1185_D}
\end{figure}

The figures \ref{A1185_G} and \ref{A1185_D} show the sky distribution
in Abell 1185 of giant galaxies $M_{r}$$<$-19.5 and low-luminosity
galaxies -19.5$<$$M_{r}$$<$-18, respectively. Both figures, also show the
sky distribution of star-forming and passive galaxies. 

In Figure \ref{A1185_G}, it can be seen the main concentration of
giant galaxies from the virial region of ABELL~1185 around
RA$\sim$167.75 deg DEC$\sim$28.5 framed by the dashed circle. In the
same way, there are evident galaxy agglomerations around the virial
region of ABELL~1185 with less strucutural entity than ABELL~1185,
except for the group of galaxies in the south side around
RA$\sim$167.8 deg DEC$\sim$27.5. We check the redshift distribution of
galaxies around this location and find an evident dynamical structure
around $z$=0.034. This aggregate of galaxies, showing a strikingly
high fraction of passive giant galaxies, can be linked with the "bare"
massive-cluster cores identified by
\citet{Poggianti_et_al_2006}. \citet{Poggianti_et_al_2006} propose, as
a hypothesis, that systems close to more massive structures, thus
embedded in a massive superstructure, have a different galactic
content than completely isolated galaxy systems of similar mass. They
suggest these objects lived in regions that were very dense at high
redshift but failed to acquire star-forming galaxies at later times,
possibly due to the characteristics of their surrounding supercluster
environment. On the other hand, the maxima in the sky distribution of
passive giant galaxies trace the central position of the main
structures as ABELL~1185 and the ``bare core'' at the south side,
while star-forming galaxies occupy these regions with a more spread
distribution, following the general trend for clustering depending on
spectral type founds in astrophysical observations and simulations
\citep[e.g.,][]{Madgwick_et_al_2003,Springel_et_al_2005_Nat}.

We plot the sky distribution of low-luminosity
\hbox{-19.5$<$$M_{r}$$<$-18} galaxies in Figure \ref{A1185_D}. These
galaxies show a more continuous sky distribution around the central
region of ABEL 1185 connecting this region with the structures in the
south, east and west side of the cluster. This is in good agreement
with a less clustered low-luminosity population as suggested the
literature
\citep[e.g.,][]{Norberg_et_al_2002,Springel_et_al_2005_Nat}. The
star-forming galaxies occupy both the densest regions and less dense
regions, but the passive galaxies seem to inhabit preferably the
central region of the structures avoiding the field environment in the
same way as observed by \citet{Haines_et_al_2006}.

\section{Summary}
\label{sec:conclusions}

We expose the main results and conclusions of this paper in this
itemized summary:

\begin{itemize}

\item We compile a sample of galaxies which inhabits in clusters
  showing a broad range of cluster properties ($\sigma_{c}$,
  morphology, etc). This galaxy sample is oberved down to the
  luminosity frontier between giant and dwarf galaxies by the Main
  Galaxy Sample of SDSS and other galaxy surveys from UV to FIR. We
  build a spectrophotometric catalogue for this cluster galaxy sample
  with a detailed photometry for each galaxy in order to be accurate
  for spectral template fitting.

\item The clusters from the sample with X-ray detections or confident
  upper limits are consistent with the X-ray luminosity vs. cluster
  velocity dispersion $L_{X}$$\propto$$\sigma_{c}^{4.4}$ trend found
  by \citet{Mahdavi&Geller_2001}. The clustes with no X-ray fluxes in
  the literature can be reconciled with the $L_{X}$-$\sigma_{c}$ trend
  assuming an upper limit in X-ray luminosity of 10$^{41}$ erg
  s$^{-1}$, except for the case of WBL 205, a cluster with clear
  evidences of the presence of dynamical substructures.

\item The galaxy density $\Sigma_{5}$ distribution of virial regions
  are biased to higher densities with respect to the $\Sigma_{5}$
  distribution of the outskirts. The $\Sigma_{5}$ distribution of
  massive clusters (virial regions and the outskirts) shows similar
  ranges than the low-mass clusters, but they have higher averages of
  $\Sigma_{5}$ than the low-mass clusters and present a highest
  density tail which is missing in the low-mass clusters. The
  $\Sigma_{5}$ distribution of virial regions of massive clusters is
  statistically distinguishable, up to a $\sim$96 \% of probability,
  from the corresponding distribution for low-mass clusters. The
  overlapping of distributions of $\Sigma_{5}$ between virial regions
  and their outskirts at highest densities suggests the presence of
  galaxy structures in the outskirts as massive as the cluster cores.

\item The $\Sigma_{5}$-$r_{P}$ relation shows a more broad trend
  outside the virial region than the trend for the virial region, due
  to the presence of density peaks. Both the massive and low-mass
  clusters follow a similar trend inside the virial region, but the
  low-mass clusters avoid the highest density tail. This relation is
  well fitted by a King profile along the clustercentric radius, for
  both the massive and the low mass clusters.

\item ABELL~1185 shows clear evidences of galaxy structures around the
  virial region. In this cluster, low-luminosity star-forming galaxies
  are distributed along more spread structures than their giant
  counterparts, whereas low-luminosity passive galaxies avoid the
  low-density environment. Giant passive and star-forming galaxies
  share rather similar sky regions with passive galaxies exhibiting
  more cuspy distributions.

\end{itemize}




\section*{Acknowledgements}

J.D.H.F. thanks the Laboratoire d'Astrophysique de Marseille and
L'Osservatorio Astronomico di Padova for hospitality during the stays
to carry out part of this work. Special thanks are given to Veronique
Buat, Denis Burgarella and Bianca M$^{\b{a}}$ Poggianti for their help
and advice during the first stages of this work.

J.D.H.F. acknowledges financial support from the Spanish Ministerio de
Ciencia e Innovaci\'on under the FPI grant BES-2005-7570. We also
acknowledge funding by the Spanish PNAYA project ESTALLIDOS (grants
AYA2007-67965-C03-02, AYA2010-21887-C04-01) and project CSD2006 00070
``1st Science with GTC'' from the CONSOLIDER 2010 program of the
Spanish MICINN.

This publication has made use of the following resources:

\begin{itemize}

\item the NASA/IPAC Extragalactic Database (NED) which is operated by
  the Jet Propulsion Laboratory, California Institute of Technology,
  under contract with the National Aeronautics and Space
  Administration.

\item the Sloan Digital Sky Survey (SDSS) database. Funding for the
  Sloan Digital Sky Survey (SDSS) and SDSS-II has been provided by the
  Alfred P. Sloan Foundation, the Participating Institutions, the
  National Science Foundation, the U.S. Department of Energy, the
  National Aeronautics and Space Administration, the Japanese
  Monbukagakusho, and the Max Planck Society, and the Higher Education
  Funding Council for England. The SDSS Web site is
  http://www.sdss.org/. \\ The SDSS is managed by the Astrophysical
  Research Consortium (ARC) for the Participating Institutions. The
  Participating Institutions are the American Museum of Natural
  History, Astrophysical Institute Potsdam, University of Basel,
  University of Cambridge, Case Western Reserve University, The
  University of Chicago, Drexel University, Fermilab, the Institute
  for Advanced Study, the Japan Participation Group, The Johns Hopkins
  University, the Joint Institute for Nuclear Astrophysics, the Kavli
  Institute for Particle Astrophysics and Cosmology, the Korean
  Scientist Group, the Chinese Academy of Sciences (LAMOST), Los
  Alamos National Laboratory, the Max-Planck-Institute for Astronomy
  (MPIA), the Max-Planck-Institute for Astrophysics (MPA), New Mexico
  State University, Ohio State University, University of Pittsburgh,
  University of Portsmouth, Princeton University, the United States
  Naval Observatory, and the University of Washington.

\item the Galaxy Evolution Explorer (GALEX), which is a NASA mission
  managed by the Jet Propulsion Laboratory and launched in 2003
  April. We gratefully acknowledge NASA's support for the
  construction, operation, and science analysis for the GALEX mission,
  developed in cooperation with the Centre National d'Etudes Spatiales
  of France and the Korean Ministry of Science and Technology.

\item the Two Micron All Sky Survey (2MASS), which is a joint project
  of the University of Massachusetts and the Infrared Processing and
  Analysis Center at the California Institute of Technology, funded by
  the National Aeronautics and Space Administration and the National
  Science Foundation.

\item the NASA/IPAC Infrared Science Archive, which is operated by the
  Jet Propulsion Laboratory, California Institute of Technology, under
  contract with the National Aeronautics and Space Administration.

\end{itemize}


\bibliographystyle{apj} 
\bibliography{/media/_/home/jonatan/Tesis/muestra_GALEX/Sumario/Bibliografia_v2}

\begin{thebibliography}{75}
\expandafter\ifx\csname natexlab\endcsname\relax\def\natexlab#1{#1}\fi

\bibitem[{{Abazajian} {et~al.}(2004){Abazajian}, {Adelman-McCarthy},
  {Ag{\"u}eros}, {Allam}, {Anderson}, {Anderson}, {Annis}, {Bahcall}, {Baldry},
  {Bastian}, {Berlind}, {Bernardi}, {Blanton}, {Bochanski}, {Boroski},
  {Briggs}, {Brinkmann}, {Brunner}, {Budav{\'a}ri}, {Carey}, {Carliles},
  {Castander}, {Connolly}, {Csabai}, {Doi}, {Dong}, {Eisenstein}, {Evans},
  {Fan}, {Finkbeiner}, {Friedman}, {Frieman}, {Fukugita}, {Gal}, {Gillespie},
  {Glazebrook}, {Gray}, {Grebel}, {Gunn}, {Gurbani}, {Hall}, {Hamabe},
  {Harris}, {Harris}, {Harvanek}, {Heckman}, {Hendry}, {Hennessy}, {Hindsley},
  {Hogan}, {Hogg}, {Holmgren}, {Ichikawa}, {Ichikawa}, {Ivezi{\'c}}, {Jester},
  {Johnston}, {Jorgensen}, {Kent}, {Kleinman}, {Knapp}, {Kniazev}, {Kron},
  {Krzesinski}, {Kunszt}, {Kuropatkin}, {Lamb}, {Lampeitl}, {Lee}, {Leger},
  {Li}, {Lin}, {Loh}, {Long}, {Loveday}, {Lupton}, {Malik}, {Margon},
  {Matsubara}, {McGehee}, {McKay}, {Meiksin}, {Munn}, {Nakajima}, {Nash},
  {Neilsen}, {Newberg}, {Newman}, {Nichol}, {Nicinski}, {Nieto-Santisteban},
  {Nitta}, {Okamura}, {O'Mullane}, {Ostriker}, {Owen}, {Padmanabhan},
  {Peoples}, {Pier}, {Pope}, {Quinn}, {Richards}, {Richmond}, {Rix}, {Rockosi},
  {Schlegel}, {Schneider}, {Scranton}, {Sekiguchi}, {Seljak}, {Sergey},
  {Sesar}, {Sheldon}, {Shimasaku}, {Siegmund}, {Silvestri}, {Smith}, {Smol{\v
  c}i{\'c}}, {Snedden}, {Stebbins}, {Stoughton}, {Strauss}, {SubbaRao},
  {Szalay}, {Szapudi}, {Szkody}, {Szokoly}, {Tegmark}, {Teodoro}, {Thakar},
  {Tremonti}, {Tucker}, {Uomoto}, {Vanden Berk}, {Vandenberg}, {Vogeley},
  {Voges}, {Vogt}, {Walkowicz}, {Wang}, {Weinberg}, {West}, {White}, {Wilhite},
  {Xu}, {Yanny}, {Yasuda}, {Yip}, {Yocum}, {York}, {Zehavi}, {Zibetti}, \&
  {Zucker}}]{SDSS_II}
{Abazajian}, K., {Adelman-McCarthy}, J.~K., {Ag{\"u}eros}, M.~A., {et~al.}
  2004, \aj, 128, 502

\bibitem[{{Abazajian} {et~al.}(2003){Abazajian}, {Adelman-McCarthy},
  {Ag{\"u}eros}, {Allam}, {Anderson}, {Annis}, {Bahcall}, {Baldry}, {Bastian},
  {Berlind}, {Bernardi}, {Blanton}, {Blythe}, {Bochanski}, {Boroski},
  {Brewington}, {Briggs}, {Brinkmann}, {Brunner}, {Budav{\'a}ri}, {Carey},
  {Carr}, {Castander}, {Chiu}, {Collinge}, {Connolly}, {Covey}, {Csabai},
  {Dalcanton}, {Dodelson}, {Doi}, {Dong}, {Eisenstein}, {Evans}, {Fan},
  {Feldman}, {Finkbeiner}, {Friedman}, {Frieman}, {Fukugita}, {Gal},
  {Gillespie}, {Glazebrook}, {Gonzalez}, {Gray}, {Grebel}, {Grodnicki}, {Gunn},
  {Gurbani}, {Hall}, {Hao}, {Harbeck}, {Harris}, {Harris}, {Harvanek},
  {Hawley}, {Heckman}, {Helmboldt}, {Hendry}, {Hennessy}, {Hindsley}, {Hogg},
  {Holmgren}, {Holtzman}, {Homer}, {Hui}, {Ichikawa}, {Ichikawa}, {Inkmann},
  {Ivezi{\'c}}, {Jester}, {Johnston}, {Jordan}, {Jordan}, {Jorgensen},
  {Juri{\'c}}, {Kauffmann}, {Kent}, {Kleinman}, {Knapp}, {Kniazev}, {Kron},
  {Krzesi{\'n}ski}, {Kunszt}, {Kuropatkin}, {Lamb}, {Lampeitl}, {Laubscher},
  {Lee}, {Leger}, {Li}, {Lidz}, {Lin}, {Loh}, {Long}, {Loveday}, {Lupton},
  {Malik}, {Margon}, {McGehee}, {McKay}, {Meiksin}, {Miknaitis}, {Moorthy},
  {Munn}, {Murphy}, {Nakajima}, {Narayanan}, {Nash}, {Neilsen}, {Newberg},
  {Newman}, {Nichol}, {Nicinski}, {Nieto-Santisteban}, {Nitta}, {Odenkirchen},
  {Okamura}, {Ostriker}, {Owen}, {Padmanabhan}, {Peoples}, {Pier}, {Pindor},
  {Pope}, {Quinn}, {Rafikov}, {Raymond}, {Richards}, {Richmond}, {Rix},
  {Rockosi}, {Schaye}, {Schlegel}, {Schneider}, {Schroeder}, {Scranton},
  {Sekiguchi}, {Seljak}, {Sergey}, {Sesar}, {Sheldon}, {Shimasaku}, {Siegmund},
  {Silvestri}, {Sinisgalli}, {Sirko}, {Smith}, {Smol{\v c}i{\'c}}, {Snedden},
  {Stebbins}, {Steinhardt}, {Stinson}, {Stoughton}, {Strateva}, {Strauss},
  {SubbaRao}, {Szalay}, {Szapudi}, {Szkody}, {Tasca}, {Tegmark}, {Thakar},
  {Tremonti}, {Tucker}, {Uomoto}, {Vanden Berk}, {Vandenberg}, {Vogeley},
  {Voges}, {Vogt}, {Walkowicz}, {Weinberg}, {West}, {White}, {Wilhite},
  {Willman}, {Xu}, {Yanny}, {Yarger}, {Yasuda}, {Yip}, {Yocum}, {York},
  {Zakamska}, {Zehavi}, {Zheng}, {Zibetti}, \& {Zucker}}]{SDSS_I}
{Abazajian}, K., {Adelman-McCarthy}, J.~K., {Ag{\"u}eros}, M.~A., {et~al.}
  2003, \aj, 126, 2081

\bibitem[{{Abell}(1958)}]{Abell_1958}
{Abell}, G.~O. 1958, {\apjs}, 3, 211

\bibitem[{{Adelman-McCarthy} {et~al.}(2008){Adelman-McCarthy}, {Ag{\"u}eros},
  {Allam}, {Allende Prieto}, {Anderson}, {Anderson}, {Annis}, {Bahcall},
  {Bailer-Jones}, {Baldry}, {Barentine}, {Bassett}, {Becker}, {Beers}, {Bell},
  {Berlind}, {Bernardi}, {Blanton}, {Bochanski}, {Boroski}, {Brinchmann},
  {Brinkmann}, {Brunner}, {Budav{\'a}ri}, {Carliles}, {Carr}, {Castander},
  {Cinabro}, {Cool}, {Covey}, {Csabai}, {Cunha}, {Davenport}, {Dilday}, {Doi},
  {Eisenstein}, {Evans}, {Fan}, {Finkbeiner}, {Friedman}, {Frieman},
  {Fukugita}, {G{\"a}nsicke}, {Gates}, {Gillespie}, {Glazebrook}, {Gray},
  {Grebel}, {Gunn}, {Gurbani}, {Hall}, {Harding}, {Harvanek}, {Hawley},
  {Hayes}, {Heckman}, {Hendry}, {Hindsley}, {Hirata}, {Hogan}, {Hogg}, {Hyde},
  {Ichikawa}, {Ivezi{\'c}}, {Jester}, {Johnson}, {Jorgensen}, {Juri{\'c}},
  {Kent}, {Kessler}, {Kleinman}, {Knapp}, {Kron}, {Krzesinski}, {Kuropatkin},
  {Lamb}, {Lampeitl}, {Lebedeva}, {Lee}, {Leger}, {L{\'e}pine}, {Lima}, {Lin},
  {Long}, {Loomis}, {Loveday}, {Lupton}, {Malanushenko}, {Malanushenko},
  {Mandelbaum}, {Margon}, {Marriner}, {Mart{\'{\i}}nez-Delgado}, {Matsubara},
  {McGehee}, {McKay}, {Meiksin}, {Morrison}, {Munn}, {Nakajima}, {Neilsen},
  {Newberg}, {Nichol}, {Nicinski}, {Nieto-Santisteban}, {Nitta}, {Okamura},
  {Owen}, {Oyaizu}, {Padmanabhan}, {Pan}, {Park}, {Peoples}, {Pier}, {Pope},
  {Purger}, {Raddick}, {Re Fiorentin}, {Richards}, {Richmond}, {Riess}, {Rix},
  {Rockosi}, {Sako}, {Schlegel}, {Schneider}, {Schreiber}, {Schwope}, {Seljak},
  {Sesar}, {Sheldon}, {Shimasaku}, {Sivarani}, {Smith}, {Snedden}, {Steinmetz},
  {Strauss}, {SubbaRao}, {Suto}, {Szalay}, {Szapudi}, {Szkody}, {Tegmark},
  {Thakar}, {Tremonti}, {Tucker}, {Uomoto}, {Vanden Berk}, {Vandenberg},
  {Vidrih}, {Vogeley}, {Voges}, {Vogt}, {Wadadekar}, {Weinberg}, {West},
  {White}, {Wilhite}, {Yanny}, {Yocum}, {York}, {Zehavi}, \&
  {Zucker}}]{Adelman-McCarthy_et_al_2008}
{Adelman-McCarthy}, J.~K., {Ag{\"u}eros}, M.~A., {Allam}, S.~S., {et~al.} 2008,
  \apjs, 175, 297

\bibitem[{{Andreon} \& {Ettori}(1999)}]{Andreon&Ettori_1999}
{Andreon}, S. \& {Ettori}, S. 1999, \apj, 516, 647

\bibitem[{{Andreon} {et~al.}(2006){Andreon}, {Quintana}, {Tajer}, {Galaz}, \&
  {Surdej}}]{Andreon_et_al_2006}
{Andreon}, S., {Quintana}, H., {Tajer}, M., {Galaz}, G., \& {Surdej}, J. 2006,
  \mnras, 365, 915

\bibitem[{{Baldwin} {et~al.}(1981){Baldwin}, {Phillips}, \&
  {Terlevich}}]{BPT_1981}
{Baldwin}, J.~A., {Phillips}, M.~M., \& {Terlevich}, R. 1981, \pasp, 93, 5

\bibitem[{{Balogh} {et~al.}(2004){Balogh}, {Eke}, {Miller}, {Lewis}, {Bower},
  {Couch}, {Nichol}, {Bland-Hawthorn}, {Baldry}, {Baugh}, {Bridges}, {Cannon},
  {Cole}, {Colless}, {Collins}, {Cross}, {Dalton}, {de Propris}, {Driver},
  {Efstathiou}, {Ellis}, {Frenk}, {Glazebrook}, {Gomez}, {Gray}, {Hawkins},
  {Jackson}, {Lahav}, {Lumsden}, {Maddox}, {Madgwick}, {Norberg}, {Peacock},
  {Percival}, {Peterson}, {Sutherland}, \& {Taylor}}]{Balogh_et_al_2004}
{Balogh}, M., {Eke}, V., {Miller}, C., {et~al.} 2004, \mnras, 348, 1355

\bibitem[{{Beers} {et~al.}(1995){Beers}, {Kriessler}, {Bird}, \&
  {Huchra}}]{Beers_et_al_1995}
{Beers}, T.~C., {Kriessler}, J.~R., {Bird}, C.~M., \& {Huchra}, J.~P. 1995,
  \aj, 109, 874

\bibitem[{{Bekki} {et~al.}(2002){Bekki}, {Couch}, \&
  {Shioya}}]{Bekki_et_al_2002}
{Bekki}, K., {Couch}, W.~J., \& {Shioya}, Y. 2002, \apj, 577, 651

\bibitem[{{Bell} {et~al.}(2003){Bell}, {McIntosh}, {Katz}, \&
  {Weinberg}}]{Bell_et_al_2003}
{Bell}, E.~F., {McIntosh}, D.~H., {Katz}, N., \& {Weinberg}, M.~D. 2003, \apjs,
  149, 289

\bibitem[{{Bertin} \& {Arnouts}(1996)}]{Bertin&Arnouts_1996}
{Bertin}, E. \& {Arnouts}, S. 1996, \aaps, 117, 393

\bibitem[{{Binggeli} {et~al.}(1988){Binggeli}, {Sandage}, \&
  {Tammann}}]{LF_review}
{Binggeli}, B., {Sandage}, A., \& {Tammann}, G.~A. 1988, \araa, 26, 509

\bibitem[{{Biviano} {et~al.}(1997){Biviano}, {Katgert}, {Mazure}, {Moles}, {den
  Hartog}, {Perea}, \& {Focardi}}]{Biviano_et_al_1997}
{Biviano}, A., {Katgert}, P., {Mazure}, A., {et~al.} 1997, \aap, 321, 84

\bibitem[{{Blanton} {et~al.}(2003){Blanton}, {Hogg}, {Bahcall}, {Baldry},
  {Brinkmann}, {Csabai}, {Eisenstein}, {Fukugita}, {Gunn}, {Ivezi{\'c}},
  {Lamb}, {Lupton}, {Loveday}, {Munn}, {Nichol}, {Okamura}, {Schlegel},
  {Shimasaku}, {Strauss}, {Vogeley}, \& {Weinberg}}]{Blanton_et_al_2003}
{Blanton}, M.~R., {Hogg}, D.~W., {Bahcall}, N.~A., {et~al.} 2003, \apj, 594,
  186

\bibitem[{{Blanton} {et~al.}(2005){Blanton}, {Schlegel}, {Strauss},
  {Brinkmann}, {Finkbeiner}, {Fukugita}, {Gunn}, {Hogg}, {Ivezi{\'c}}, {Knapp},
  {Lupton}, {Munn}, {Schneider}, {Tegmark}, \& {Zehavi}}]{Blanton_et_al_2005}
{Blanton}, M.~R., {Schlegel}, D.~J., {Strauss}, M.~A., {et~al.} 2005, \aj, 129,
  2562

\bibitem[{{Boselli} \& {Gavazzi}(2006)}]{Boselli&Gavazzi_2006}
{Boselli}, A. \& {Gavazzi}, G. 2006, \pasp, 118, 517

\bibitem[{{Cardelli} {et~al.}(1989){Cardelli}, {Clayton}, \&
  {Mathis}}]{Cardelli_et_al_1989}
{Cardelli}, J.~A., {Clayton}, G.~C., \& {Mathis}, J.~S. 1989, \apj, 345, 245

\bibitem[{{Chilingarian} \& {Zolotukhin}(2011)}]{Chilingarian&Zolotukhin_2011}
{Chilingarian}, I. \& {Zolotukhin}, I. 2011, ArXiv e-prints

\bibitem[{{Colless} {et~al.}(2001){Colless}, {Dalton}, {Maddox}, {Sutherland},
  {Norberg}, {Cole}, {Bland-Hawthorn}, {Bridges}, {Cannon}, {Collins}, {Couch},
  {Cross}, {Deeley}, {De Propris}, {Driver}, {Efstathiou}, {Ellis}, {Frenk},
  {Glazebrook}, {Jackson}, {Lahav}, {Lewis}, {Lumsden}, {Madgwick}, {Peacock},
  {Peterson}, {Price}, {Seaborne}, \& {Taylor}}]{2dFGRS}
{Colless}, M., {Dalton}, G., {Maddox}, S., {et~al.} 2001, \mnras, 328, 1039

\bibitem[{{Cowie} \& {Songaila}(1977)}]{Cowie&Songaila_1977}
{Cowie}, L.~L. \& {Songaila}, A. 1977, \nat, 266, 501

\bibitem[{{{Cox}, A. N.}(2000)}]{Cox_2000}
{{Cox}, A. N.}, ed. 2000, {Allen's Astrophysical Quantities} (Springer-Verlag
  New York, Inc.)

\bibitem[{{{Cutri} et al.}(2001)}]{Cutri_et_al_2001}
{{Cutri} et al.} 2001, {Explanatory Supplement to the 2MASS Second Incremental
  Data Release},
  {\it{http://www.ipac.caltech.edu/2mass/releases/second/doc/explsup.html}}

\bibitem[{{De Propris} {et~al.}(2004){De Propris}, {Colless}, {Peacock},
  {Couch}, {Driver}, {Balogh}, {Baldry}, {Baugh}, {Bland-Hawthorn}, {Bridges},
  {Cannon}, {Cole}, {Collins}, {Cross}, {Dalton}, {Efstathiou}, {Ellis},
  {Frenk}, {Glazebrook}, {Hawkins}, {Jackson}, {Lahav}, {Lewis}, {Lumsden},
  {Maddox}, {Madgwick}, {Norberg}, {Percival}, {Peterson}, {Sutherland}, \&
  {Taylor}}]{De_Propris_et_al_2004}
{De Propris}, R., {Colless}, M., {Peacock}, J.~A., {et~al.} 2004, \mnras, 351,
  125

\bibitem[{{Ellingson} {et~al.}(2001){Ellingson}, {Lin}, {Yee}, \&
  {Carlberg}}]{Ellingson_et_al_2001}
{Ellingson}, E., {Lin}, H., {Yee}, H.~K.~C., \& {Carlberg}, R.~G. 2001, \apj,
  547, 609

\bibitem[{{Fairley} {et~al.}(2002){Fairley}, {Jones}, {Wake}, {Collins},
  {Burke}, {Nichol}, \& {Romer}}]{Fairley_et_al_2002}
{Fairley}, B.~W., {Jones}, L.~R., {Wake}, D.~A., {et~al.} 2002, \mnras, 330,
  755

\bibitem[{{Finlator} {et~al.}(2000){Finlator}, {Ivezi{\'c}}, {Fan}, {Strauss},
  {Knapp}, {Lupton}, {Gunn}, {Rockosi}, {Anderson}, {Csabai}, {Hennessy},
  {Hindsley}, {McKay}, {Nichol}, {Schneider}, {Smith}, {York}, \& {the SDSS
  Collaboration}}]{Finlator_et_al_2000}
{Finlator}, K., {Ivezi{\'c}}, {\v Z}., {Fan}, X., {et~al.} 2000, \aj, 120, 2615

\bibitem[{{Finn} {et~al.}(2005){Finn}, {Zaritsky}, {McCarthy}, {Poggianti},
  {Rudnick}, {Halliday}, {Milvang-Jensen}, {Pell{\'o}}, \&
  {Simard}}]{Finn_et_al_2005}
{Finn}, R.~A., {Zaritsky}, D., {McCarthy}, Jr., D.~W., {et~al.} 2005, \apj,
  630, 206

\bibitem[{{Fukugita} {et~al.}(1996){Fukugita}, {Ichikawa}, {Gunn}, {Doi},
  {Shimasaku}, \& {Schneider}}]{Fukugita_et_al_1996}
{Fukugita}, M., {Ichikawa}, T., {Gunn}, J.~E., {et~al.} 1996, \aj, 111, 1748

\bibitem[{{Goto}(2005)}]{Goto_2005}
{Goto}, T. 2005, \mnras, 357, 937

\bibitem[{{Gunn} \& {Gott}(1972)}]{Gunn&Gott_1972}
{Gunn}, J.~E. \& {Gott}, III, J.~R. 1972, \apj, 176, 1

\bibitem[{{Haines} {et~al.}(2006){Haines}, {La Barbera}, {Mercurio},
  {Merluzzi}, \& {Busarello}}]{Haines_et_al_2006}
{Haines}, C.~P., {La Barbera}, F., {Mercurio}, A., {Merluzzi}, P., \&
  {Busarello}, G. 2006, \apjl, 647, L21

\bibitem[{{Henry} \& {Arnaud}(1991)}]{Henry&Arnaud_1991}
{Henry}, J.~P. \& {Arnaud}, K.~A. 1991, \apj, 372, 410

\bibitem[{{Hern{\'a}ndez-Fern{\'a}ndez}(2011)}]{Hernandez-Fernandez_Tesis}
{Hern{\'a}ndez-Fern{\'a}ndez}, J.~D. 2011, PhD thesis, {Universidad de Granada}

\bibitem[{{{Hern{\'a}ndez-Fern{\'a}ndez} et
  al.}(2011{\natexlab{a}})}]{Hernandez-Fernandez_et_al_2011_NUVr}
{{Hern{\'a}ndez-Fern{\'a}ndez} et al.} 2011{\natexlab{a}}, {{\it Disentangling
  the role of environmental processes in galaxy clusters} (submitted)}

\bibitem[{{{Hern{\'a}ndez-Fern{\'a}ndez} et
  al.}(2011{\natexlab{b}})}]{Hernandez-Fernandez_et_al_2011}
{{Hern{\'a}ndez-Fern{\'a}ndez} et al.} 2011{\natexlab{b}}, {{\it Galaxy
  projected distribution in a unbiased sample of nearby clusters} (submitted)}

\bibitem[{{Jarrett} {et~al.}(2000){Jarrett}, {Chester}, {Cutri}, {Schneider},
  {Skrutskie}, \& {Huchra}}]{Jarrett_et_al_2000}
{Jarrett}, T.~H., {Chester}, T., {Cutri}, R., {et~al.} 2000, \aj, 119, 2498

\bibitem[{{Joint Iras Science}(1994)}]{IRAS_PSC}
{Joint Iras Science}, W.~G. 1994, VizieR Online Data Catalog, 2125, 0

\bibitem[{{Kaviraj} {et~al.}(2007){Kaviraj}, {Kirkby}, {Silk}, \&
  {Sarzi}}]{Kaviraj_E+A}
{Kaviraj}, S., {Kirkby}, L.~A., {Silk}, J., \& {Sarzi}, M. 2007, \mnras, 382,
  960

\bibitem[{{Kennicutt}(1998)}]{Kennicutt_1998}
{Kennicutt}, Jr., R.~C. 1998, \araa, 36, 189

\bibitem[{{Kewley} {et~al.}(2005){Kewley}, {Jansen}, \&
  {Geller}}]{Kewley_et_al_2005}
{Kewley}, L.~J., {Jansen}, R.~A., \& {Geller}, M.~J. 2005, \pasp, 117, 227

\bibitem[{{King}(1966)}]{King_1966}
{King}, I.~R. 1966, \aj, 71, 64

\bibitem[{{King}(1972)}]{King_1972}
{King}, I.~R. 1972, \apjl, 174, L123

\bibitem[{{Lupton} {et~al.}(2001){Lupton}, {Gunn}, {Ivezi{\'c}}, {Knapp}, \&
  {Kent}}]{Lupton_et_al_2001}
{Lupton}, R., {Gunn}, J.~E., {Ivezi{\'c}}, Z., {Knapp}, G.~R., \& {Kent}, S.
  2001, in Astronomical Society of the Pacific Conference Series, Vol. 238,
  Astronomical Data Analysis Software and Systems X, ed. {F.~R.~Harnden Jr.,
  F.~A.~Primini, \& H.~E.~Payne}, 269

\bibitem[{{{Lupton}, R.}(2005)}]{Lupton_2005}
{{Lupton}, R.} 2005, {Sloan to Johnson Photometric Transformations},
  {\it{http://www.sdss.org/DR6/algorithms/sdssUBVRITransform.html\#Rodgers2005%
}}

\bibitem[{{Madgwick} {et~al.}(2003){Madgwick}, {Hawkins}, {Lahav}, {Maddox},
  {Norberg}, {Peacock}, {Baldry}, {Baugh}, {Bland-Hawthorn}, {Bridges},
  {Cannon}, {Cole}, {Colless}, {Collins}, {Couch}, {Dalton}, {De Propris},
  {Driver}, {Efstathiou}, {Ellis}, {Frenk}, {Glazebrook}, {Jackson}, {Lewis},
  {Lumsden}, {Peterson}, {Sutherland}, \& {Taylor}}]{Madgwick_et_al_2003}
{Madgwick}, D.~S., {Hawkins}, E., {Lahav}, O., {et~al.} 2003, \mnras, 344, 847

\bibitem[{{Mahdavi} {et~al.}(2000){Mahdavi}, {B{\"o}hringer}, {Geller}, \&
  {Ramella}}]{Mahdavi_et_al_2000}
{Mahdavi}, A., {B{\"o}hringer}, H., {Geller}, M.~J., \& {Ramella}, M. 2000,
  \apj, 534, 114

\bibitem[{{Mahdavi} \& {Geller}(2001)}]{Mahdavi&Geller_2001}
{Mahdavi}, A. \& {Geller}, M.~J. 2001, \apjl, 554, L129

\bibitem[{{Margoniner} {et~al.}(2001){Margoniner}, {de Carvalho}, {Gal}, \&
  {Djorgovski}}]{Margoniner_et_al_2001}
{Margoniner}, V.~E., {de Carvalho}, R.~R., {Gal}, R.~R., \& {Djorgovski}, S.~G.
  2001, \apjl, 548, L143

\bibitem[{{Martin} {et~al.}(2005){Martin}, {Fanson}, {Schiminovich},
  {Morrissey}, {Friedman}, {Barlow}, {Conrow}, {Grange}, {Jelinsky},
  {Milliard}, {Siegmund}, {Bianchi}, {Byun}, {Donas}, {Forster}, {Heckman},
  {Lee}, {Madore}, {Malina}, {Neff}, {Rich}, {Small}, {Surber}, {Szalay},
  {Welsh}, \& {Wyder}}]{Martin_et_al_2005}
{Martin}, D.~C., {Fanson}, J., {Schiminovich}, D., {et~al.} 2005, \apjl, 619,
  L1

\bibitem[{{Mart{\'{\i}}nez} {et~al.}(2002){Mart{\'{\i}}nez}, {Zandivarez},
  {Dom{\'{\i}}nguez}, {Merch{\'a}n}, \& {Lambas}}]{Martinez_et_al_2002}
{Mart{\'{\i}}nez}, H.~J., {Zandivarez}, A., {Dom{\'{\i}}nguez}, M.,
  {Merch{\'a}n}, M.~E., \& {Lambas}, D.~G. 2002, \mnras, 333, L31

\bibitem[{{Mateo}(1998)}]{Mateo_1998}
{Mateo}, M.~L. 1998, \araa, 36, 435

\bibitem[{{Mihos}(2004)}]{Mihos_2004}
{Mihos}, J.~C. 2004, Clusters of Galaxies: Probes of Cosmological Structure and
  Galaxy Evolution, 277

\bibitem[{{Mobasher} {et~al.}(2003){Mobasher}, {Colless}, {Carter},
  {Poggianti}, {Bridges}, {Kranz}, {Komiyama}, {Kashikawa}, {Yagi}, \&
  {Okamura}}]{Mobasher_et_al_2003}
{Mobasher}, B., {Colless}, M., {Carter}, D., {et~al.} 2003, \apj, 587, 605

\bibitem[{{Moore} {et~al.}(1996){Moore}, {Katz}, {Lake}, {Dressler}, \&
  {Oemler}}]{Moore_et_al_1996}
{Moore}, B., {Katz}, N., {Lake}, G., {Dressler}, A., \& {Oemler}, A. 1996,
  \nat, 379, 613

\bibitem[{{Moore} {et~al.}(1998){Moore}, {Lake}, \& {Katz}}]{Moore_et_al_1998}
{Moore}, B., {Lake}, G., \& {Katz}, N. 1998, \apj, 495, 139

\bibitem[{{Moore} {et~al.}(1999){Moore}, {Lake}, {Quinn}, \&
  {Stadel}}]{Moore_et_al_1999}
{Moore}, B., {Lake}, G., {Quinn}, T., \& {Stadel}, J. 1999, \mnras, 304, 465

\bibitem[{{Moshir} {et~al.}(1993){Moshir}, {Copan}, {Conrow}, {McCallon},
  {Hacking}, {Gregorich}, {Rohrbach}, {Melnyk}, {Rice}, \&
  {Fullmer}}]{Moshir_1993}
{Moshir}, M., {Copan}, G., {Conrow}, T., {et~al.} 1993, VizieR Online Data
  Catalog, 2156, 0

\bibitem[{{Moustakas} {et~al.}(2006){Moustakas}, {Kennicutt}, \&
  {Tremonti}}]{Moustakas_et_al_2006}
{Moustakas}, J., {Kennicutt}, Jr., R.~C., \& {Tremonti}, C.~A. 2006, \apj, 642,
  775

\bibitem[{{Neugebauer} {et~al.}(1984){Neugebauer}, {Habing}, {van Duinen},
  {Aumann}, {Baud}, {Beichman}, {Beintema}, {Boggess}, {Clegg}, {de Jong},
  {Emerson}, {Gautier}, {Gillett}, {Harris}, {Hauser}, {Houck}, {Jennings},
  {Low}, {Marsden}, {Miley}, {Olnon}, {Pottasch}, {Raimond}, {Rowan-Robinson},
  {Soifer}, {Walker}, {Wesselius}, \& {Young}}]{Neugebauer_et_al_1984}
{Neugebauer}, G., {Habing}, H.~J., {van Duinen}, R., {et~al.} 1984, \apjl, 278,
  L1

\bibitem[{{Norberg} {et~al.}(2002){Norberg}, {Baugh}, {Hawkins}, {Maddox},
  {Madgwick}, {Lahav}, {Cole}, {Frenk}, {Baldry}, {Bland-Hawthorn}, {Bridges},
  {Cannon}, {Colless}, {Collins}, {Couch}, {Dalton}, {De Propris}, {Driver},
  {Efstathiou}, {Ellis}, {Glazebrook}, {Jackson}, {Lewis}, {Lumsden},
  {Peacock}, {Peterson}, {Sutherland}, \& {Taylor}}]{Norberg_et_al_2002}
{Norberg}, P., {Baugh}, C.~M., {Hawkins}, E., {et~al.} 2002, \mnras, 332, 827

\bibitem[{{Obri{\'c}} {et~al.}(2006){Obri{\'c}}, {Ivezi{\'c}}, {Best},
  {Lupton}, {Tremonti}, {Brinchmann}, {Ag{\"u}eros}, {Knapp}, {Gunn},
  {Rockosi}, {Schlegel}, {Finkbeiner}, {Ga{\'c}e{\v s}a}, {Smol{\v c}i{\'c}},
  {Anderson}, {Voges}, {Juri{\'c}}, {Siverd}, {Steinhardt}, {Jagoda},
  {Blanton}, \& {Schneider}}]{Obric_et_al_2006}
{Obri{\'c}}, M., {Ivezi{\'c}}, {\v Z}., {Best}, P.~N., {et~al.} 2006, \mnras,
  370, 1677

\bibitem[{{Poggianti} {et~al.}(2006){Poggianti}, {von der Linden}, {De Lucia},
  {Desai}, {Simard}, {Halliday}, {Arag{\'o}n-Salamanca}, {Bower}, {Varela},
  {Best}, {Clowe}, {Dalcanton}, {Jablonka}, {Milvang-Jensen}, {Pello},
  {Rudnick}, {Saglia}, {White}, \& {Zaritsky}}]{Poggianti_et_al_2006}
{Poggianti}, B.~M., {von der Linden}, A., {De Lucia}, G., {et~al.} 2006, \apj,
  642, 188

\bibitem[{{Quilis} {et~al.}(2000){Quilis}, {Moore}, \&
  {Bower}}]{Quilis_et_al_2000}
{Quilis}, V., {Moore}, B., \& {Bower}, R. 2000, Science, 288, 1617

\bibitem[{{Rines} {et~al.}(2003){Rines}, {Geller}, {Kurtz}, \&
  {Diaferio}}]{CAIRNS_I}
{Rines}, K., {Geller}, M.~J., {Kurtz}, M.~J., \& {Diaferio}, A. 2003, \aj, 126,
  2152

\bibitem[{{Rines} {et~al.}(2005){Rines}, {Geller}, {Kurtz}, \&
  {Diaferio}}]{CAIRNS_III}
{Rines}, K., {Geller}, M.~J., {Kurtz}, M.~J., \& {Diaferio}, A. 2005, \aj, 130,
  1482

\bibitem[{{S{\'e}rsic}(1963)}]{Sersic_1963}
{S{\'e}rsic}, J.~L. 1963, Boletin de la Asociacion Argentina de Astronomia La
  Plata Argentina, 6, 41

\bibitem[{{Smail} {et~al.}(1998){Smail}, {Edge}, {Ellis}, \&
  {Blandford}}]{Smail_et_al_1998}
{Smail}, I., {Edge}, A.~C., {Ellis}, R.~S., \& {Blandford}, R.~D. 1998, \mnras,
  293, 124

\bibitem[{{Springel} {et~al.}(2005){Springel}, {White}, {Jenkins}, {Frenk},
  {Yoshida}, {Gao}, {Navarro}, {Thacker}, {Croton}, {Helly}, {Peacock}, {Cole},
  {Thomas}, {Couchman}, {Evrard}, {Colberg}, \&
  {Pearce}}]{Springel_et_al_2005_Nat}
{Springel}, V., {White}, S.~D.~M., {Jenkins}, A., {et~al.} 2005, \nat, 435, 629

\bibitem[{{Strateva} {et~al.}(2001){Strateva}, {Ivezi{\'c}}, {Knapp},
  {Narayanan}, {Strauss}, {Gunn}, {Lupton}, {Schlegel}, {Bahcall}, {Brinkmann},
  {Brunner}, {Budav{\'a}ri}, {Csabai}, {Castander}, {Doi}, {Fukugita}, {Gy{\H
  o}ry}, {Hamabe}, {Hennessy}, {Ichikawa}, {Kunszt}, {Lamb}, {McKay},
  {Okamura}, {Racusin}, {Sekiguchi}, {Schneider}, {Shimasaku}, \&
  {York}}]{Strateva_et_al_2001}
{Strateva}, I., {Ivezi{\'c}}, {\v Z}., {Knapp}, G.~R., {et~al.} 2001, \aj, 122,
  1861

\bibitem[{{Strauss} {et~al.}(2002){Strauss}, {Weinberg}, {Lupton}, {Narayanan},
  {Annis}, {Bernardi}, {Blanton}, {Burles}, {Connolly}, {Dalcanton}, {Doi},
  {Eisenstein}, {Frieman}, {Fukugita}, {Gunn}, {Ivezi{\'c}}, {Kent}, {Kim},
  {Knapp}, {Kron}, {Munn}, {Newberg}, {Nichol}, {Okamura}, {Quinn}, {Richmond},
  {Schlegel}, {Shimasaku}, {SubbaRao}, {Szalay}, {Vanden Berk}, {Vogeley},
  {Yanny}, {Yasuda}, {York}, \& {Zehavi}}]{Strauss_et_al_2002}
{Strauss}, M.~A., {Weinberg}, D.~H., {Lupton}, R.~H., {et~al.} 2002, \aj, 124,
  1810

\bibitem[{{Struble} \& {Rood}(1991)}]{Struble&Rood_1991}
{Struble}, M.~F. \& {Rood}, H.~J. 1991, \apjs, 77, 363

\bibitem[{{Toniazzo} \& {Schindler}(2001)}]{Toniazzo&Schindler_2001}
{Toniazzo}, T. \& {Schindler}, S. 2001, \mnras, 325, 509

\bibitem[{{Wilman} {et~al.}(2005){Wilman}, {Balogh}, {Bower}, {Mulchaey},
  {Oemler}, {Carlberg}, {Morris}, \& {Whitaker}}]{Wilman_et_al_2005}
{Wilman}, D.~J., {Balogh}, M.~L., {Bower}, R.~G., {et~al.} 2005, \mnras, 358,
  71

\bibitem[{{Zabludoff} \& {Mulchaey}(1998)}]{Zabludoff&Mulchaey_1998}
{Zabludoff}, A.~I. \& {Mulchaey}, J.~S. 1998, \apj, 496, 39

\end{thebibliography}

\end{document}